\documentclass[a4paper,fleqn,usenatbib]{mnras}

\usepackage{newtxtext,newtxmath}

\usepackage[T1]{fontenc}
\usepackage{ae,aecompl}

\usepackage{graphicx}	
\usepackage{amsmath}	
\usepackage{amssymb}	

\def\loa{low-$\alpha\,$}
\def\hia{high-$\alpha\,$}
\def\lz{$L_{\rm z}\,$}
\def\jz{$J_{\rm z}\,$}
\def\jr{$J_{\rm r}\,$}

\newbox\grsign \setbox\grsign=\hbox{$>$} \newdimen\grdimen \grdimen=\ht\grsign
\newbox\simlessbox \newbox\simgreatbox
\setbox\simgreatbox=\hbox{\raise.5ex\hbox{$>$}\llap
     {\lower.5ex\hbox{$\sim$}}}\ht1=\grdimen\dp1=0pt
\setbox\simlessbox=\hbox{\raise.5ex\hbox{$<$}\llap
     {\lower.5ex\hbox{$\sim$}}}\ht2=\grdimen\dp2=0pt
\def\simgreater{\mathrel{\copy\simgreatbox}}
\def\simless{\mathrel{\copy\simlessbox}}
\newbox\simppropto
\setbox\simppropto=\hbox{\raise.5ex\hbox{$\sim$}\llap
     {\lower.5ex\hbox{$\propto$}}}\ht2=\grdimen\dp2=0pt

\title[Massive early accretion]{Evidence from APOGEE for the presence of 
a major building block of the halo buried in the inner Galaxy
}

\author[Horta et al.]
{Danny Horta$^1$,\thanks{E-mail: D.HortaDarrington@2018.ljmu.ac.uk}
Ricardo P. Schiavon$^1$,
J. Ted Mackereth$^{2,3,4}$,
Joel Pfeffer$^1$,
\newauthor
Andrew C. Mason$^1$,
Shobhit Kisku$^1$,
Francesca Fragkoudi$^5$,
Carlos Allende Prieto$^{6,7}$,
\newauthor
Katia Cunha$^{8,9}$,
Sten Hasselquist$^{10,11}$,
Jon Holtzman$^{12}$,
Steven R. Majewski$^{13}$,
\newauthor
David Nataf$^{14}$,
Robert W. O'Connell$^{13}$,
Mathias Schultheis$^{15}$,
Verne V. Smith$^{16}$
\bigskip
\\
$^1$Astrophysics Research Institute, Liverpool John Moores University, 146 Brownlow Hill, 
Liverpool, L3 5RF UK\\
$^{2}$School of Astronomy and Astrophysics, University of Birmingham, Edgbaston, Birmimgham, B15 2TT, UK\\
$^{3}$ Canadian Institute for Theoretical Astrophysics, University of Toronto, 60 St. George Street, Toronto, ON, M5S 3H8, Canada \\
$^{4}$ Dunlap Institute for Astronomy and Astrophysics, University of Toronto, 50 St. George Street, Toronto, ON M5S 3H4, Canada \\
$^5$Max-Planck-Institut f\"{u}r Astrophysik, Karl-Schwarzschild-Str. 1, 85741 Garching, Germany \\
$^6$ Instituto de Astrof\'isica de Canarias, E-38200 La Laguna,Tenerife,
Spain \\
$^7$ 16 Departamento de Astrof\'isica, Universidad de La Laguna, E-38206
La Laguna, Tenerife, Spain \\
$^8$ University of Arizona, Tucson, AZ 85719, USA \\
$^9$ Observat\'orio Nacional, S\~ao Crist\'ov\~ao, Rio de Janeiro, Brazil
\\
$^{10}$ Department of Physics \& Astronomy, University of Utah, Salt Lake
City, UT, 84112, USA \\
$^{11}$ NSF Astronomy and Astrophysics Postdoctoral Fellow \\
$^{12}$ New Mexico State University, Las Cruces, NM 88003, USA\\
$^{13}$ Department of Astronomy, University of Virginia, Charlottesville, VA 22904-4325, USA\\
$^{14}$ Center for Astrophysical Sciences and Department of Physics and Astronomy,
The Johns Hopkins University,
Baltimore, MD 21218 \\
$^{15}$ Universit\'e C\^ote d'Azur, Observatoire de la C\^ote d'Azur, 
Laboratoire Lagrange, CNRS, Blvd de l'Observatoire, F-06304 Nice, France \\
$^{16}$ National Optical Astronomy Observatories, Tucson, AZ 85719, USA
}

\date{Accepted XXX. Received YYY; in original form ZZZ}

\pubyear{2020}

\begin{document}
\label{firstpage}
\pagerange{\pageref{firstpage}--\pageref{lastpage}}
\maketitle

\begin{abstract} 

We report evidence from APOGEE for the presence of a new metal-poor
stellar structure located within $\sim$4~kpc of the Galactic centre.
Characterised by a chemical composition resembling those of low
mass satellites of the Milky Way, this new inner Galaxy structure
(IGS) seems to be chemically and dynamically detached from more
metal-rich populations in the inner Galaxy.  We conjecture that
this structure is associated with an accretion event that likely
occurred in the early life of the Milky Way.  Comparing the mean
elemental abundances of this structure with predictions from
cosmological numerical simulations, we estimate that the progenitor
system had a stellar mass of $\sim5\times10^8M_\odot$, or approximately
twice the mass of the recently discovered Gaia-Enceladus/Sausage
system.  We find that the accreted:{\it in situ} ratio within our
metal-poor ([Fe/H]$<$--0.8) bulge sample is somewhere between 1:3
and 1:2, confirming predictions of cosmological numerical simulations
by various groups.

\end{abstract}

\begin{keywords}
Galaxy: formation -- Galaxy: halo -- Galaxy: chemistry and kinematics
\end{keywords}



\section{Introduction}

According to the prevailing formation paradigm, galaxy mass assembly
takes place in great measure through the accretion of low mass
structures.  It follows naturally that the accretion history of a
galaxy plays a key role in defining how its various components are
organized today.  Our own Galaxy is no exception.  Much of the
accretion activity of the Milky Way has happened in the early stages
of its formation, and the footprints are frozen in the chemo-dynamical
record of its halo stellar populations.  Since the seminal papers
by \cite{ELS} and \cite{SearleZinn1978}, many groups sought to
characterise halo stellar populations, associating them with {\it
in situ} formation or an accretion origin.  Detection of substructure
in phase space has worked very well for the identification of ongoing
and/or recent accretion events
\citep[e.g.,][]{Ibata1994,Helmi1999,Majewski2003,Belokurov2006}.  However, phase
mixing makes the identification of early accretion activity that
occurred during the first few Gyr of the Milky Way's life more
difficult, requiring additional information, usually in the form
of detailed chemistry
\cite[e.g.,][]{NissenSchuster2010,Hayes2018,Mackereth2019}.

The combination of data from the \emph{Gaia} satellite \citep{Gaiadr2}
and chemistry from massive high resolution spectroscopic surveys
\citep[e.g.,][]{GES,Majewski2017,GALAH} is transforming this field.
In the past few years, several distinct satellite accretions have
been suggested by various groups, including the Gaia-Enceladus/Sausage
system \citep[][]{Haywood2018,Helmi2018,Belokurov2018,Mackereth2019},
Sequoia \citep{Myeong2019}, the {\it Kraken} \citep[][]{Kruijssen2019b},
and Thamnos 1 and 2 \citep{Koppelman2019}.  More recently, a number
of additional structures were identified by \cite{Naidu2020} on the
basis of data collected as part of the H3 survey \citep{Conroy2019}.
While the reality of some of those events still needs to be fully
established \citep[e.g.,][]{JeanBaptiste2017,Koppelman2020}, it
certainly is clear that we are just scratching the surface and the
field is ripe for exciting new findings in the near future.

The central few kpc of the Galactic halo are obviously extremely
important when it comes to retelling the early accretion history
of the Milky Way and discerning the contribution of {\it in situ}
formation to the stellar halo mass.  Assuming a single power law
density profile with exponent $\alpha=-2.96$
\citep[e.g.,][]{Horta2020b,Iorio2018} with spherical symmetry,
roughly 50\% of the halo stellar mass (out to $R_{GC}\sim$30~kpc)
is located within $\sim$3~kpc of the Galactic centre.  Moreover the
inner few kpc of the Galaxy is the region one would expect to host
most of the early {\it in situ} halo star formation, including the
oldest stars in the Galaxy \citep[e.g.,][]{Tumlinson2010,Savino2020},
as well as the remnants of early accretion events driven there by
dynamical friction \citep[e.g.,][]{Tremaine1975,Pfeffer2020}.
Observational access to inner halo populations is however quite
difficult.  They inhabit a region of the Milky Way that by convention
is referred to as the Galactic bulge, whose stellar population
content consists of a superposition of structures that are far more
densely populated than the halo itself, including the thick and
thin disks, and the bar
\citep[e.g.,][]{Ness2013a,Ness2013b,Rich2013,Nataf2017,Barbuy2018}.
Moreover, observations of the Galactic bulge are further hampered
by severe extinction towards the inner several degrees from the
Galactic centre.

Since the groundbreaking work by \cite{Rich1988}, a truly herculean
effort has been invested by several groups towards mapping the
stellar population content of the Galactic bulge, mostly based on
optical spectroscopy
\cite[e.g.,][]{McWilliam1994,Ibata1994,Zoccali2006,Fulbright2007,
Howard2008,Gonzalez2011, RichOriglia2012,
GarciaPerez2013,Rojas2014,Ryde2016,Rojas2017} \footnote{This is of
course an incomplete list.  For a more comprehensive account of
previous observational work see reviews by \cite{Rich2013},
\cite{Nataf2017} and \cite{Barbuy2018}.}.  However, until very
recently, the detailed chemistry of the {\it metal-poor} population
inhabiting the inner few kpc of the Galactic centre has been poorly
characterized, with most studies being based on small samples with
metallicities below [Fe/H]~$\sim-1$.  With the advent of the Apache
Point Observatory Galactic Evolution Experiment \citep{Majewski2017},
detailed chemistry (combined with precision radial velocities) has
recently been obtained for a statistically significant sample of
metal-poor stars within a few kpc of the Galactic centre.  A number
of studies resulted from that database.  \cite{Nidever2012} identified
cold velocity peaks in the radial velocity distribution of stars
within $\sim20^\circ$ of the Galactic centre, which may be associated
with resonant bar orbits \citep{Molloy2015}.  \cite{GarciaPerez2013}
identified some of the most metal-poor stars known in the Galactic
bulge; \cite{Zasowski2016} showed that the kinematics of metal-poor
bulge stellar populations is dynamically hotter than that of their
metal-rich counterparts; \cite{Schultheis2017} found evidence for
the presence of a young component among the bulge stellar populations,
in agreement with previous work; \cite{Schiavon2017} discovered a
large number of field stars likely resulting from the destruction
of an early population of globular clusters; \cite{GarciaPerez2018}
studied the metallicity distribution function (MDF) of bulge
populations, and its spatial variation; \cite{Zasowski2019} scrutinised
the detailed chemistry of bulge populations, showing that they seem
to be consistent with a single evolutionary track, across a wide
metallicity range; \cite{Rojas2019} suggests the presence of a
bimodal distribution of bulge stellar populations in the Mg-Fe
plane, and \cite{Rojas2020} argues for a bulge MDF that is characterised
by only three metallicity peaks; \cite{Hasselquist2020} presented
the age distribution of several thousand bulge stars, showing the
presence of a sizeable population with moderately young ages; last,
but not least, \cite{Queiroz2020b} made a detailed
characterization of bulge stellar populations on the basis of their
kinematics and chemistry.

In this paper we report evidence based on APOGEE spectroscopy,
distances based on \emph{Gaia} DR2, and EAGLE numerical simulations,
for the presence of a metal-poor stellar population within $\sim$4~kpc
of the Galactic centre, that is both chemically and dynamically
distinct from its co-spatial, more metal-rich counterparts.  We
argue that this population is likely the remnant of a massive
accretion event that probably occurred in the early stages of the
Milky Way formation.  Due to its location at the heart of the Galaxy,
we name this system the Inner Galaxy Structure (IGS).  In
Section~\ref{sec:data} we briefly describe the data upon which this
work is based, and our sample selection criteria.  In
Section~\ref{sec:chem_distinct} we discuss the chemical criteria
to distinguish accreted from {\it in situ} populations. In
Section~\ref{sec:subiom} we discuss the identification of substructure
in integrals of motion (IoM) space, arguing that the newly identified
structure is dynamically detached from other components of the inner
Galaxy.  Section~\ref{sec:chemprop} presents a discussion of the
chemical properties of the new structure, where we provide evidence
that it is chemically detached from the other populations inhabiting
the inner Galaxy.  In Section~\ref{sec:discussion} we discuss the
origins of the metal-poor stellar populations providing an assessment
of the contribution of accreted and {\it in situ} populations to
the stellar mass budget of the inner halo.  We also estimate physical
properties of the putative satellite progenitor of the IGS, such
as mass and star formation history.  Our conclusions are summarised
in Section~\ref{sec:conclusions}.

\section{Data and sample} \label{sec:data}

This paper is based on a combination of data from the SDSS/APOGEE
survey \citep{Blanton2017,Majewski2017}, data release 16
\citep[DR16,][]{DR16}, with 6D phase information and orbital parameters
inferred from \emph{Gaia} DR2 distances and proper motions
\citep{Gaiadr2} and APOGEE radial velocities \citep{Nidever2015}.
We make use of the distances from \cite{Leung2019b} for the APOGEE
DR16 data, generated using the \texttt{astroNN} python package
\citep[for a full description see][]{Leung2019a}. These distances
are determined using a previously trained \texttt{astroNN} neural-network,
which predicts stellar luminosity from spectra using a training set
comprising of stars with APOGEE spectra and \textit{Gaia} DR2
parallax measurements \citep{Gaiadr2}. The model is able to
simultaneously predict distances and account for the parallax offset
present in \textit{Gaia}-DR2, producing high precision, accurate
distance estimates for APOGEE stars, which match well with external
catalogues and standard candles.  Orbital parameters were calculated
using the publicly available code \texttt{galpy}
\footnote{https://github.com/jobovy/galpy}
\citep{Bovy2015,MackerethBovy2018}, and employing a \cite{McMillan2017}
potential.  We checked the quality of the Gaia astrometry for the
whole sample finding that, save for a very small number of outliers,
the renormalized unit weight error (RUWE) for our bulge sample is
within the safe recommended region (RUWE$<$1.44)\footnote{See
discussion in {\tt https://www.cosmos.esa.int/web/gaia/dr2-known-issues}},
which speaks for the reliability of our orbital parameters.

APOGEE data are based on observations collected with twin high
resolution multi-fiber spectrographs \citep{Wilson2019} attached
to the 2.5~m Sloan telescope at Apache Point Observatory \citep{Gunn2006}
and the du Pont 2.5~m telescope at Las Campanas Observatory
\citep{BowenVaughan1973}.  Elemental abundances are derived from
automatic analysis of stellar spectra using the ASPCAP abundance
pipeline \citep{GarciaPerez2016}.  The spectra themselves were
reduced by adoption of a customized pipeline \citep{Nidever2015}.
For a detailed description of the APOGEE data, see \citet{Jonsson2020} for DR16 and \cite{Holtzman2015,Holtzman2018}
and \cite{Jonsson2018} for previous data releases.  For details on
targeting procedures, see \cite{Zasowski2017}.

\begin{figure}
	\includegraphics[width=\columnwidth]{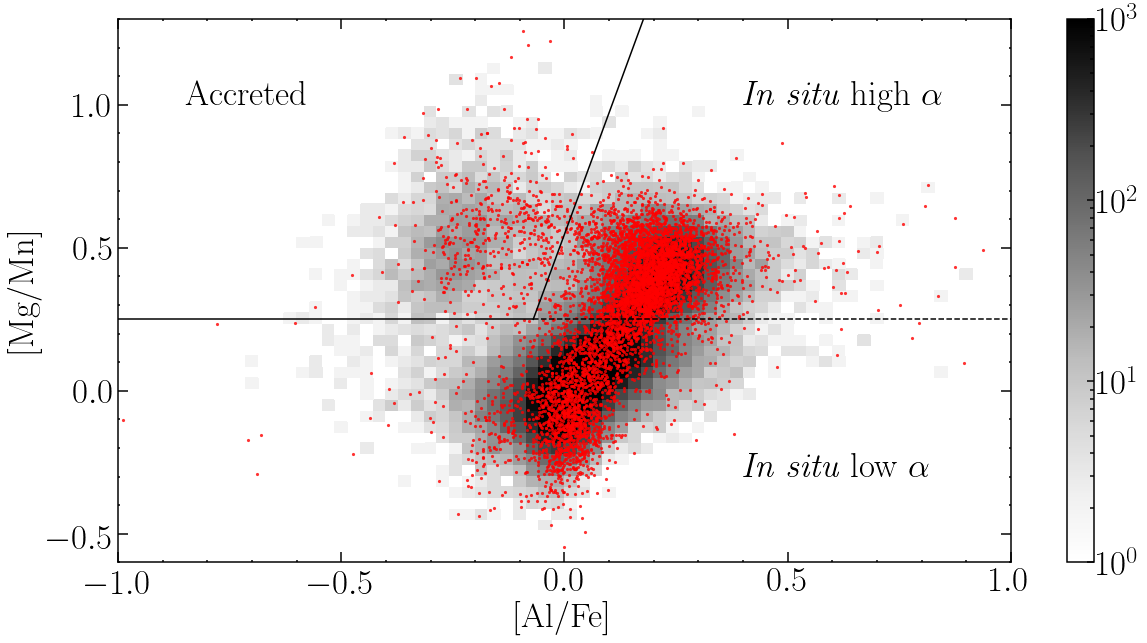}
    \caption{Parent and bulge sample displayed in the [Mg/Mn] vs. [Al/Fe] plane.  
The 2D histogram shows the entire parent sample, and the red dots highlight 
stars located at $R_{\rm GC}<4$~kpc.  The black solid line
defines our criterion to distinguish {\it in situ} from {\it accreted} 
populations.  The dotted line further splits the latter group between low- and
high-$\alpha$ stars.  See details in Section~\ref{sec:data}.}
    \label{fig:accdef}
\end{figure}



The parent sample upon which our work is based is defined as follows.
Stars were selected from the DR16 catalog that match the following
criteria: $4000<T_{\rm eff}<6000$~K, $1<\log{g}< 3$, [Fe/H]$>-1.7$,
S/N$>$70, $d_{\rm err}/d<0.2$ (where $d$ and $d_{\rm err}$ are
distance and its uncertainty).  The S/N and stellar parameter
criteria are designed to maximise the quality of the elemental
abundances, minimise systematic effects at low and high $T_{\rm
eff}$, minimise/eliminate contamination of the sample by AGB and
nearby dwarf stars at low and high $\log g$, respectively, and
remove stars with very low metallicity, for which abundances of
elements such as Mn and N are uncertain.  We also remove stars
for which ASPCAP did not provide reliable abundances for Fe, Mg,
Al, and Mn, and any stars with {\tt STAR\_BAD} flags set.  Finally,
6,022 globular cluster stars, identified following the procedure
described by \citet{Horta2020a}, were also removed from consideration.
The resulting parent sample contains 144,490 stars with high quality
elemental abundances and phase space information, 6,350 of which
are located within $R_{\rm GC}=4$~kpc of the Galactic centre.

\section{Chemical distinction between accreted and {\it in situ}
populations} \label{sec:chem_distinct}

Our first task is to distinguish stars within our parent sample
that were formed {\it in situ} from those that were likely accreted.
For that purpose we display the parent sample on the [Mg/Mn] vs.
[Al/Fe] plane in Figure~\ref{fig:accdef}.  \cite{Das2020} showed
that stars in the APOGEE DR14 sample occupied three major distinct
loci on this plane \citep[see also][]{Hawkins2015}.  Most of the
stars occupy two large concentrations which, in the DR16 data, are
centered at ([Al/Fe],[Mg/Mn]) $\sim$ (0,0) and (0.3,0.4), corresponding
to the low- and high-$\alpha$ disk populations, respectively
\citep[e.g.,][]{Bovy2012,Bensby2014,Nidever2014,Hayden2015,Mackereth2017,
Queiroz2020a}.  A third group of stars populates a more diffuse
``blob'' apparent in Figure~\ref{fig:accdef}, which is centered
roughly at ([Al/Fe],[Mg/Mn]) = (--0.2,0.5), which \cite{Das2020}
associate with an accreted origin.  That association was originally
proposed by \cite{Hawkins2015}, who identified accreted halo stars
as those with low [$\alpha$/Fe], and whose radial velocities deviate
strongly from those of the bulk of disk stars at the same Galactic
longitude.  The solid line in Figure~\ref{fig:accdef} defines the
locus occupied by accreted stars in that chemical plane (top left
corner).  According to \cite{Das2020},  stars to the right and
below that line are thus deemed to have formed {\it in situ}.  The
latter population is further divided into two sub-classes characterized
by low and high $\alpha$-element abundances, as indicated by the
dotted line in Figure~\ref{fig:accdef}.  In total our {\it in situ}
and {\it accreted} samples contain 141,514 and 2,976 stars,
respectively.

The red points on top of the 2D histogram in Figure~\ref{fig:accdef}
include only stars within 4~kpc of the Galactic centre.  A sizable
sub-sample of these so-called bulge stars are located in the accreted
locus of chemical space, suggesting that the bulge may host an
accreted population.  Our sample contains 463 supposedly accreted
stars within 4~kpc of the Galactic centre, modulo a contamination
by stars formed {\it in situ}, which we estimate by different means
in Sections~\ref{sec:chemevol} and \ref{sec:mp_orbs}.


\subsection{Predictions from chemical evolution models} \label{sec:chemevol}

Before proceeding with our analysis, it is interesting to examine
the association of abundance patterns and stellar origin on the
basis of standard Galactic chemical evolution modelling.  In this
way we hope to ground the empirical definition of accreted vs. {\it
in situ} chemistry on a theoretical basis.  

\begin{figure}
	\includegraphics[width=\columnwidth]{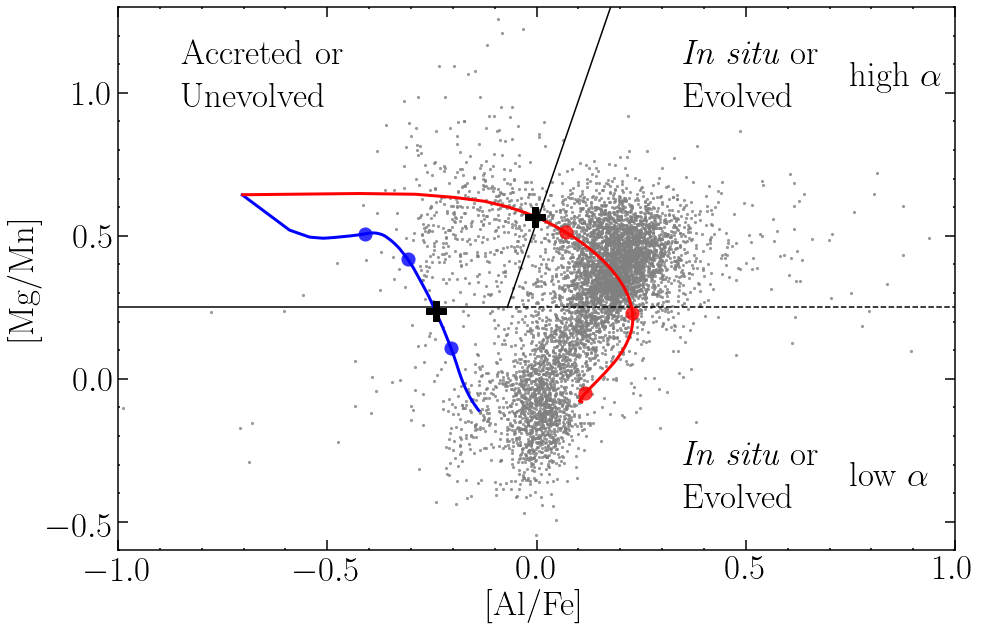}
    \caption{APOGEE sample for stars with $R_{\rm GC}<4$~kpc compared
with chemical evolution models.  The red line shows a model for a
Milky Way like galaxy and the blue line the model for a dwarf
satellite.   Both models start from the same chemical composition,
and initially occupy the region of parameter space usually ascribed
to ``Accreted'' stellar populations, but which also hosts {\it in
situ} populations that are chemically unevolved.  The filled circles
indicate the positions along the evolution after $t=$~0.3, 1, and
5~Gyr after the beginning of chemical evolution.  The black cross
in both curves indicates the point at which [Fe/H]=--0.8.  Evolution
of the massive galaxy is a lot faster, but the star formation rate
is much higher, so the two systems are expected to produce similar
total stellar mass within the ``Accreted'' area of the diagram.  }
    \label{fig:modelcomp}
\end{figure}

Figure~\ref{fig:modelcomp} displays the bulge population on the
same chemical plane as Figure~\ref{fig:accdef}, but now overlaid
by two chemical evolution models starting from the same initial
chemical composition, calculated using the flexCE package by
\cite{Andrews2017}.  The red line shows the ``fiducial'' model
calculated by those authors, with parameters chosen to match the
properties of the stellar populations in the solar neighbourhood.
We adopt it as a proxy for the behaviour of an {\it in situ}
population.  The blue line represents a model calculated by Hasselquist
et al.\ (2020, in prep.) to match the properties of the
Gaia-Enceladus/Sausage system, which we therefore choose to mimic
the expected behaviour of accreted populations.  The parameters
adopted for each model are listed in Table~\ref{tab:chemod}.  While
the fiducial model adopts a standard exponentially decaying inflow
law, for the accreted population gas inflow follows a dependence
of the $te^{-t}$ type.  Both models adopt a \cite{Kroupa2001} IMF
with stellar mass ranging from 0.1 to 100~$M_\odot$.  In both models
the distribution of time delays before the occurrence of SN Ia is
an exponential with timescale 1.5~Gyr and a minimum delay time of
150~Myr. Tests were performed to verify that the overall mass scaling
is not important for chemical evolution.  The deciding factors in
fact are the ratio between initial and inflow masses, the outflow
mass loading factor, and the star formation efficiency.

\begin{table}
\centering
\caption{Parameters adopted for chemical evolution models in
Figure~\ref{fig:modelcomp}}
\label{tab:chemod}
\begin{tabular}{lcc} 
	\hline
	\hline
	Parameter & {\it In Situ} & Accreted \\
	\hline
	Initial gass mass & $2\times10^{10} M_\odot$ & $3\times10^{9} M_\odot$	\\
	Inflow mass scale  & $3.5\times10^{11} M_\odot$ & $6\times10^{10} M_\odot$	\\
	Outflow mass loading factor & 2.5 & 6 \\
	Star formation efficiency & $1.5\times10^{-9}$yr$^{-1}$ &
$1.0\times10^{-10}$yr$^{-1}$ \\
	Exponential inflow timescale & 6.0 Gyr & 2.5 Gyr  \\
	\hline
\end{tabular}
\end{table}
%
%

The filled circles along the model lines indicate the positions at
evolutionary time $t$~=~300~Myr, 1~Gyr, and 5~Gyr.  The black cross
on both curves indicates the point at which the models reach
[Fe/H]=--0.8.  The models are far from a perfect match to the data,
but provide a reasonable qualitative description of the main trends,
which suffices for our purposes.  

The first point to be taken from this model comparison is that the
``Accreted'' region of chemical space in Figure~\ref{fig:accdef}
is inhabited by old metal-poor stars from both models, suggesting that
both {\it in situ} and accreted stars share that area of chemical
space.  In fact, the accreted stars in halo samples are located in
that particular locus of Figure~\ref{fig:accdef} not due to an
intrinsic evolutionary property of dwarf galaxies, but rather because
their star formation was quenched at the moment when they were
disrupted while merging into the Milky Way.  As a result, all the
stars ever formed by such accreted systems inhabit the {\it unevolved}
region of Figure~\ref{fig:accdef}.  The natural implication of this
conclusion is that the ``Accreted'' region of the chemical plane
in Figure~\ref{fig:accdef} is likely to also contain stars that
were in fact formed {\it in situ}.

Secondly, one can try to exploit the different timescales and star
formation rates of the two models to attempt a zero-th order estimate
of the expected contribution of {\it in situ} and accreted populations
to our sample within the ``Accreted'' region of chemical space.  It
is clear that chemical enrichment proceeds at a much faster pace
in the {\it in situ} model, whose gas leaves the ``Accreted'' region
of the chemical plane at $t\sim250$~Myr.  Conversely, the gas in
the low mass galaxy remains within the ``Accreted'' locus for about
$\sim$3~Gyr.  On the other hand, the star formation rate at the
early stages of the evolution of the high mass galaxy, according
to the models, is of the order of $\sim20~M_\odot$~yr$^{-1}$, whereas
in the case of the dwarf galaxy it is $\sim0.3~M_\odot$~yr$^{-1}$.
Therefore, considering the characteristic star formation rates and
timescales for the two types of systems, one would expect that the
total stellar mass associated with {\it in situ} and accreted
populations in the ``Accreted'' region of Figure~\ref{fig:accdef}
to be of the same order of magnitude.  In Section~\ref{sec:ins_vs_acc}
we provide a more accurate estimate of the contribution of {\it in
situ} and accreted populations to the stellar mass budget of the
metal-poor bulge, based on their distributions in the $E-L_{\rm z}$
plane.

We conclude that, on the basis of position on the [Mg/Mn] vs. [Al/Fe]
plane, it is in principle impossible to completely separate {\it
in situ} from accreted stars.  However, we argue that the clumpy
nature of the distribution of stellar populations in
Figure~\ref{fig:accdef} speaks in favour of a different origin for
at least part of the stellar populations in the ``Accreted'' region
of that plane, and those in the {\it in situ} region.  In
Section~\ref{sec:mp_orbs} we show that by imposing selection criteria
based on orbital parameters it is possible to keep the contamination
of our accreted sample by stars formed {\it in situ} to an acceptable
level.


\section{Detection of Substructure in IoM Space} \label{sec:subiom}

\begin{figure}
	\includegraphics[width=\columnwidth]{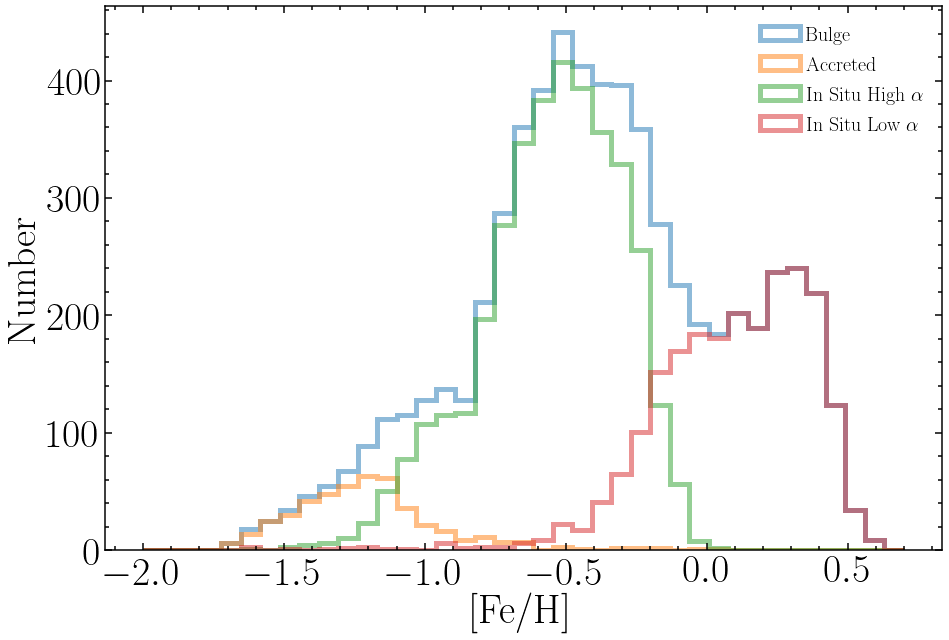}
\caption{Raw metallicity distribution functions of accreted and
{\it in situ} populations within $R_{\rm GC}<$~4~kpc, selected as described
in the text.  The accreted
population straddles the metal-poor end of the bulge MDF, with
[Fe/H]$<-0.8$.  It overlaps strongly with {\it in situ} populations,
particularly in the high metallicity end.  }
    \label{fig:MDFs}
\end{figure}

\subsection{Raw MDFs and the definition of our metal-poor sample} \label{sec:rMDFs}

In this section we examine the orbital properties of accreted and
{\it in situ} stellar populations, as defined in the previous
section.  We begin this exercise by focusing on the orbital properties
of {\it metal-poor} populations, with an eye towards establishing
whether the chemically defined accreted and {\it in situ} groups
also differ in their orbital properties.  For that purpose, we must
first define where we draw the line between metal-poor and metal-rich
populations.

We begin by examining the metallicity distribution of the sub-populations
defined in Figure~\ref{fig:accdef} in order to determine the [Fe/H]
threshold defining ``metal-poor'' stars for the purposes of our
exercise.  The {\it raw} metallicity distribution functions of the
accreted, high- and low-$\alpha$ populations are displayed in
Figure~\ref{fig:MDFs}, together with that for the entire $R_{\rm
GC}<4$~kpc population.  We stress that this {\it raw} MDF is not
corrected for selection effects, so it is presented here just as a
rough guide to the relative metallicities covered by the populations
defined in Figure~\ref{fig:accdef}.  The MDFs of the high- and
low-$\alpha$ populations cover the thin disk, bar, and thick disk
structures identified in previous works
\citep[e.g.,][]{Ness2013a,Rojas2017,GarciaPerez2018,Rojas2020}.
The MDF of the accreted population peaks at [Fe/H]$\sim$--1.2, and
overlaps substantially with that of the high-$\alpha$ population,
extending all the way to [Fe/H]$\sim$--0.8 with a tail towards higher
metallicity.  As a compromise between obtaining a good representation
of the accreted and high-$\alpha$ populations, without pushing too
far into the metal-rich regime, we adopt an [Fe/H]=--0.8 cutoff.
Henceforth, when referring to the "metal-poor" bulge, unless otherwise
noticed, we mean stars with [Fe/H]$<$--0.8.  In Section~\ref{sec:subchem}
we go beyond a simple MDF approach in order to better explore the
chemical complexity of bulge stellar populations.  Since our goal
in this Section is simply to contrast orbital properties of accreted
and {\it in situ} stars within the same metallicity regime, a
straight [Fe/H] cutoff should suffice.

\subsection{Orbital properties of metal-poor stars} \label{sec:mp_orbs}

\begin{figure}
	\includegraphics[width=\columnwidth]{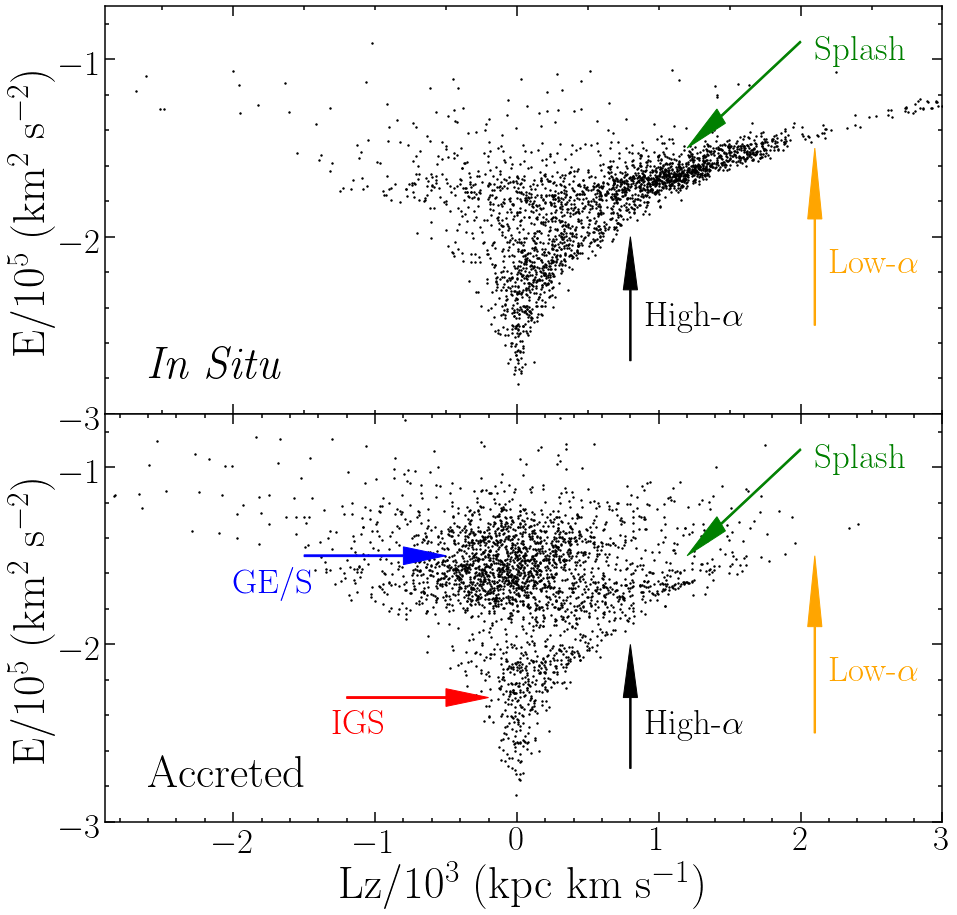}
\caption{Distribution of metal-poor ([Fe/H]$<$--0.8) bulge stars
in the energy-angular momentum plane.  The {\it top panel} shows
{\it in situ} stars and their accreted counterparts are displayed
in the {\it bottom panel}.  Arrows indicate the positions of the
main structures visible on this plane: thin disk (orange arrow), thick
disk (black arrow), {\it Splash} (green arrow), Gaia-Enceladus/Sausage
(GE/S, blue arrow), and the inner Galaxy Structure (IGS, red
arrow).  The low- and high-$\alpha$ disks and the {\it Splash} show prominently
in the {\it in situ} population (top) panel, whereas the GE/S and
IGS can be clearly distinguished in the accreted population (bottom
panel).}
    \label{fig:ELz_mp}
\end{figure}

In Figure~\ref{fig:ELz_mp} metal-poor stars ([Fe/H]$<$--0.8) are
displayed on the energy vs angular momentum plane.  In the top panel
stars in the {\it in situ} locus of Figures~\ref{fig:accdef} and
\ref{fig:modelcomp} are displayed, whereas accreted stars are shown
in the bottom panel.  The [Fe/H]$<$--0.8 cutoff leaves us with a
sub-sample of 2,688 accreted stars and 2,488 {\it in situ} stars.
Of the latter, only 179 belong to the low-$\alpha$ group, as expected
given the MDF of the low-$\alpha$ disk \citep[e.g.,][]{Hayden2015}.
Arrows of different colours are placed on the positions of the main
{\it in situ} and accreted structures identifiable in the $E-L_{\rm
z}$ plane, which stand out in the top and bottom panels, respectively.
The positioning of the arrows in both panels are exactly the same,
to guide the eye.

One can readily spot significant differences between the two panels,
which at face value validate the chemical distinction between
accreted and {\it in situ} populations.  Features associated with
well known {\it in situ} populations show much more prominently in
the top panel, and nearly vanish in the bottom panel.  The main
{\it in situ} structures identifiable are the {\it low-$\alpha$/thin
disk}, indicated by the {\it orange} arrow, {\it high-$\alpha$/thick
disk} stars are indicated by the {\it black} arrow.  An extended
prograde ``branch'' at about $E/10^5\sim-1.75$~km$^2$s$^{-2}$, which
seems to be associated with the ``Splash'' population identified
by \cite{Belokurov2019}, is indicated by the {\it green} arrow.
Because of the focus on the metal-poor end of the parent sample,
these structures are relatively weak, but in the Appendix the loci
of these populations on the IoM place are shown in better detail.

The clumpy distribution of accreted populations in IoM space is
clear even under a casual visual inspection of the bottom panel of
Figure~\ref{fig:ELz_mp}.  By far the most prominent substructure
is the large ``blob'' centered around $L_{\rm z}\sim$~0~km~s$^{-1}$kpc
and $E/10^5\sim-1.5{\rm km^2}~{\rm s^{-2}}$ ({\it blue arrow}) which
is associated with the Gaia-Enceladus/Sausage (GE/S) system
\citep{Haywood2018,Belokurov2018,Helmi2018,Mackereth2019}, and is
largely absent in the top panel where {\it in situ} populations
are displayed.  Additional substructures can be discerned in the
bottom panel, which are absent in the top panel of Figure~\ref{fig:ELz_mp}.
The focus of this paper is a low energy clump centred at $L_{\rm
z}/10^3\sim$0.1~km~s$^{-1}$kpc ({\it red arrow}) with energies in
the interval $-2.6\simless E/10^5\simless-2.0~{\rm km^2s^{-2}}$,
whose position coincides with that of the low energy (L-E) family
of globular clusters identified by \cite{Massari2019}.  The bulk
of the stars associated with this group are located within $\sim$4~kpc
of the Galactic centre, so we refer to it as the Inner Galaxy
Structure (IGS).

An energy gap is clearly visible in the bottom panel at about
$E/10^5\sim-1.9$~km$^2$s$^{-2}$, which separates stars belonging
to the GE/S and the {\it Splash} from stars belonging to the IGS.
In the top panel, that gap is filled by stars belonging to the thick
disk and the {\it Splash}, which present a predominantly smooth
distribution, much as expected from numerical simulations of Milky
Way-like disks undergoing satellite accretion
\citep[e.g.,][]{JeanBaptiste2017}.  This energy gap is not entirely
empty in the accreted sample, which is partly due to measurement
errors and partly due to the fact that {\it in situ} structures,
in particular the high-$\alpha$ disk and the {\it Splash} make a
small, but noticeable, contribution to the data in the bottom panel.
This is to be expected on the basis of our discussion in
Section~\ref{sec:chem_distinct}, where we showed that standard
chemical evolution models predict {\it in situ} contamination in
our accreted sample.  Under this interpretation of the data, the
distribution of the chemically-defined accreted sample in $E-L_{\rm
z}$ space consists of two overlapping populations: one of them
consists of various clumps associated with accreted structures,
most prominently the GE/S and the IGS, and the other is comprised
of a predominantly smooth, metal-poor, residual contamination
associated with high-$\alpha$ disk stars, which include both the
unperturbed disk and the {\it Splash}.

These results suggest that the IGS is dynamically detached from
other metal-poor populations in the halo and disk. This is an
important result, so it is vital that the reality of this energy
gap is firmed up quantitatively.  To do that we measure the ratio
between the number of stars with energies within the gap and those
below it in both the accreted and {\it in situ} samples.  We consider
stars within the gap to have energies given by $-2.00 < E/10^5 <
-1.85$~km$^2$~s$^{-2}$ and those below it have $E/10^5 <
-2.00$~km$^2$~s$^{-2}$.  The ratio in the {\it in situ} sample is
0.40$\pm$0.04 and in the accreted sample 0.27$\pm$0.04, configuring
a difference that is significant at the 2$\sigma$ level.  We thus
conclude that the gap is real, and an indication of the presence
of real substructure in the distribution of accreted stars in the
low energy locus of IoM space.

Finally, it is worth mentioning that a relatively small number of
stars can be seen scattered at high energy levels
($E/10^5\simgreater-1.6$~km$^2$~s$^{-2}$) among the {\it in
situ} population (top panel), on a wide range of angular momenta.
These are predominantly low-$\alpha$ metal-poor stars and are
likely accreted contaminants in the {\it in situ} sample (c.f.
Figure~\ref{fig:modelcomp}). 

\begin{figure}
	\includegraphics[width=\columnwidth]{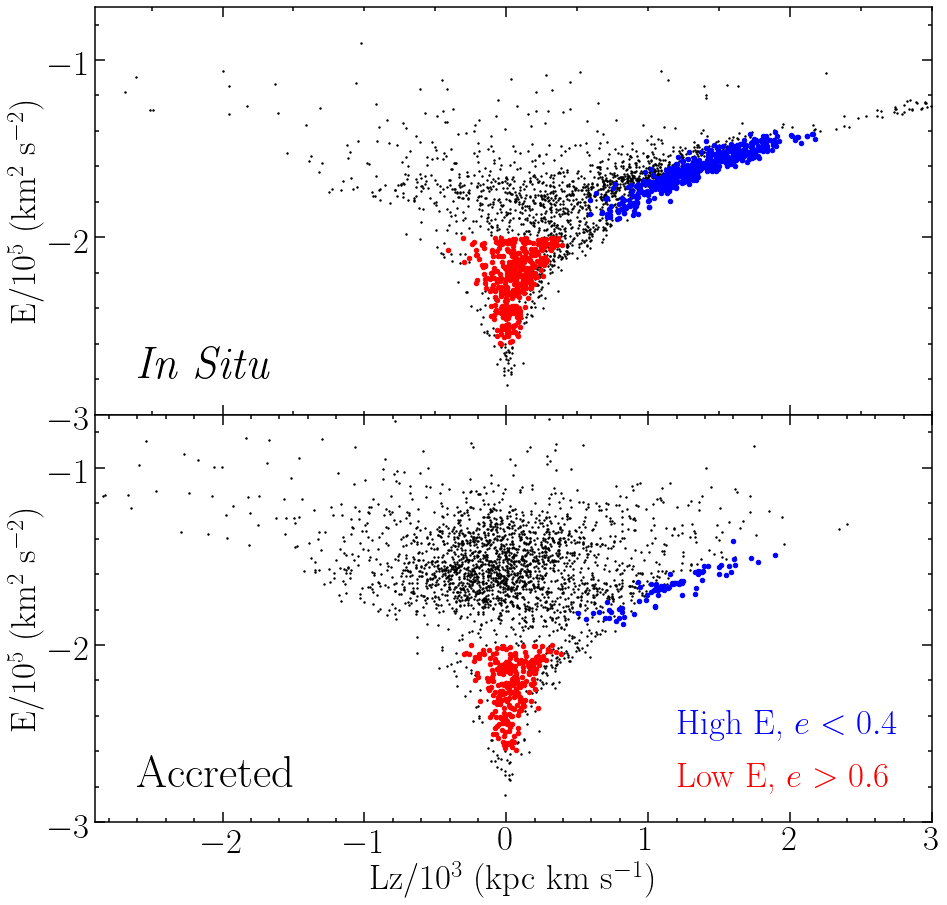}
\caption{Subsamples used to estimate the contamination of the IGS
sample by {\it in situ} stars.  In both panels, red symbols mark
stars located within the locus in orbital space defined by the IGS
stars.  Blue symbols indicate stars in a reference region of orbital
space occupied by high-$\alpha$ stars alone.  The ratio between
numbers of red and blue stars in the {\it in situ} population ({\it
top panel}) is used to infer the number of {\it in situ} contaminants
among the IGS stars (red symbols in the {\it bottom panel}).  We
estimate that somewhere between 22 and 40\% of stars in our IGS
sample are actually high-$\alpha$ disk contaminants.  }
    \label{fig:ELz_cont}
\end{figure}

\subsection{IGS definition and contamination by {\it in
situ} stars} \label{sec:contamination}

 As pointed out in the previous section, the IGS corresponds
to a low energy substructure visually discernible in IoM space.  In
order to define our IGS sample in an easily reproducible manner,
we initially ran an algorithm to identify data clustering in
multi-parameter space.  The algorithm of choice was HDBSCAN
\citep{Campelo2013} and the parameters adopted were energy, vertical
and radial action ($J_{\rm r}$ and $J_{\rm z}$), angular momentum
($L_{\rm z}$), and eccentricity.  We ran HDBSCAN on several
combinations of these parameters, varying the input parameters {\it
minimum cluster size} and {\it minimum samples} in such a way that
the structures visible in action space could be detected.  However,
the specific subsamples associated by HDBSCAN to the various
substructures are very sensitive to the input parameters adopted.
Therefore, we do not feel confident that HDBSCAN-selected samples
represent the distribution of the properties of the accreted systems
in action space in a statistically robust fashion.

We thus adopt simple, easily reproducible criteria to establish IGS
membership.  Stars belonging to the IGS are those belonging to the
``Accreted'' region of chemical space in Figure~\ref{fig:accdef},
which meet the following simple orbital energy and eccentricity
criteria:

\begin{itemize}

\item $-2.60 < E_{\rm IGS}/10^5 < -2.00$~km$^2$~s$^{-2}$

\item $e_{\rm IGS} > 0.6$

\end{itemize}

\noindent The energy criterion restricts the sample to the clump that is
visually identified in Figure~\ref{fig:ELz_mp}.  The eccentricity
criterion is aimed at minimising contamination by stars belonging
to the inner high-$\alpha$ disk (see Figure~\ref{fig:action_alpha}).  

As discussed in the previous section, examination of
Figure~\ref{fig:ELz_mp} reveals a discernible signature of {\it in
situ} structures  in the distribution of accreted stars in the
bottom panel of Figure~\ref{fig:ELz_mp}, which overlap the IGS.
Those are metal-poor {\it in situ} stars that inhabit the ``Accreted''
locus of the chemical plane in Figure~\ref{fig:accdef}, as discussed
in Section~\ref{sec:chemevol}.  We can use those {\it in situ}
features that are present in the distribution of accreted stars in
IoM as a means to estimate the contamination of {\it in situ} stars
in our IGS sample.

To assess that contamination, we use sub-samples of the chemically
defined ''Accreted'' and {\it in situ} samples, displayed in
Figure~\ref{fig:ELz_cont}, overlaid on the same data originally
displayed in Figure~\ref{fig:ELz_mp}.  We assume that the energy
distribution of stars in the high-$\alpha$ disk is smooth \citep[as
predicted by simulations, see, e.g.,][]{JeanBaptiste2017} and can
be estimated from the distribution of {\it in situ} stars in the
$E-L_{\rm z}$ plane (upper panel of Figure~\ref{fig:ELz_cont}).  We
further assume that that energy distribution is the same as that
of the contaminants in the accreted sample (bottom panel of
Figure~\ref{fig:ELz_cont}).  If those assumptions are correct, one
can calculate the ratio in the {\it in situ} sample between the
numbers of low energy stars in the IGS region, $n$(low E)$_{\it
InS}$ ({\it red symbols in the top panel} of Figure~\ref{fig:ELz_cont}),
and those in a reference high energy locus that is free from accreted
stars, $n$(high E)$_{\it InS}$ ({\it blue symbols in the top panel} of
Figure~\ref{fig:ELz_cont}).  That ratio can then be used to convert
the number of high energy disk stars in the accreted sample, $n$(high
E)$_{\it Acc}$ ({\it blue symbols in the bottom panel} of
Figure~\ref{fig:ELz_cont}), into the number of contaminating disk
stars in the IGS sample, $N_{\rm cont}$, as follows:

\begin{equation} 
N_{\rm cont} \,=\,
\left[\frac{n({\rm low\,E})}{n({\rm high\, E})}\right]_{\it InS}
\times \,\,\, n({\rm high \,E})_{\rm Acc} 
\end{equation} 

That number can then be compared to the actual number of stars
meeting the above defined criteria for the IGS in the accreted
region ({\it red symbols in the bottom panel} of Figure~\ref{fig:ELz_cont})
to estimate the contamination fraction.

For that purpose we must select a region of the $E-L_{\rm z}$ space
that provides a pure sample of {\it in situ} stars belonging
predominantly to the high-$\alpha$ disk.  We choose such a locus
by picking stars with eccentricity $<$ 0.4 and in the energy range
$-1.9 < E/10^5 < -1.4$~km$^2$~s$^{-2}$.  Stars in this reference
region and in the IGS locus defined above are displayed as blue and
red dots in Figure~\ref{fig:ELz_cont}.

\begin{table*}
\centering
\caption{Numbers for the assessment of the contamination of our IGS sample
by {\it in situ} stars.}
\label{tab:contamination}
\begin{tabular}{lccccc} 
	\hline
	\hline
	[Fe/H]$_{\rm cut}$ & $n$(low E)$_{\it InS}$ & $n$(high E)$_{\it InS}$ & 
                      $n$(low E)$_{\rm Acc}$ & $n$(high E)$_{\rm Acc}$ &
        $\%_{\rm cont}$ \\
	\hline
	--0.8 & 370 & 516 & 244 & 74 & 22\% \\
	--1.0 & 153 & 116 & 226 & 68 & 40\% \\
	\hline
\end{tabular}
\end{table*}

By proceeding in this fashion, we obtain the numbers listed in
Table~\ref{tab:contamination}, which mean that approximately 22\%
of our IGS sample actually consist of high-$\alpha$ disk stars.  A
fundamental systematic uncertainty in this procedure concerns the
metallicity cut adopted for the accreted and {\it in situ} samples.
The calculation above was based on the [Fe/H]=--0.8 definition of
metal-poor stars adopted throughout this paper.  However, if a
slightly more conservative cut is adopted at, say, [Fe/H]=--1.0,
the inferred contamination goes up to 40\%.  This is due to how the
numbers in the {\it in situ} population, particularly the reference
disk stars, depend on metallicity.  The angular momentum distribution
of disk stars in our sample is strongly dependent on metallicity.
At lower metallicity, the $L_{\rm z}$ distribution is shifted towards
lower values, leading to a higher $[n({\rm low E})/n({\rm high})]_{\it
In\, situ}$, and thus a higher number of estimated contaminants
($N_{\rm cont}$).


We therefore conclude that the contamination of our IGS sample by
high-$\alpha$ disk stars ranges between roughly 22 and 40\%.


%
%
%
%


\subsection{Orbital properties of bulge stars} \label{sec:IOM_bulge}

\begin{figure}
\includegraphics[width=\columnwidth]{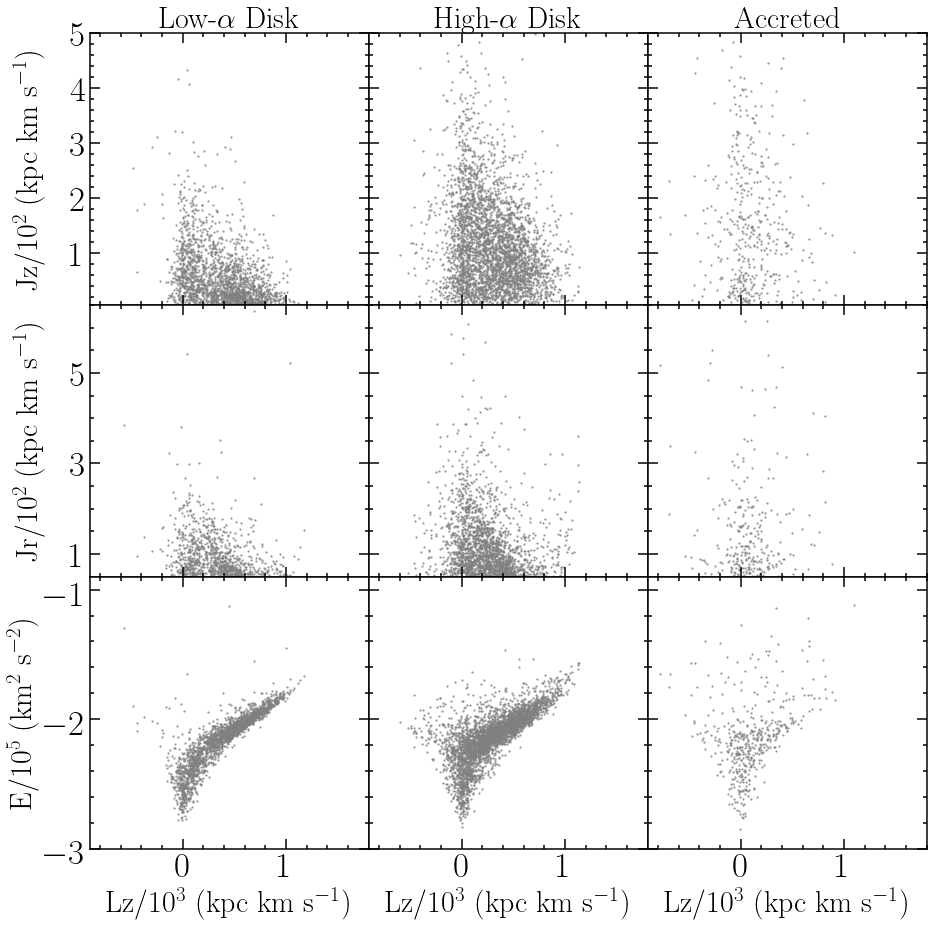}
\caption{Distribution of low-$\alpha$ (left panels), high-$\alpha$
(middle panels), and accreted stars (right panels) in action space.
Only stars within $R_{\rm GC}<4$~kpc are shown.  Chemically defined
accreted populations in the bulge have on average lower angular
momentum, and higher radial and vertical actions.
}
\label{fig:bulge_action}
\end{figure}

The bulk of the stars belonging to the substructure identified in
Section~\ref{sec:mp_orbs} are contained within 4~kpc of the Galactic
centre.  Therefore, it is useful to inspect the orbital behaviour
of {\it in situ} and accreted stars within that central volume of
the Galaxy, with an eye towards determining whether the properties
of the accreted population mimic those of their {\it in situ}
counterparts, or whether they are distinct and cannot be derived
from the latter.  Figure~\ref{fig:bulge_action} displays high- and
low-$\alpha$ {\it in situ} stars of all metallicities with $R_{\rm
GC}<4$~kpc, as well as their accreted counterparts on various IoM
spaces.  Plotted as a function of angular momentum ($L_{\rm z}$)
are the vertical action ($J_{\rm z}$, top panels), radial action
($J_{\rm r}$, middle panels), and total energy ($E$, bottom panels).

The shift to a focus on the inner 4~kpc causes obvious changes in
the distribution of stellar populations.  The clearest one is the
near complete absence of stars with $L_{\rm
z}/10^3\simgreater1$~kpc~km~s$^{-1}$ in the inner Galaxy, which is
understandable on purely geometric grounds.  Interestingly, {\it
Splash} populations are nearly absent at $R_{\rm GC}<4$~kpc.
Of the three main {\it in situ} structures apparent in Figure \ref{fig:ELz_mp},
one can
only find a mix of the low $L_{\rm z}$ high-$\alpha$ populations
and the extension of the low-$\alpha$ towards lower $L_{\rm z}$.
The former also manifests itself by the presence of stars with large
$J_{\rm z}$ in the top middle panel.  From the \loa disk through the
\hia disk to the accreted population, there is a consistent trend
towards a larger fraction of the stars having larger vertical and
radial actions, as well as lower angular momentum.  Accreted
populations are dominated by hot, low \lz kinematics.

\begin{figure}
\includegraphics[width=\columnwidth]{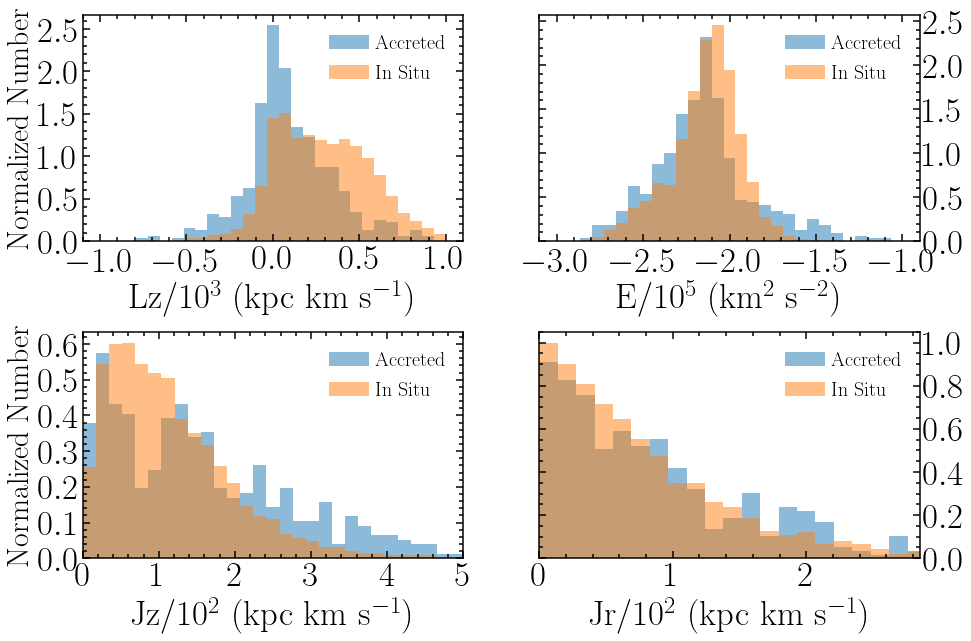}
\caption{Comparison between the distributions of high-$\alpha$ {\it in situ} and
accreted inner Galaxy populations in action variables.  Statistically
significant differences can be seen in $L_{\rm z}$ and $J_{\rm z}$,
whereby accreted stars are dynamically hotter, being distributed
towards smaller angular momentum ($L_{\rm z}$) and larger vertical
action ($J_{\rm z}$).  The energy distributions differ substantially
due to the presence of GE/S stars with E$>-2\times10^5$km$^2$s$^{-2}$.
Removal of that contamination does not change the results.
}
    \label{fig:acthist}
\end{figure}

The distribution of stars in the accreted population on all planes
differs markedly from that of their {\it in situ} counterparts.  To
gain further confidence on the reality of this critical result, we
display in Figure~\ref{fig:acthist} histograms comparing the
distributions of action variables of the accreted and {\it in situ}
inner Galaxy populations.  We limit the comparison to the \hia
population only, since the differences between accreted and \loa
stars can be easily recognised by eye on Figure~\ref{fig:bulge_action}.

Substantial differences can be seen in the distributions of all
integrals of motion.  This is particulary obvious when comparing
the 1D distributions of each integral of motion separately, as shown
in Figure~\ref{fig:acthist}.  We perform Kolmogorov-Smirnoff (KS)
tests comparing the distributions of accreted and {\it in situ}
populations in those variables, finding that the two samples are
not extracted from the same parent population in \lz
($p$-value=$1\times10^{-34}$), \jz ($6\times10^{-15}$) \jr
($7\times10^{-7}$), and $E$ ($5\times10^{-6}$).  This result persists,
with decreased statistical significance, when only metal-poor stars
are considered, with $p$-value$<$0.01 in all cases.  To isolate
stars that are only associated with the IGS, we repeat the KS tests
for $L_{\rm z}$, $J_{\rm z}$, and $J_{\rm r}$, by restricting the
comparison only to stars with $E<-2\times10^5$ km$^2$ s$^{-2}$.
The results are unchanged except for the case of $J_{\rm r}$, for
which the KS test cannot rule out the null hypothesis.

We conclude that the distribution of accreted stars in the inner
Galaxy differs from those of major {\it in situ} components in
action space.  This result agrees with the findings by \cite{Ness2013b}
and \cite{Arentsen2020a}, who show that metal-poor populations are
characterised by hot kinematics \citep[see also][]{Minniti1996},
even though without detailed chemical compositions they were unable
to distinguish {\it in situ} from accreted populations.  In the
most detailed investigation of the kinematics of bulge stellar
populations to date, \cite{Ness2013b} found that the metal-poor
component of the bulge (${\rm [Fe/H]}\simless-1$) does not partake in
its cylindrical rotation, being characterised by slightly lower
rotation and higher velocity dispersion \citep[see also][]{Zasowski2016}.

We point out that the differences between accreted and \hia {\it
in situ} populations are reduced when only metal-poor ([Fe/H]$<$--0.8),
low-energy stars ($E/10^5<-2$~km$^2$s$^{-2}$) are considered.  In
that case, the only variable for which a residual statistically
significant difference is found is $L_{\rm z}$ ($p$-value=0.004).
This is an interesting result. Such similarity in action space
is partly due to the fact that the contamination of the accreted
sample by {\it in situ} stars is not negligible.  Recall that in
Section~\ref{sec:contamination} we showed that the IGS sample is
contaminated at the 22-40\% level.  The {\it in situ} contamination
for the accreted sample displayed in Figure~\ref{fig:acthist} is
higher, since unlike the IGS sample, it includes stars of all
eccentricities.  Nevertheless, taken at face value, this result
suggests that, if indeed the IGS is the remnant of an accretion
event, it has had the time to mix well with its co-spatial {\it in
situ} population, which suggests a very early accretion event.  We
discuss this and other possible interpretations of the data in
Section~\ref{sec:origin} .


In summary, we conclude that the chemically defined accreted
populations in the bulge are not dynamically associated with their
{\it in situ} counterparts.  They are dynamically hotter than the
{\it in situ} population typically possessing low angular momentum
and being distributed towards large values of vertical and radial
action.  We take these results as suggestive of an accretion origin.
In the next section we discuss the chemical properties of the IGS.

\begin{figure}
	\includegraphics[width=\columnwidth]{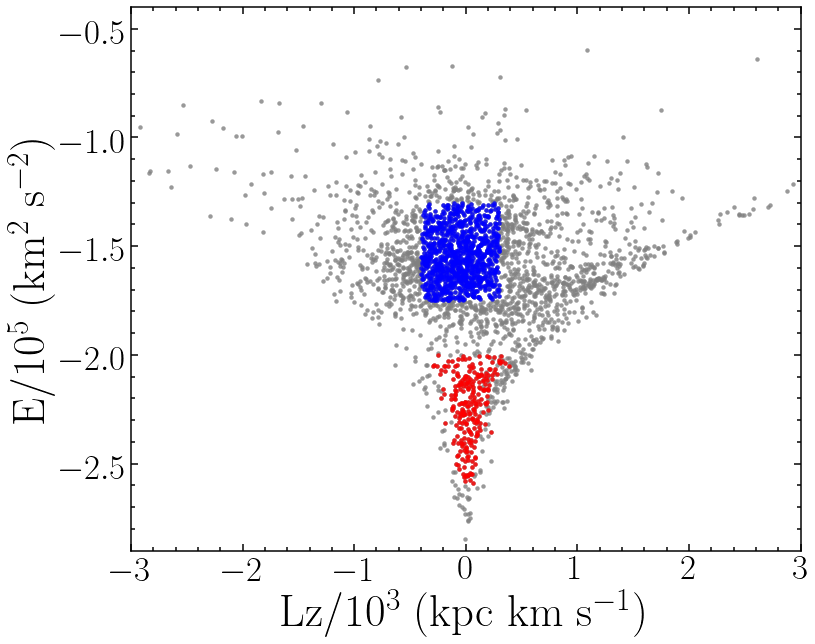}
\caption{Same as bottom panel of Figure~\ref{fig:ELz_mp}, 
identifying stars deemed to be members of the GE/S system (blue dots) and
the IGS (red dots).}
    \label{fig:GELE}
\end{figure}


\begin{figure}
	\includegraphics[width=\columnwidth]{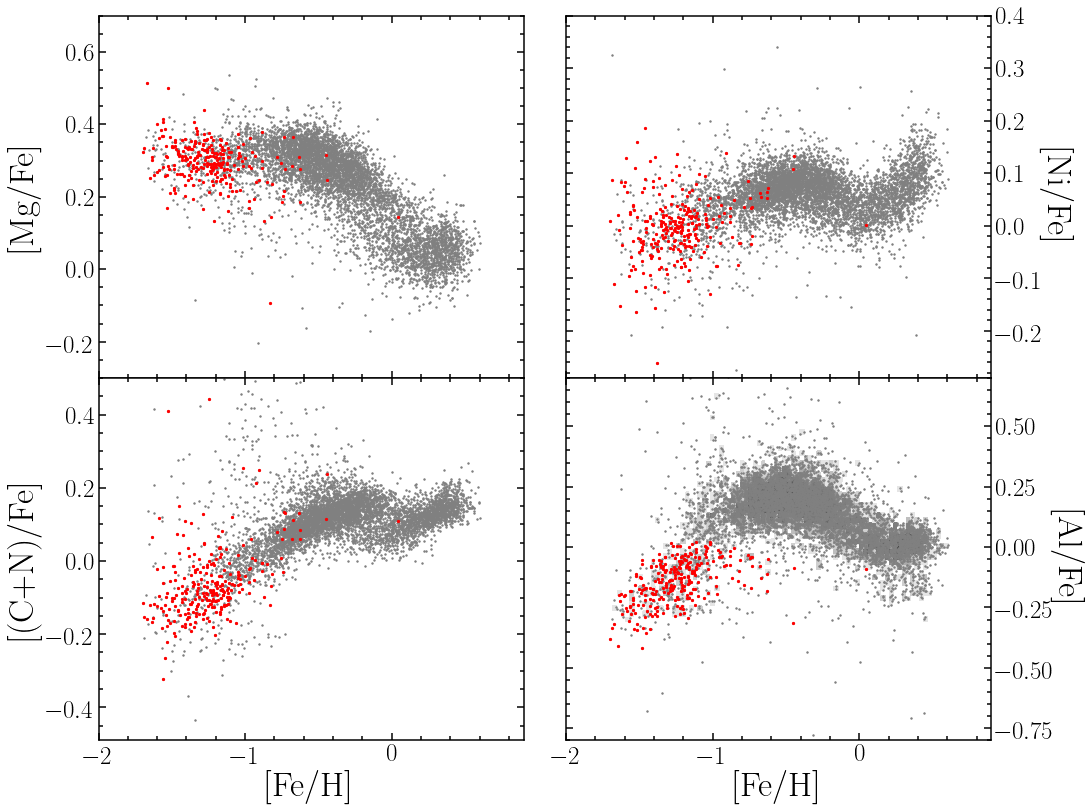}
    \caption{Comparison between stars from the IGS (red dots) with
their bulge counterparts (gray dots).  The IGS has an abundance
pattern that is characteristic of those of dwarf satellites of the Milky
Way, yet they seem to form a single sequence with the remainder of the
bulge populations, even though they all are associated with different
Galactic components.
}
    \label{fig:abunds}
\end{figure}

\section{Chemical Properties} \label{sec:chemprop}

In the previous section we investigated the orbital properties of
the IGS, concluding that they are distinct from those of co-spatial
{\it in situ} populations, which supports an accretion origin.  In
this section we examine the chemical compositions of the stars from
the IGS alongside those of their inner Galaxy counterparts.  Our
main objective is to check whether detailed chemistry is consistent
or not with an accretion scenario.  In Figure~\ref{fig:GELE} we
display again the accreted stars shown in the bottom panel of
Figure~\ref{fig:ELz_mp}, this time marking stars deemed to be members
of the IGS and GE/S systems as red and blue dots, respectively.
The IGS sample is defined in Section~\ref{sec:contamination}.  We
consider members of the GE/S system to be those stars with $-0.4 <
L_{\rm z}/10^3 < 0.3$~kpc~km~s$^{-1}$ and $-1.75 < E/10^5 <
-1.3$~km$^2$~s$^{-2}$.  According to these criteria, our samples
contain 259 candidate members of the IGS and 1,026 associated with
the GE/S system.  These definitions are henceforth adopted in our
discussion of the chemical properties of the IGS.

In Figure~\ref{fig:abunds} we contrast the stars from the IGS with
their {\it in situ} $R_{\rm GC}<4$~kpc counterparts in chemical
space.  We limit our comparison to a few elements that sample a
range of nucleosynthetic pathways, with Mg being mostly contributed
by SN II, C and N by massive and AGB stars, and Ni by SN Ia
\citep[e.g.,][]{Chiappini2003,Nomoto2013}.  The latter element is
additionally interesting due to it being typically depressed in
dwarf satellites of the Milky Way \cite[e.g.,][]{Shetrone2003}.
For more comprehensive studies of the detailed abundance patterns
of bulge stars, we refer the reader to the excellent studies by
\cite{Schultheis2017} and \cite{Zasowski2019} and the review by
\cite{Barbuy2018}.

Two main results emerge from this comparison: {\it (i)} the stars
belonging to the IGS have an abundance pattern that is typical of
dwarf galaxies, characterised by low abundances of elements such
as Al (by construction), C, N, and Ni
\citep[e.g.,][]{Hayes2018,Mackereth2019,Helmi2020,Das2020}); {\it
(ii)} the abundance pattern of the IGS apparently behaves as a
natural extension of the bulge field population towards [Fe/H]~$<-1$,
a result that has also been noted by \cite{Schultheis2017} and
\cite{Zasowski2019}.  To be clear, this extension of the trends
towards metal-poor populations concerns only the {\it locus} occupied
by stars in the various chemical planes, not the relative counts.
The latter are examined in Section~\ref{sec:subchem}.  Similar
behaviour is seen in the data presented by
\cite{Queiroz2020b}\footnote{Note that \cite{Queiroz2020a}
report the presence of an $\alpha$-bimodality in the distribution
of bulge stars {\it on the high metallicity end} of the $\alpha$-Fe plane,
at variance with our results.  We suspect this discrepancy may
result from the adoption of different distances by our two studies.},
who examine the spatial variation of the stellar distribution on
the $\alpha$-Fe plane across the Galactic disk and bulge.  Naively,
one would consider this seeming continuity between the IGS and its
{\it in situ} counterparts surprising, given the alleged accreted
nature of the IGS.  In Section~\ref{sec:eagle} we show that continuity
between the chemical compositions of the IGS and its more metal-rich
bulge counterparts is actually {\it expected} on the basis of
numerical simulations.

Despite this apparent continuity, it is worth noting that
slight deviations of this pattern are apparent in elements such as
Mg, Ni, and Si (not shown), for which the mean of the abundances
of IGS stars is lower than those of their {\it in situ} metal-poor
counterparts.  In all cases this result does not have a high
statistical significance, as the differences in mean abundance
ratios are slighlty smaller than the standard deviation of the
distributions.  Considering the three elements in aggregate, we get
${\rm \langle[(Mg+Si+Ni)/Fe]\rangle_{IGS}=0.54\pm0.11}$ and ${\rm
\langle[(Mg+Si+Ni)/Fe]\rangle_{\it in\, situ}=0.68\pm0.17}$ when
only stars with $-1.1<{\rm [Fe/H]}<-0.9$ and $R_{GC}<4$~kpc are
considered.  That a larger difference is not found may not at all
be surprising, given that the IGS sample is contaminated by {\it
in situ} stars at the 22--40\% level (c.f.
Section~\ref{sec:contamination}).


It is also worth pointing out that IGS stars with [Fe/H]$>$--1.0
show a variety of different behaviours.  Some of them follow the
same trend as the \hia disk in Mg, Ni, and C+N, whereas others
display a behaviour more similar to those of dwarf satellites, with
low [Mg/Fe], [Ni/Fe], and [(C+N)/Fe].  By construction, all of them
display low [Al/Fe].  We hypothesise that these metal-rich IGS stars
are a mixture of thick disk contaminants (whose low [Al/Fe] ratios
would be difficult to understand), thin disk contaminants, and
genuine IGS members.


In summary, the results presented in this Section and
Sections~\ref{sec:mp_orbs} and \ref{sec:IOM_bulge} show that a
metal-poor stellar population located in the inner Galaxy shares
chemical and IoM properties with well known halo accreted systems.
Figure~\ref{fig:abunds} suggests on the other hand that the abundance
pattern of this same structure is consistent with sharing a common
chemical evolution path with its bulge counterparts potentially
associated with other co-spatial Galactic components.  Yet the
analysis in Sections~\ref{sec:mp_orbs} and \ref{sec:IOM_bulge}
suggests that the IGS is dynamically detached from the rest of the
bulge (Figures~\ref{fig:bulge_action} and \ref{fig:acthist}).  This
is a puzzling result.  In the next subsections we look into the
theoretical expectations for the chemical evolution of Milky Way-like
galaxies that underwent comparable accretion events
(Section~\ref{sec:eagle}) and inspect the data for the presence of
a chemical connection between the accreted and {\it in situ}
populations in the inner Galaxy.

\begin{figure*}
	\includegraphics[width=\textwidth]{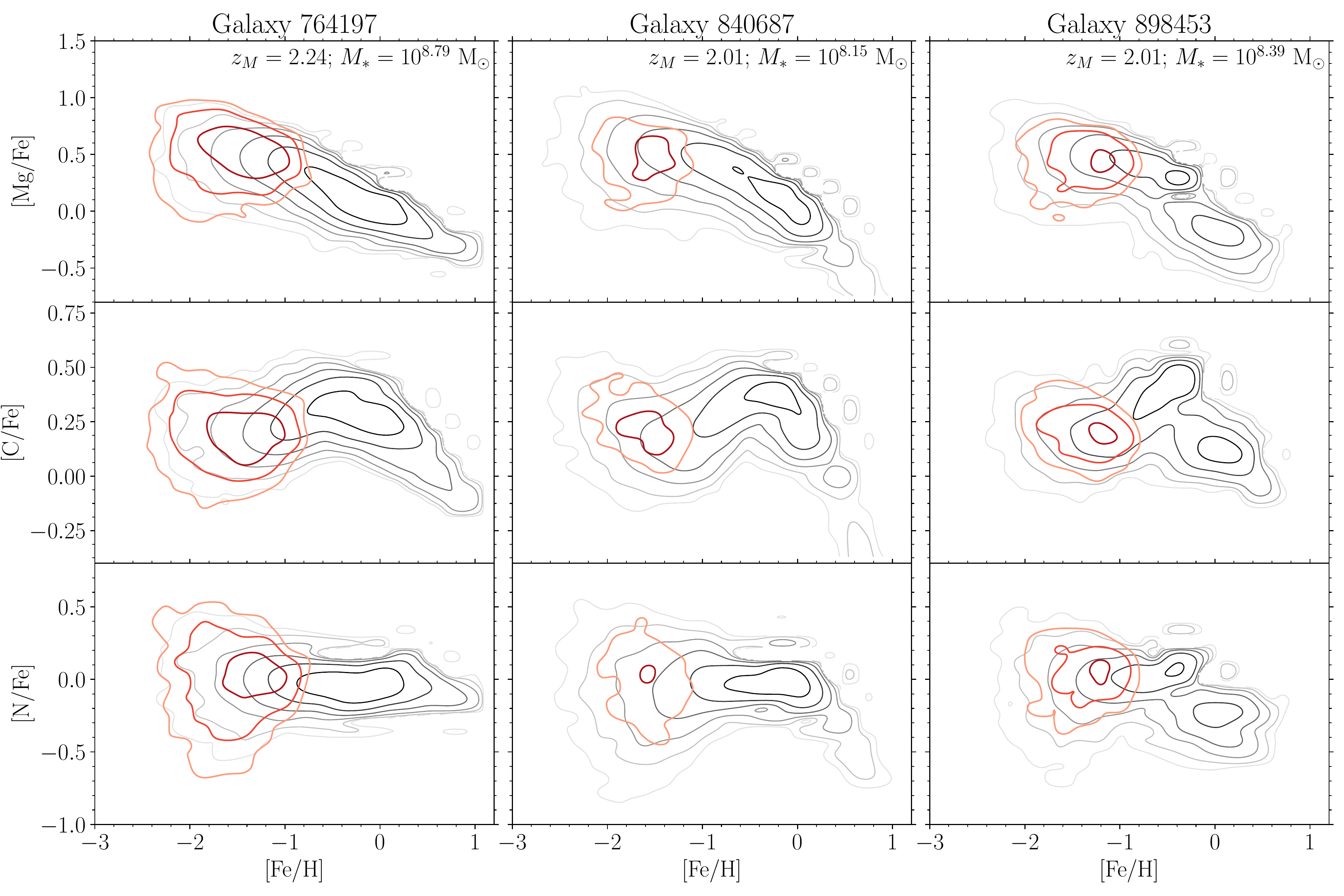}
    \caption{Chemical compositions of stellar particles in Milky Way-like
galaxies in the EAGLE simulations.  Shown are three examples of a sample of
simulated galaxies selected with masses
similar to that of the Milky Way, which underwent a massive accretion event at
$z \ge 2$.  Red countours indicate the distribution of stars
belonging to the accreted system, and black contours indicate the
distribution of stars formed {\it in situ}.  Contouring starts at 10 stars per bin,
increasing logarithmically in 0.5 dex steps.  In all systems, for all
abundances tracked by EAGLE, the accreted stars possess chemical
compositions that are in good agreement with those of
the {\it in situ} populations of same metallicity.  
}
    \label{fig:eagleabunds}
\end{figure*}

\subsection{Predictions from Numerical Simulations} \label{sec:eagle}

The discussion above naturally prompts us to enquire into the
theoretical predictions for the chemical evolution of Milky Way-like
Galaxies with accretion histories similar to that seemingly implied
by our results.  For that purpose we resort to the predictions from
the suite of cosmological numerical simulations by the Evolution
and Assembly of GaLaxies and their Environments (EAGLE) project
\citep{Schaye2015,Crain2015}.  The question guiding this analysis
is the following: what are the chemical properties of Milky Way
like galaxies undergoing an important early accretion event?  More
specifically, do we expect the chemical properties of the stars
belonging to the accreted system to differ substantially from those
of the {\it in situ} population?

The EAGLE simulations provide excellent material for this enquiry.
The large volumes included in the various sets of simulations make
possible robust statistical analyses of the evolutionary outcomes
for a broad gamut of final galaxy properties, and at a wide enough
range of resolutions to address the needs of different types of
investigations.  To maximise resolution we choose to study the
$25^3$~cMpc$^3$ L025N752-Recal simulation, whose volume is large
enough to yield a substantial number of galaxies with Milky Way-like
masses and accretion histories similar to that suggested by our
results.  The chemical evolution prescriptions of the simulations
track only elements relevant to the thermal balance of the interstellar
medium, yet this is sufficient for our purposes, as predictions for crucial
elements Fe, C, N, and Mg are included.  For more details on the
simulations we refer the reader to \cite{Schaye2015} and \cite{Crain2015}.

The analysis was conducted as follows.  We selected galaxies with
Milky Way-like masses within the interval $7\times10^{11} <
M_{200}/M_\odot < 3\times10^{12}$ and looked for those that underwent
an accretion of a satellite with stellar mass $M_\star/M_\odot \ge
10^8$ at $z \ge 2$ (see Section~\ref{sec:properties} for an estimate
of the mass of the IGS progenitor).  We found a total of 15 accreted
galaxies with such a profile (34\% of the host galaxy sample within
that halo mass range).  The particles for the accreted galaxies
were selected when they were at their peak stellar mass.  For three
of those galaxies we display chemical composition data in
Figure~\ref{fig:eagleabunds}.  The data are displayed in the form
of contour maps indicating the distribution of stars on abundance
planes, with {\it in situ} and accreted populations shown in gray
and red, respectively.  {\it In situ} populations in this context
are defined as stellar particles that were bound to a halo in the
main galaxy branch at the snapshot prior to star formation.  The
abundance planes shown are Mg-Fe, C-Fe, and N-Fe, which track the
contribution of SN II, SN Ia, and AGB stars, sampling a range of
chemical enrichment timescales.

The answer to the question posed above immediately strikes the eye
upon inspection of Figure~\ref{fig:eagleabunds}.  For all three
galaxies, and all three elemental abundances, the abundance ratios
of the accreted populations are in good agreement with those of the
{\it in situ} populations at same metallicity.  This is the case
for {\it all} the 15 galaxies meeting our selection criteria.  It
is also the case for all the other elemental abundances tracked by
EAGLE.  This result is also fully consistent with the analytical
model presented in Figure~\ref{fig:modelcomp}, which shows that the
early chemical evolution of a Milky Way-like galaxy and its satellite
are similar.  We conclude therefore that the approximate continuity
between the abundances of the accreted and {\it in situ} populations
in Figure~\ref{fig:abunds} is in fact not surprising, but actually
an expectation from theory, at least for the elements whose chemical
evolutions are modelled by EAGLE.

This is a very important result, the implications of which configure
the proverbial double-edged sword.  On one hand the theoretical
predictions for the {\it locus} of accreted populations in chemical
space cannot rule out the accreted nature of the IGS.  On the other,
it cannot ascertain it either, as the apparent continuity between
the chemistry of the IGS and its {\it in situ} populations could
alternatively be a by-product of pure {\it in situ} chemical
evolution.


To find our way out of this conundrum we invoke relative number
counts in chemical space.  In trying to understand the apparent
chemical continuity between the accreted and {\it in situ} populations
in Figure~\ref{fig:abunds} we ask the following question: is the
distribution of stellar data on abundance planes consistent with
the accreted population being an additional structure, or do the
number counts suggest that it is just the tail end of the distribution
of the {\it in situ} populations?  In the next section we address
this question in a quantitative fashion.


\subsection{Substructure in Chemical Space} \label{sec:subchem}

\begin{figure*}
\includegraphics[width=\textwidth]{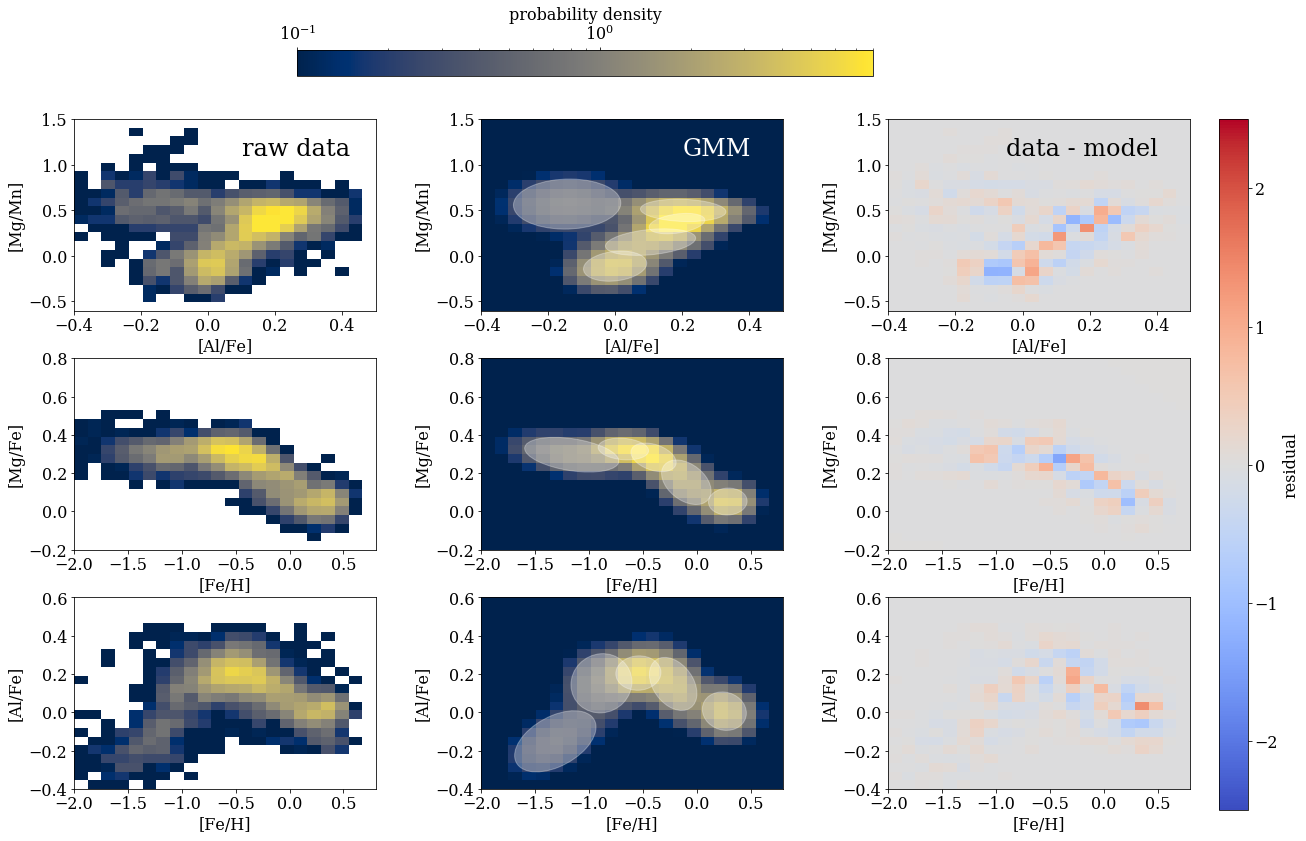}
\caption{{\it Left Panels:} Distribution of stars with $R_{\rm GC}<4$~kpc
in various chemical planes; {\it middle panels:} Gaussian
Mixture Modelling (GMM) of substructure of the data distribution
on the same chemical planes; {\it right panels:} residuals from GMM
fit.  A satisfactory match to the density distribution
of data points is obtained by adopting five {\it detached yet
overlapping} substructures.  Each of those correspond to peaks in
the metallicity distribution function of bulge populations
\citep{Ness2013a}, after accounting for zero point differences in
[Fe/H].  }
\label{fig:chem_sub}
\end{figure*}

\subsubsection{Matching the distribution with multiple components}
\label{sec:gmm_match}

In this Section we tackle the problem of understanding the apparent
chemical continuity between the IGS and its more metal-rich inner
Galaxy counterparts (Figure~\ref{fig:abunds}) from a purely
observational perspective.  We center our approach around the
following question: can the stellar density distribution in chemical
space be described by a discrete number of {\it detached} substructures?
\cite{Ness2013a} showed that the bulge MDF can be well modelled by
a combination of five components, which they ascribe, in order of
increasing metallicity, to the inner halo, the metal-poor thick
disk, the bulk of the thick disk itself, the bar, and the thin disk.
Some of these peaks can be identified in our raw MDF in
Figure~\ref{fig:MDFs}, although a one-to-one mapping with the
\cite{Ness2013a} MDF is quite difficult due to differences in the
two metallicity scales and in the relative strengths of the various
peaks (recall that our MDF is not corrected for selection effects),
and the uncertainties associated with multiple gaussian fitting in
1D space.  These difficulties are particularly important towards
low metallicity, where the statistics is not as robust and the
results of abundance analysis may be afflicted by important
systematics.



The exceptional quality of our detailed chemical compositions comes
in handy in this juncture.  By considering the distribution of our
sample in higher dimensional space we hope to gain additional
leverage in distinguishing the different components making up the
stellar population mix of the inner Galaxy.  The stellar distribution
on some chemical planes, such as [Mg/Mn]-[Al/Fe] and [Al/Fe]-[Fe/H],
show a clumpy distribution which suggests that the data may be
amenable to such an approach.  In other words, one should be able
to model the stellar density distribution in chemical space as a
combination of overlapping structures, which are not necessarily
connected to one another by a coherent chemical evolution path.  To
test this hypothesis we check whether the stellar density distribution
in a few chemical planes can be reproduced by a Gaussian Mixture
Modelling (GMM) approach.  Obviously a Gaussian distribution is
only a zero-th order approximation to the stellar density distributions
in chemical space.  A massive stellar system such as, for instance,
the Galactic disk, underwent star formation long enough that the
run of the number of stars with metallicity and abundance ratios
is strongly determined by the star formation history.  Conversely,
in a less massive system whose star formation was quenched during
an early accretion to a massive halo, the distribution of stellar
metallicities should be narrower and thus the GMM approach may work
better.

We are particulary concerned with modelling the metal-poor end of
the metallicity distribution to check whether the IGS is detached
from its more metal-rich peers in the inner Galaxy.  The key measure
of chemical detachment of the IGS from its more metal-rich counterparts
in the inner Galaxy is provided by the ability of a GMM approach
to account for the density distribution of the IGS, so that no
important residual connection between that and the other populations
is left after model subtraction (although {\it some} residual should
be expected given the contamination by {\it in situ} populations
discussed in Section~\ref{sec:chem_distinct}).

The sample for analysis is displayed in the left panels of
Figure~\ref{fig:chem_sub}.  It consists of all stars within $R_{\rm
GC}<4$~kpc, but removing outliers with [Al/Fe]$>$+0.5, as their
abundances are affected by the N-rich-star phenomenon \citep{Schiavon2017}.
We also removed relatively metal-poor stars with low [Mg/Fe]
([Fe/H]$<$--0.5 and [Mg/Fe]$<$+0.15), which have high orbital energy
and are not associated with the IGS, but rather with the GE/S system
(see Section~\ref{sec:ins_vs_acc} and Figure~\ref{fig:ins_vs_acc}).
The stars thus selected are shown in three chemical planes, where
the 2D histograms are cast in terms of probability density.  On the
top row, stars are displayed in the [Mg/Mn] vs [Al/Fe] plane, and
on the middle and bottom rows their distribution can be seen on the
Mg-Fe and Al-Fe planes, respectively.  According to our definition
(Figure~\ref{fig:accdef}), accreted stars occupy the upper left
quadrant of the [Mg/Mn]-[Al/Fe] plane (top left panel of
Figure~\ref{fig:chem_sub}).  On the other planes, accreted stars
are located in the low metallicity end of the distribution
([Fe/H]$\simless$--1).  We fit a separate GMM to the distribution of
these stars on each of the three chemical planes.  The ideal number of
components was decided by calculating the Bayesian Information
Criterion (BIC) for each fit.  We found that the best match is found
when either 4 or 5 components are adopted.  We allow the mean and
covariances of each component to be entirely free with no informative
or uninformative priors. The GMMs for the 5 component case are
overlaid on the data in the middle panels of Figure~\ref{fig:chem_sub}.
The standard deviation of the residuals around the fits are 0.23,
0.19, and 0.17 for the top, middle, and bottom panels, respectively.

%

The residuals after subtraction of the best fit models and the data
are displayed in the right panels. The residuals are relatively
small, and slightly larger towards higher metallicity, which reflects
the much larger stellar density in that regime.  When normalized
by the Poisson fluctuations per bin, the residuals are homogeneously
distributed across the planes, and are everywhere smaller than $\sim
2\sigma$ of the Poisson uncertainty in each bin.  These numbers
suggest that the distribution of the stellar density in the locus
of accreted stars is well accounted for by the GMM approach, showing
that they can be matched by a simple combination of detached, yet
overlapping structures.  Interestingly, the GMM fit to the stellar
distribution in the Mg-Fe plane highlights the fact that the lowest
metallicity component has a slightly lower [Mg/Fe] than the next
one in order of increasing metallicity.  This result supports the
notion that the most metal-poor component is dominated by an accreted
stellar population, as one would expect [Mg/Fe] to be constant in
an {\it in situ} population within the metallicity interval involved.
Also quite importantly, even though the GMM match was performed
independently to the data on the three chemistry planes, the
associations of stars to a given group were largely consistent
across the three models, which attests for the reality of these
groups.

%
%

\subsubsection{Breaking the MDF degeneracy with additional chemistry}
\label{sec:MMDF}

The five-component match identifies peaks located at [Fe/H]~$\sim$~--1.3,
--0.8, --0.5, and --0.2 and +0.3, which are very roughly consistent
with the five MDF peaks identified by \cite{Ness2013a}, after
allowing for a $\sim$0.2-0.4~dex difference between the zero points
of the metallicity scales of the APOGEE and ARGOS surveys \citep[see,
for instance, the positions of the peaks in the APOGEE bulge MDF
determined by][]{GarciaPerez2018,Rojas2020}.  The bulge MDF has
been analysed exhaustively in recent work based, e.g., on the ARGOS
\citep{Ness2013a}, Gaia-ESO \citep{Rojas2017}, and APOGEE
\citep{Schultheis2017,GarciaPerez2018,Rojas2020} surveys.  The number of peaks
in the bulge MDF, which in turn constrains the number of physically
distinct components cohabiting the bulge, have been subject to
debate in the recent literature \citep[see discussion by][]{Fragkoudi2018}.

We do not aim at a detailed analysis of the topic in this paper.
However, we wish to highlight the fact that the consideration of
additional abundances helps untangling the two most metal-poor
populations in our sample.  The task of disentangling these populations
in the MDF of Figure~\ref{fig:MDFs} is far from straightforward,
as the distributions of these components overlap strongly in
one-dimensional metallicity space.  However, consideration of
additional elemental abundances lifts the degeneracy between these
components, bringing them to sharp relief on elemental abundance
planes such as those in the middle and bottom rows in
Figure~\ref{fig:chem_sub}.

\bigskip

The analysis presented in Section~\ref{sec:subchem} leads to a few
important conclusions: {\it (i)} the distribution of the numbers
of stars over various chemical planes can be modelled by the
combination of overlapping yet detached sub-structures.  This is
particularly true in the low metallicity end, where residuals are
the lowest, adding further support to the notion that the IGS has
an accreted origin;  {\it (ii)} the fit in the Mg-Fe suggests that
the [Mg/Fe] distribution of the lowest metallicity component is
consistent with an accretion origin; and {\it (iii)} the GMM method
consistently picks out the same substructures in various chemical
planes, involving the abundances of Fe, Mg, Al, and Mn.  The
metallicities of these substructures are a reasonably good match
to those of well known peaks in the MDF of the Galactic bulge
\citep{Ness2013a}.  This latter result validates the application
of the GMM method to describe the distribution of stellar data in
chemical space.

\section{Discussion} \label{sec:discussion}

We start this section by briefly summarising our results before
proceeding with a discussion of their implications.  In
Sections~\ref{sec:subiom} and \ref{sec:chemprop} we discussed
evidence for the presence of a remnant of an accretion event in the
inner Galaxy, on the basis of the existence of substructure in
orbital and chemical space.  While the whole body of evidence for
accretion is strong, when considered in isolation, our interpretation
of either dynamical or chemical evidence may reasonably be called
into question.

The evidence from the orbital side hinges on two main results: {\it
(i)} the statistically significant differences in orbital properties
between the accreted stars and their more metal-rich {\it in situ}
counterparts in the inner Galaxy (Figures~\ref{fig:bulge_action}
and \ref{fig:acthist}); {\it (ii)} the existence of an energy gap
separating the IGS from other metal-poor halo substructure.  The
presence of this energy gap seems to be immune to selection effects,
since it is absent in chemically defined metal-poor {\it in situ}
populations (Figure~\ref{fig:ELz_mp}), whose spatial distribution,
uncorrected for selection effects, is the same as that of their
accreted counterparts.


It is important to note, however, that when $R_{\rm GC}<4$~kpc
stellar populations are considered in isolation, there is a smooth
dependence of orbital properties on metallicity, which could be
accommodated by a fully {\it in situ} scenario.  On the other hand
such smooth behaviour would also be expected in the accreted case,
given that the orbital and chemical properties of substructures
overlap strongly due to a combination of observational error and
cosmic variance.

On the chemistry side, evidence for accretion relies on {\it (i)}
the similarity between the chemical composition of the IGS and those
of satellites of the Milky Way and other accreted systems
(Figure~\ref{fig:le_vs_ge}), and {\it (ii)} the presence of clumpiness
of the stellar distribution in various chemical spaces
(Figures~\ref{fig:accdef} and \ref{fig:chem_sub}).  Regarding {\it
(i)}, we showed that the chemistry of unevolved {\it in situ}
populations are also similar to that of their accreted counterparts
(Figure~\ref{fig:modelcomp}).  Moreover, there is an apparent
continuity between the chemical compositions of the IGS and its
bulge counterparts (Figure~\ref{fig:abunds}).  However we also
showed that similarity between the chemical compositions of accreted
and {\it in situ} populations is an actual expectation from
cosmological numerical simulations (Figure~\ref{fig:eagleabunds}).
Regarding {\it (ii)} we showed that the data in chemical space can
be well matched by the combination of detached substructures
(Figure~\ref{fig:chem_sub}).  However, one cannot rule out that
such clumpiness is the result of a bursty history of star formation
and chemical enrichment.

Despite the above caveats, when considered in aggregate, chemical
and orbital information suggest the presence of a previously
unidentified structure located in inner Galaxy, the IGS.  We
hypothesise that this structure is the remnant of a galaxy accreted
to the Milky Way in its very early life.  In the next Subsections we
discuss the properties of the hypothetical progenitor of the IGS.
In Section~\ref{sec:origin} we discuss the origin of the metal-poor
populations in the Galactic bulge, in Section~\ref{sec:properties}
we discuss the properties of the putative progenitor of the IGS;
and in Section~\ref{sec:scenario} we present a speculative scenario
for the early evolution of the Milky Way halo.

\begin{figure}
\includegraphics[width=\columnwidth]{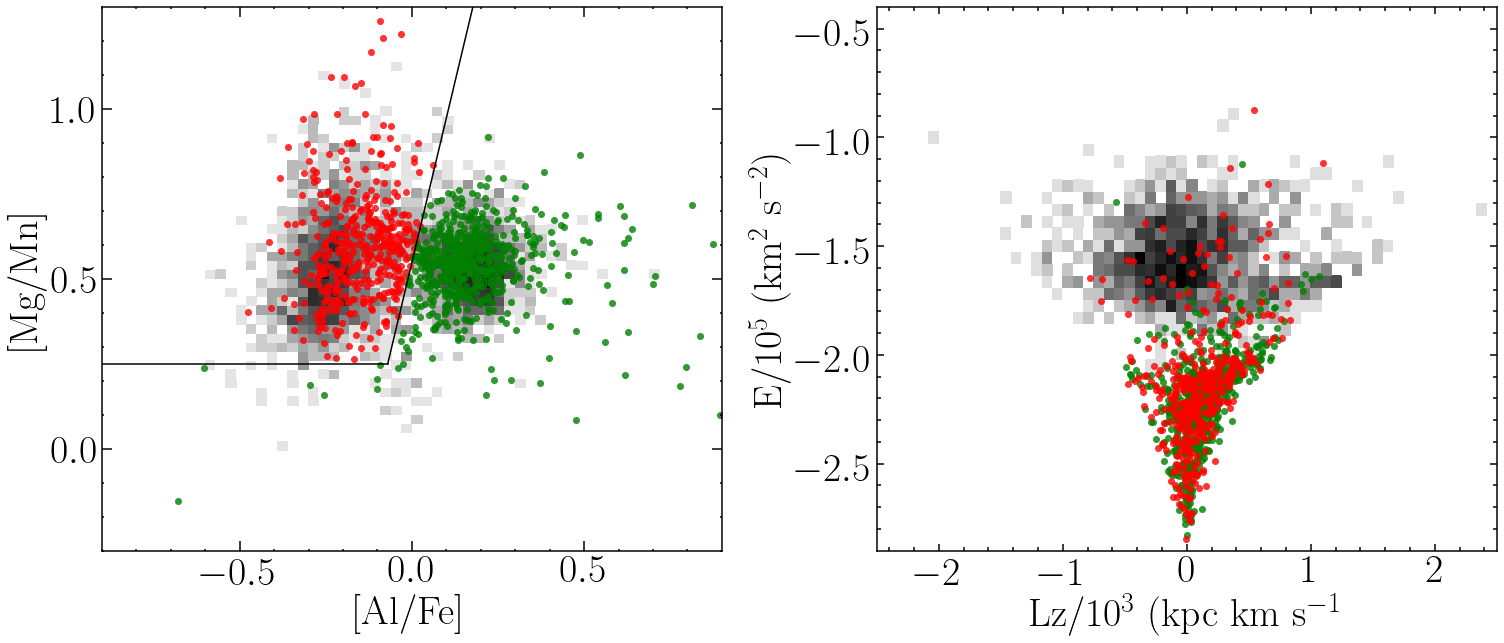}
\caption{{\it Left panel:} Data for metal-poor stars ([Fe/H]~$<-0.8$)
are displayed on the chemistry plane used to distinguish accreted
from {\it in situ} populations.  The dividing line is the same as
shown in Figure~\ref{fig:accdef}.  The 2D histogram shows metal-poor
stars at $R_{\rm GC}>4$~kpc, and the red (green) dots represent
accreted ({\it in situ}) stars in the inner Galaxy.  A clear bimodal
distribution in [Al/Fe] is visible in the data for both the inner and outer Galaxy.  
{\it Right panel:} Same stars displayed on the energy-angular momentum
plane.  The data for the outer Galaxy are dominated by the GE/S and {\it
Splash} systems.  In the inner Galaxy, the accreted sample is dominated by
the IGS, but about 1/10 of the sample is consistent with an association
with the GE/S system ($E>-1.85\times10^5$~km$^2$~s$^{-2}$).  The {\it in
situ} sample is dominated by thick disk stars, although a small fraction
seems to belong to the {\it Splash}.  Note that the distributions of accreted and
{\it in situ} stars in IoM space are different at a
statistically significant level (see Section~\ref{sec:mp_orbs}).
}
    \label{fig:ins_vs_acc}
\end{figure}

\subsection{The origin of metal-poor bulge stars} \label{sec:origin}

In this Section we estimate the contribution of accreted and {\it in situ}
stars to the metal-poor star content of the Galactic bulge, and contrast
that estimate with predictions from fully {\it in situ} chemical evolution
models.

\subsubsection{Contribution by accreted and {\it in situ} formation to the
stellar mass budget} \label{sec:ins_vs_acc}

We begin by assessing the relative contribution of accreted and
{\it in situ} stars to the metal-poor stellar mass budget of the
inner Galaxy.  The data shown in Figure~\ref{fig:ins_vs_acc} may
help shed light on this question. We display metal-poor stars
([Fe/H]~$<-0.8$) on the [Mg/Mn] vs. [Al/Fe] plane, which we used
in Section~\ref{sec:data} to distinguish accreted from {\it in situ}
populations (Figure~\ref{fig:accdef}).  The underlying 2D histogram
shows the metal-poor populations located at $R_{\rm GC}>4$~kpc.
The dividing line defining the border between accreted (upper left
quadrant) and {\it in situ} populations is also shown.  Inner Galaxy
stars in the accreted ({\it in situ}) region of the diagram are
shown in red (green).  A clear bimodal distribution of [Al/Fe] data
for the stars in the inner Galaxy can be seen.  In total, there are
430 stars in the accreted area of this chemical plane, and 602 stars
in the {\it in situ} region.  Recall, moreover, that in
Section~\ref{sec:mp_orbs} we estimated that the chemically-defined
sample of metal-poor accreted stars within the IGS region of $E-L_{\rm
Z}$ space was contaminated by thick disk stars at the 22--40\%
level.  Accounting for that additional factor, {\it we estimate
that approximately 25--33\% of the stars with [Fe/H]$<$--0.8 in our
inner Galaxy sample have an accreted origin.}

The right panel of Figure~\ref{fig:ins_vs_acc} shows the distribution
of accreted and {\it in situ} stars on the $E-L_{\rm z}$ plane.
The accreted star sample is dominated by low energy stars, mostly
associated to the IGS.  About 10\% of the accreted stars are located
at high energies and are likely stars associated to the GE/S system
(large gray blob) which are currently crossing the inner Galaxy.
Notably, very few {\it in situ} stars (green points) have such high
energies.  They predominantly occupy the low energy region of the
plane, with a number of stars positioned along the {\it Splash} (to
the right of GE/S, see green arrow in Figure~\ref{fig:ELz_mp}).


Other studies have spotted duplexity in the properties of metal-poor
bulge samples, which may be attributed to the presence of a mix of
accreted and {\it in situ} populations.  \cite{Pietrukowicz2015}
identified two major sequences in the period-amplitude plane for a
large sample of fundamental-mode RR Lyr variable stars.  The two
populations differ slightly in metallicity, with the more metal-poor
population outnumbering their metal-rich counterparts by approximately
a factor of 4.  Those authors suggest that this duality is indicative
of merging activity in the early history of the Milky Way bulge.
\cite{Kunder2020} obtained radial velocities for a few thousand RR
Lyrae stars from the \cite{Pietrukowicz2015} sample and determined
the presence of two kinematically distinct populations.  One
population follows bar-like orbits and the other one is more centrally
concentrated, being characterised by hotter kinematics.  They suggest
that the latter may result from an accretion event that preceded
the formation of the bar.


On the theory side, cosmological numerical simulations predict
relatively intense accretion activity in the first few Gyr of the
lives of massive galaxies.  They have reached a degree of sophistication
that enables a quantitative confrontation with our data.  Early
numerical simulations by \cite{Kobayashi2011} propose that the bulge
was fully formed by the merger of galaxies with $M\sim5\times10^9~M_\odot$
at $z\simgreater3$.  On the other hand, the more recent EAGLE
simulations provide more detailed information that can be compared
with our data.  For instance, in the simulated galaxies whose
abundances are displayed in Figure~\ref{fig:eagleabunds}, the
fraction of accreted stars within 4~kpc of the centre for [Fe/H]~$<-1$
varies from 0.10 to 0.55.  Along the same lines, the analysis of
FIRE simulations by \cite{ElBadry2018} finds that same fraction
within 1~kpc of simulated Milky Way-like galaxy centres to be around
0.5.  More recently, \cite{Fragkoudi2020} analysed high resolution
data for Milky Way-like galaxies from the Auriga suite of cosmological
numerical simulations.  They found that the fraction of the stellar
mass with [Fe/H]~$<-1$ within 4~kpc of the centre that has an
accretion origin to vary between 0.13 and 0.8.  Moreover, for the
two simulated galaxies matching the properties of the Milky Way
best, that fraction was smaller than 0.4.

Considering all the uncertainties on the observational and theoretical
fronts, and taking into account the relative arbitrariness of the
spatial and [Fe/H] limits adopted in such comparisons, our results
broadly confirm the predictions by numerical simulations from various
groups.  Obviously, better statistics on both the observational and
numerical simulation fronts will be required for such data to provide
stringent constraints on models.


\subsubsection{A Fully {\it in situ} bulge?}

It is encouraging that the main tentative result presented in this
paper does not fall into a theoretical void, that it is in consonance
with expectations from cosmological numerical simulations.  Recourse
to theoretical predictions to validate an observational result is
nonetheless methodologically, and indeed philosophically, unsound.
Thus, before discussing the properties of the putative progenitor
of the IGS, it behooves us to mention theoretical predictions by
various groups which match data for bulge stellar populations without
resort to satellite accretion.  

The history of modelling the chemo-kinematic properties of bulge
stellar populations is decades long, so that paying tribute to all
the excellent work done in the past is beyond the scope of this
paper.  For that we refer the reader to Section~4 of the review by
\cite{Barbuy2018}.  We restrict our discussion to recent work by
\cite{Matteucci2019,Matteucci2020}, because to our knowledge it is
the only instance of a detailed comparison betwen model predictions
and multiple chemical compositions for a large sample based on
APOGEE data.  Their analytical model for the history of star formation
and chemical enrichment of the bulge stellar populations provides
a good match to the MDF \citep{Matteucci2019} and the abundances
of O, Mg, Si, Ca, Al, K, Mn, Cr, and Ni.  While such models resort
to {\it ad hoc} assumptions about gas infall for different Galactic
components, and some non-negligible amount of yield adjustment,
they match the observed data very well.  The implication is that
the fraction of the stellar mass within the bulge that is due to
accretion is {\it zero} at all metallicities.  In that scenario,
the entirety of the bulge stellar populations, including those
deemed {\it accreted} by our criteria described in Figure~\ref{fig:accdef},
were actually formed {\it in situ}.


In light of our discussion in Section~\ref{sec:chem_distinct}, where
we showed that {\it in situ} and accreted populations share a similar
path in chemical space (Figure~\ref{fig:modelcomp}), the results
by \cite{Matteucci2019,Matteucci2020} are not surprising.  As pointed
out in Section~\ref{sec:discussion}, orbital information such as
that discussed in Section~\ref{sec:subiom} is indeed required for
that distinction to be made.  Moreover, recall that according to
our criteria there are 463 accreted stars out of a total sample of
6,350 stars within 4~kpc of the Galactic centre.  Assuming a
contamination at the 22--40\% level of the accreted sample by {\it
in situ} thick disk stars, the total accreted stellar mass amounts
to only $\sim$4--5\% of our sample\footnote{Note however that our
data are not corrected for selection effects, so this is an order
of magnitude estimate}.  Therefore, as far as the chemical evolution
of the bulge is concerned, Matteucci's work offers, at face value
and without consideration of the Milky Way's insertion into the
cosmological context, a viable {\it in situ} formation scenario
that can possibly explain the chemical compositions of the vast majority
of the stellar populations therein.  Notwithstanding this remarkable
success, accounting for the considerable morphological and kinematic
differences between the overlapping components within a few kpc of
the Galactic centre under a single unified chemodynamical evolutionary
picture remains a challenge.

While the IGS contributes a very small fraction of the stellar mass
of the Galactic bulge, it is a major contributor to the stellar
mass budget of the stellar {\it halo}.  Therefore, it is important to give
further consideration to the reality of its discrimination as a separate
system, which is the topic of the next sub-section.



\subsubsection{Core of Gaia-Enceladus/Sausage?} \label{sec:ges}

Recent studies based on N-body simulations
\citep[e.g.,][]{JeanBaptiste2017,Koppelman2020} have shown that
substructure in kinematic and orbital planes cannot be ascribed to
individual accretion events in a straightforward manner.  In
particular, \cite{JeanBaptiste2017} simulated the evolution of
orbital properties of accreted and {\it in situ} populations in the
event of accretion by single and multiple satellites with varying
merger mass ratio.  The main conclusions from that work have
consequences for the interpretation of our results.

Firstly, \cite{JeanBaptiste2017} show that distributions on the
$E-L_{\rm z}$ plane that resemble qualitatively those presented in
Figures~\ref{fig:ELz_mp} and \ref{fig:GELE} can result from the
accretion of one single satellite with a 1:10 mass ratio.  It can
be seen from their Figure 2 that several Gyr after the occurrence
of the accretion event, satellite remnants are distributed across
a large range of energies and angular momenta, spanning the loci
occupied by both the IGS and GE/S in Figure~\ref{fig:GELE}.  Most
of the substructure in orbital space in their simulations is
associated to the accreted populations, including low energy clumps
resembling that characterising the IGS in Figure~\ref{fig:ELz_mp}.
By the same token, the distribution of the {\it in situ} population
is a lot smoother and strikingly similar to the distribution in the
upper panel of Figure~\ref{fig:ELz_mp} (minus the Splash), and in the
bottom panel of Figure~\ref{fig:action_alpha} in the Appendix.  Their
simulations also show that accretion moves stars formed {\it in
situ} into high energy and low angular momentum orbits, where
accreted stars are usually found.  In fact, {\it in situ} stars
contribute to some of the substructure on the $E-L_{\rm z}$ plane,
although not at the same level as the accreted populations, which
dominate the substrucutre, particularly at low energy.  Along the
same lines, \cite{Koppelman2020} analysed a set of N-body simulations
that match the observed properties of the GE/S debris today.
According to those simulations, the core of the accreted satellite
spirals in and is dissolved, with some of the stars acquiring
slightly prograde orbits.

On the basis of such simulations one might reasonably argue in
favour of a scenario in which the GE/S and the IGS are both part
of the same progenitor.  In that situation, one could conceive that
the core of GE/S, being more massive and denser than the outskirts,
could have sunk into the inner Galaxy under the effect of dynamical
friction, leaving its less bound stellar tenants behind to be
observed in higher energy orbits today, where GE/S stars are found
(c.f. Figure~\ref{fig:ELz_mp}).

It is not easy to reconcile such a scenario with the relative
distribution of IGS and GE/S data in chemical space in
Figure~\ref{fig:le_vs_ge}.  Stars belonging to the IGS on average
present higher mean [Mg/Fe] and lower mean [Fe/H] than their GE/S
counterparts.  As pointed out in Section~\ref{sec:properties}, the
IGS lacks a substantial ``knee-shin'' component to its distribution
in the Mg-Fe plane, unlike the GE/S.  Note that these differences
exist despite the fact that the chemical criteria defining both
samples are identical.  The IGS is clearly less chemically evolved
than the GE/S.  Therefore, if the IGS was indeed the core of the
GE/S, that hypothetical progenitor system would be characterised
by a positive metallicity gradient in the early universe.  Galaxies
in the nearby and intermediate redshift universe are predominantly
characterised by negative metallicity gradients
\citep[e.g.,][]{Stott2014,Goddard2017}.  That is also the behaviour
of satellites of the Milky Way such as the Magellanic Clouds
\citep[e.g.,][]{Cioni2009} and the Sagittarius Dwarf
\citep[e.g.,][]{Hayes2020}.  This is understood as being due to the
fact that star formation in galaxies takes place in an inside-out
manner, since their central regions evolve more quickly than their
outskirts, due to availability of larger amounts of gas for star
formation.  On the other hand, higher redshift galaxies with positive
metallicity gradient have been identified, an occurrence that may
be caused by strong winds associated with intense star formation
in the centres of dwarf galaxies at $z\sim2$ \citep[e.g.,][]{Wang2019}.
It is not clear however how prevalent that phenomenon is.

Secondly, it is worth highlighting additional results presented by
\cite{JeanBaptiste2017} where simulated Milky Way-like galaxies
undergo 1:10 mergers with more than one satellite.  Those lead to
very thorough mixing of the stars from different satellites in
orbital space, particularly in the low energy region of the $E-L_{\rm
z}$ plane, where the IGS was identified.  This suggests that our
accreted sample should include stars from more than one satellite
in the low energy regime.  Under that light our result above is
surprising.  Even if the IGS and the GE/S were indeed separate
systems, we would expect that {\it some} of our low energy accreted
stars show an abundance pattern consistent with an association with
the GE/S.  On the contrary, it is only the high energy stars
($E/10^5\simgreater-2$~km$^2$s$^{-2}$) in our bulge accreted sample
(c.f. Figure~\ref{fig:ins_vs_acc}) that possess abundances that
are consistent with an association with the GE/S.  

We conclude that it is unlikely, though not impossible, that the
IGS is the core of the Gaia Enceladus/Sausage system.  That would
require that the progenitor of the Gaia Enceladus/Sausage system
have a slightly positive metallicity gradient and have a core that
is less chemically evolved than its outskirts, which is uncommon.

\subsection{The Properties of the IGS Progenitor}
\label{sec:properties}

In the previous Sections we showed that the inner $\sim$4~kpc of
the Milky Way hosts a stellar population with the following properties:
{\it (i)} it presents a chemical composition that resembles those
of accreted systems \cite[e.g.,][]{Hayes2018,Mackereth2019,Helmi2020};
{\it (ii)} it is detected as substructure in IoM space; {\it (iii)}
on average it presents slower rotation, higher vertical action, and
slightly higher radial action than metal-rich bulge populations;
{\it (iv)} it is chemically detached from its metal-rich counterparts.
In aggregate these properties suggest the presence of an accreted
stellar population within 4~kpc of the Galactic centre.  In this
Section we use the observed properties of the remnant population
to estimate the mass and history of the putative progenitor system.
This is achieved through comparison with data for a better known
accreted system (GE/S) and the predictions of the EAGLE simulations.

\begin{figure}
	\includegraphics[width=\columnwidth]{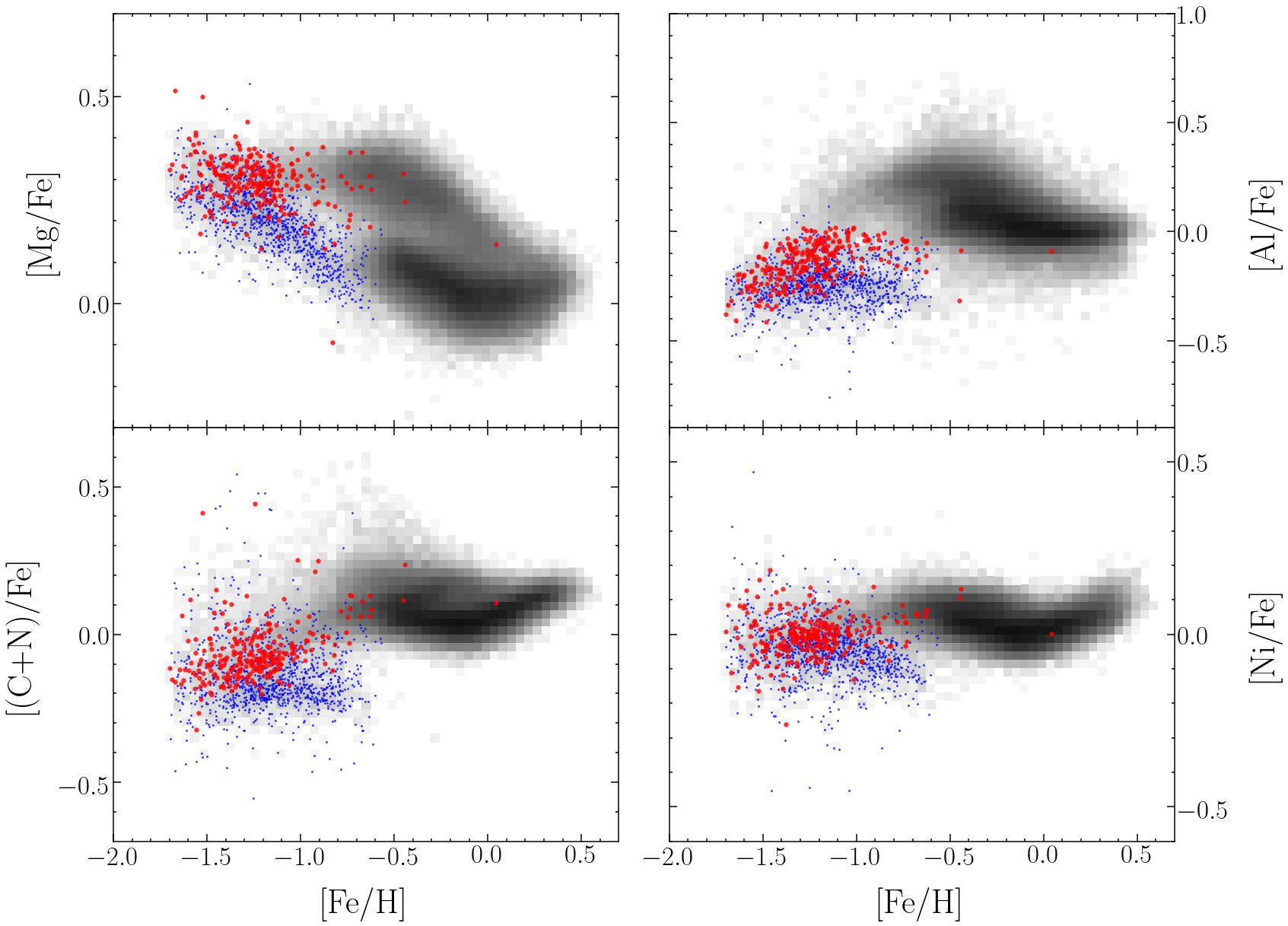}
\caption{Comparison between the IGS (red symbols) and
Gaia-Enceladus/Sausage (GE/S, blue symbols) in chemical space.  The
data for the parent sample are also plotted (gray), to mark the
positions of the low and high-$\alpha$ disks in these diagrams.
The IGS is characterized by higher average [Mg/Fe],
[(C+N)/Fe], and [Ni/Fe] than GE/S while having approximately the
same mean [Fe/H].  This suggests that the progenitor of the IGS
was slightly more massive than GE/S.  It is also noteworthy
from the top left panel that the locus occupied by the IGS
lacks the declining [Mg/Fe] sequence that is visible in the case of GE/S at
[Fe/H]$\sim$-1.2 (the ``shin'' of the $\alpha$-Fe relation).  This may
indicate an early truncation of star formation.
}
    \label{fig:le_vs_ge}
\end{figure}

In Figure~\ref{fig:le_vs_ge} we compare the chemical properties of
the IGS with those of the GE/S system
\citep{Haywood2018,Belokurov2018,Helmi2018,Mackereth2019}, a massive
satellite \citep[$M_\star\sim10^{8.5}{\rm M_\odot}$][]{MackerethBovy2020},
that was accreted to the Milky Way about 8-10 Gyr ago.  For details
on the selection of GE/S stars, see Figure~\ref{fig:GELE} and related
discussion. We ignore, for the sake of simplicity, any selection
effects that may bias the relative distribution of chemical properties
of the observed stellar populations associated with the two galaxies.
In other words, we assume that the samples displayed in
Figure~\ref{fig:GELE} represent the bulk of the populations of the
two systems.  Consequently the conclusions drawn from comparison
of the IGS and GE/S should be considered qualitative, and subject
to revision should that assumption be demonstrated to be incorrect.
With that caveat in mind, Figure~\ref{fig:le_vs_ge} displays stars
from the two systems on various abundance planes.  The IGS is
represented by red symbols, the GE/S as blue symbols, and the parent
sample, which is dominated by stars from the high- and \loa disks,
is plotted as gray-scale 2D histograms.  The data show that stars
belonging to the IGS have slightly higher average Mg, Ni, C, and N
abundances than those of GE/S at [Fe/H]~$\sim-1$.  This result
suggests that the progenitor galaxy of the IGS was likely more
massive than GE/S.  We recall that \cite{Mackereth2019} used
predictions from the EAGLE numerical simulations to estimate the
stellar mass of GE/S.  In particular, they showed that the mean
[Fe/H] and [Mg/Fe] of the stellar particles associated with dwarf
simulated galaxies were good indicators of stellar mass \citep[see
Figure 11 of][]{Mackereth2019}.


We resort to that predicted relation between mass and chemistry in
order to estimate the mass of the progenitor of the IGS.
We used the scipy {\tt curve\_fit} optimisation routine to fit the data for
satellites of Milky Way-like galaxies in the L0025N075-RECAL simulation
\citep[see][for details]{Mackereth2019} with a bivariate
$M_\star$([Fe/H],[Mg/Fe]) relation.  Only galaxies with at least 20 star
particles were included in the fit, excluding all galaxies with
$M_\star<10^{7.5}{\rm M_\odot}$ or $\langle$[Fe/H]$\rangle < -2.0$.


The equation that describes the trend is the following: \begin{multline}
\log M_\star \,=\, {\rm 10.28 + 2.18\,\langle[Fe/H]\rangle +
3.60\,\langle[Mg/Fe]\rangle}  \\ {\rm
-0.30\,\langle[Fe/H]\rangle\times\langle[Mg/Fe]\rangle } \end{multline}
Where the zero-th order term has been adjusted to match the mass
inferred by \cite{MackerethBovy2020} for GE/S, $3\times10^8M_\odot$,
and adopting the following mean values: ${\rm
\langle[Fe/H]\rangle_{GE/S}=-1.20}$, ${\rm
\langle[Mg/Fe]\rangle}_{GE/S}=+0.20$.  In this way, the theoretical
relation is used in a strictly relative way, to minimise uncertainties.
From our IGS, we have mean ${\rm \langle[Fe/H]\rangle_{IGS}=-1.26}$,
${\rm \langle[Mg/Fe]\rangle_{IGS}=+0.30}$, resulting in
$M_\star\sim5\times10^8M_\odot$ for the progenitor of the IGS, or
about twice the mass of GE/S\footnote{For these estimates we lifted
our original constraint on [Fe/H]~$>-1.7$ (c.f. Section~\ref{sec:data})
to avoid biasing our mean abundance values.}.  Consequently the
progenitor of the IGS is likely to have been an important contributor
to the stellar mass budget of the Milky Way halo.

The distribution of the data for the IGS in the Mg-Fe plane may be
used to constrain the history of star formation of the progenitor,
and thus the time of accretion.  One cannot help but notice the
fact that the data for the IGS and GE/S differ in one fundamental
aspect: our IGS sample has almost no stars with [Fe/H]~$\simgreater-1$,
and some of them are likely thick disk contaminants (see
Section~\ref{sec:chemprop}).  That is the regime where the GE/S
displays a strong decline in [Mg/Fe] for increasing [Fe/H], associated
with the onset of enrichment by SN Ia, which takes place on a
timescale of $\sim 10^{8-9}$~year \citep[e.g.,][]{Maoz2012,Nomoto2013}.
Assuming we are not being deceived by an unforeseen selection effect,
the absence of this more metal-rich population in the data for the
IGS may indicate an early quenching of star formation, likely
associated with the accretion to the Milky Way at a very early time.

This result confirms theoretical expectations from the E-MOSAICS
simulations by \cite{Pfeffer2020} and the EAGLE predictions for the
chemical compositions of early accreted satellites
(Figure~\ref{fig:eagleabunds}).  \cite{Pfeffer2020} analysed the
expected orbits of globular clusters connected with accreted
satellites in the E-MOSAICS simulations, concluding that the orbits
of surviving accreted clusters tend to have lower energies if their
host satellites were more massive and/or were accreted at higher
redshifts.  Furthermore, this observed truncation at [Fe/H]~$\sim-1$
is in agreement with results from a previous study by our group
\citep{Horta2020a}, where the chemical compositions of Galactic GCs
in APOGEE were studied.  In that paper we showed that the more
metal-poor half of the GCs associated with the L-E subgroup, whose
position in IoM space coincides with that of the IGS, have chemical
compositions that are consistent with an accreted origin.  It is
possible that the more metal-rich GCs in that group have an {\it
in situ} origin, associated with the thick disk \citep[see Figure~3
from][]{Horta2020a}.

As pointed out in Section~\ref{sec:subiom}, the locus of the IGS
in IoM space matches closely that of the L-E family of
globular clusters identified by \cite{Massari2019}.  Those authors
showed that the L-E clusters followed a tight age-metallicity
relation, suggesting that they might be associated with a massive
satellite.  Indeed, both the orbital properties and time of accretion
of the progenitor of the IGS agree with the predictions
by \cite{Kruijssen2019b,Kruijssen2020} for the {\it Kraken} accretion
event.  According to those authors, the {\it
Kraken} was a satellite of the Milky Way that was accreted within the
first few Gyr of the Galaxy's existence \citep[see also][]{Forbes2020}.
The prediction is based on a detailed comparison of the ages,
metallicities, and orbital properties of the Milky Way globular
cluster system with predictions of the E-MOSAICS suite of cosmological
numerical simulations \citep{Pfeffer2018,Kruijssen2019a}.  Although
the specific identities of the globular clusters associated to the
{\it Kraken} or the L-E group are not unique, their median metallicities
are roughly similar to that of the IGS ([Fe/H]$\sim -1.5$
and $-1.24$, respectively).  The orbital properties of those globular
clusters, with apocentric distances between 3 and 5~kpc, are also
consistent with an association with the IGS.

Finally, \cite{Kruijssen2020} estimate the stellar mass of the {\it
Kraken} to be $\sim 2\times10^8 M_\odot$, or approximately 70\% of
the mass of GE/S.  Our estimate from equation (1) is that the
progenitor of the IGS is more massive than the {\it Kraken}, with
approximately twice the mass of GE/S.  Considering the uncertainties
associated with the various methods, we do not deem the absolute
differences worrying.  Further details on the physical properties
of the progenitor of the IGS, as well as a definitive association
with the {\it Kraken} will have to await a more careful examination
of both upcoming new data and predictions from improved numerical
simulations.

\subsection{A speculative scenario} \label{sec:scenario}

With the above caveats in mind, we conclude by summarising our {\it
speculative} views on the most likely story implied by the data
presented in this paper.  While the vast majority of the stars in
the Milky Way were likely formed {\it in situ}
\citep[e.g.,][]{BlandHawthorn2016}, it is generally agreed that,
under a $\Lambda$-CDM framework, galaxies such as the Milky Way
experienced intense accretion activity in their early lives
\cite[e.g.,][]{NaabOstriker2017}.  



In Section~\ref{sec:ins_vs_acc} we found that somewhere between 1/4
and 1/3 of the {\it metal-poor} stars\footnote{Which themselves
amount to only $\sim$5\% of the stellar mass budget of the Galactic
bulge \citep[see][]{Ness2013b}} in the inner Galaxy were accreted,
and the rest were formed {\it in situ}.  Our inability to perfectly
distinguish one group from the other leaves a key question in the
air: is there an evolutionary connection between the two or were
they formed and evolved in entire isolation before co-habiting the
inner few kpc of the Galaxy?

A possible sequence of events, is one whereby the most metal-poor
stars in the inner Galaxy were formed {\it in situ}, followed by
the early accretion of the progenitor of the IGS which, under the
effect of dynamical friction, took a quick dive towards the heart
of the young Milky Way.  Assuming the IGS brought with it fresh
gas\footnote{This assumes ram pressure stripping was not important,
which may be reasonable. Ram pressure stripping is proportional to
$v^2$ and since the IGS is located at the Galactic centre, the
putative merger must have occurred at a low velocity.}, it is likely
that dissipation brought it to the Galactic centre, triggering a
burst of star formation which would influence the later course of
the chemical evolution in the inner Galaxy, explaining the chemical
continuity apparent in Figure~\ref{fig:abunds}\footnote{Along similar
lines, \cite{Whitten2019} proposed that the steep age gradient of
stellar populations in the inner halo ($R_{\rm GC}<14$~kpc), inferred
from BHB star mean color gradients, can be possibly associated with
occurrence of a few dissipative mergers in the early Milky Way history.}.

%


It will be interesting to see whether such speculative scenarios
will survive scrutiny by more detailed observations and sophisticated
modeling and, if so, how it will jibe with our understanding of the
formation, evolution, and interaction between other structures
co-habiting the inner Galaxy.

\section{Conclusions} \label{sec:conclusions}

We report evidence to the existence of an accreted structure located
in the inner Galaxy (the IGS).  To reveal its existence we resorted
to a search for clustering in multi-dimensional parameter space,
based on the best available data from APOGEE and \emph{Gaia}, and
astroNN distances.  This structure consists of a metal-poor population
with detailed chemistry resembling that of low mass satellites of
the Milky Way with low abundances of Al, C, N, and Ni.  It is
chemically and dynamically detached from the more metal-rich
populations with which it shares its location in the heart of the
Galaxy, and from other accreted populations across the Milky Way
halo.  We suggest that this population is the remnant of a satellite
that was accreted early in the history of the Milky Way.  The
chemistry of this population is characterised by a plateau of
relatively high [Mg/Fe] and the absence of a ``knee'' towards higher
[Fe/H] indicating an early quenching of star formation activity,
which is consistent with a very early accretion event.  Comparison
with mean chemical composition predictions from numerical simulations,
we infer a stellar mass of $M_\star \sim 5\times10^8M_\odot$, or
approximately twice that of the Gaia-Enceladus/Sausage system.  If
indeed its existence is confirmed by further investigation, one
would be led to conclude that its progenitor was a major building
block of the Galactic halo, which has until now remained elusive
due to its location behind a curtain of extinction and foreground
crowding.  The orbital and chemical properties of the IGS, the
implied time of accretion, and the mass of its progenitor are in
good qualitative agreement with those predicted by \cite{Pfeffer2020}
and \cite{Kruijssen2020}.  We summarise below the main implications
and questions raised by this finding:

\begin{itemize}

\item According to our favoured interpretation of the data, the IGS
is a remnant of a satellite galaxy accreted to the Milky Way a long
time ago.  That such an accretion event happened in the Milky Way's
very early history is an expectation from cosmological numerical
simulations that predict an intense accretion activity in massive
galaxies at $z\sim2-3$
\citep[e.g.,][]{ElBadry2018,Pfeffer2020,Fragkoudi2020}.  The Milky
Way itself may have been the subject of unusually strong accretion
activity at those early times, which may explain observed properties
of its disk and halo populations
\citep[e.g.,][]{Mackereth2018,Schiavon2020};

\item We present evidence for an accretion event that took place
{\it before} the main structures inhabiting the region we understand
today as the Galactic bulge were formed.  Thus, the IGS should be
viewed as a major building block of the Galactic halo, rather than
a very minor component of the Milky Way bulge, to whose stellar
mass budget it contributes no more than a few percent.  The vast
majority of the stars in the bulge today were most likely formed
{\it in situ} \citep[e.g.,][]{Kunder2012,DiMatteo2016,Debattista2017,
Fragkoudi2018}.

\item The abundance ratios of stars associated with the IGS are
well aligned with the sequence occupied by more metal-rich populations
located within the inner $\sim$4~kpc of the Galactic centre.  Naively,
that would be a surprising result, as one may expect that accreted
and {\it in situ} populations might have undergone distinct early
histories of star formation and chemical enrichment.  We show that
numerical simulations actually predict such seeming continuity
between the loci occuppied by accreted and {\it in situ} populations
in all abundances tracked by the EAGLE simulations.  



\item As a logical path for further enquiry, one would like to
explore at least three possible avenues: {\it (i)} verify whether
models can match the distribution of stellar density in chemical
space, which may be a key discriminator between accretion and {\it
in situ} formation scenarios; {\it (ii)} contrast accreted and {\it
in situ} populations in other abundance planes including elements
that sample additional nucleosynthetic pathways, such as those
resulting from s- and r-process production.  {\it (iii)} contrast
the age distributions of stars in the chemo-dynamically defined
accreted and {\it in situ} populations.  All these lines of
investigation will benefit from additional data made available in
forthcoming observational efforts.



\item The combined mass of the progenitor of the IGS and the GE/S
adds up to $8\times10^8 M_\odot$.  Considering mass estimates for
other accreted systems
\citep[e.g.,][]{Koppelman2019,Forbes2020,Kruijssen2020} and dissolved
globular clusters \citep[]{Horta2020b,Schiavon2017},
the total mass in accreted satellites may be well in excess of
$\sim1.6\times10^9M_\odot$.  Although all these mass estimates are
subject to considerable zero point uncertainties, this result is,
at face value, in mild tension with independent estimates of the
total stellar mass of the Galactic halo
\citep[e.g.,][]{Deason2019,MackerethBovy2020}.

\item There seems to be an interesting connection between the finding
reported in this paper and the presence of a population of N-rich
stars in the inner Galaxy \citep{Schiavon2017}, which is generally
interpreted as resulting from the destruction of an early population
of globular clusters
\citep[e.g.,][]{Martell2010,Martell2016,Koch2019,Savino2019,Hanke2020,Trincado2020}.
The frequency of this population has been shown to be larger by an
order of magnitude within the inner few kpc of the halo than in its
outer regions \citep{Horta2020b}.  Moreover, Kisku et
al. (2020, in prep.) show that roughly 1/4 of the N-rich population
within $\sim4$~kpc of the Galactic centre has elemental abundances
consistent with an accretion origin, and orbital properties that
are similar to those of the IGS.  In future work we intend to examine
how these finding jibe together vis \`a vis our current understanding
of the accretion history of the Milky Way and models for globular
cluster formation and destruction in a cosmological context.

\end{itemize}

This paper presents an exploration of the chemical complexity of
the metal-poor bulge on the basis of a statistically significant
sample.  The window into the bulge metal-poor bulge populations
will be further widened bt observational projects taking place in
the near future.  The Pristine Inner Galaxy Survey (PIGS)
is performing AAOmega+2dF spectroscopy with the 3.9~m Anglo-Australian
Telescope of $\sim$8,000 targets selected on the basis of CaHK
photometry collected with MegaCam on the Canada-France-Hawaii
Telescope, \citep{Arentsen2020b} amassed the largest sample to date
of very metal-poor stars ($\sim$1,300 stars with [Fe/H]$<-2$) in
the inner Galaxy.  Detailed abundance analysis of such a sample
will certainly shed light onto the nature of the early populations
of the bulge.  Further into the near future, the MOONS\footnote{MOONS
is the European Southern Observatory Multi Object Optical and
Near-infrared Spectrograph for the VLT \citep{Cirasuolo2016}.}
surveys of the Milky Way \citep{Gonzalez2020}
will likely increase the sample of bulge metal-poor stars
with detailed chemical compositions by an order of magnitude.

\section*{Acknowledgements}

We thank the many people around the world whose hard work fighting
the ongoing COVID-19 pandemic has made it possible for us to remain
safe and heathy for the past several months.  The authors thank
Nate Bastian and Marie Martig for helpful discussions, and Tim
Beers, Vasily Belokurov, Christian Moni Bidin, Cristina Chiappini,
Roger Cohen, Chris Hayes, Anibal Garc\'\i a Hern\'andez, Dante
Minniti, Christian Nitschelm, and Jos\'e Fern\'andez-Trincado, for
very helpful comments on the original manuscript. DH thanks Sue, Alex and Debra for everything, and acknowledges
an STFC studentship.  RPS thanks Ingrid, Oliver, and Luke, so deeply. JTM acknowledges support from the ERC Consolidator Grant funding scheme (project ASTEROCHRONOMETRY, G.A. n. 772293), the Banting Postdoctoral Fellowship programme administered by the Government of Canada, and a CITA/Dunlap Institute fellowship. The Dunlap Institute is funded through an endowment established by the David Dunlap family and the University of Toronto.  DMN
acknowledges support from NASA under award Number 80NSSC19K0589,
and support from the Allan C. And Dorothy H. Davis Fellowship. S.H.
is supported by an NSF Astronomy and Astrophysics Postdoctoral
Fellowship under award AST-1801940.  JP gratefully acknowledges
funding from a European Research Council consolidator grant
(ERC-CoG-646928-Multi-Pop).  This research made use of
Astropy\footnote{http://www.astropy.org} a community-developed core
Python package for Astronomy, \citep{astropy:2013,astropy:2018}.

Funding for the Sloan Digital Sky Survey IV has been provided by
the Alfred P. Sloan Foundation, the U.S. Department of Energy Office
of Science, and the Participating Institutions. SDSS acknowledges
support and resources from the Center for High-Performance Computing
at the University of Utah. The SDSS web site is www.sdss.org. SDSS
is managed by the Astrophysical Research Consortium for the
Participating Institutions of the SDSS Collaboration including the
Brazilian Participation Group, the Carnegie Institution for Science,
Carnegie Mellon University, the Chilean Participation Group, the
French Participation Group, Harvard-Smithsonian Center for Astrophysics,
Instituto de Astrof\'{i}sica de Canarias, The Johns Hopkins University,
Kavli Institute for the Physics and Mathematics of the Universe
(IPMU) / University of Tokyo, the Korean Participation Group,
Lawrence Berkeley National Laboratory, Leibniz Institut f\"{u}r Astrophysik
Potsdam (AIP), Max-Planck-Institut f\"{u}r Astronomie (MPIA Heidelberg),
Max-Planck-Institut f\"{u}r Astrophysik (MPA Garching), Max-Planck-Institut
f\"{u}r Extraterrestrische Physik (MPE), National Astronomical Observatories
of China, New Mexico State University, New York University, University
of Notre Dame, Observatório Nacional / MCTI, The Ohio State University,
Pennsylvania State University, Shanghai Astronomical Observatory,
United Kingdom Participation Group, Universidad Nacional Autónoma
de México, University of Arizona, University of Colorado Boulder,
University of Oxford, University of Portsmouth, University of Utah,
University of Virginia, University of Washington, University of
Wisconsin, Vanderbilt University, and Yale University.

{\it Software:} Astropy \citep{astropy:2013,astropy:2018}, SciPy
\citep*{SciPy}, NumPy \citep{NumPy}, Matplotlib \citep{Hunter:2007},
Galpy \citep{Bovy2015,MackerethBovy2018}, TOPCAT \citep{Taylor2005}.

\section*{Data availability}
All APOGEE DR16 data used in this study is publicly available and can be found at: https: //www.sdss.org/dr16/



\bibliographystyle{mnras}
\bibliography{ref} 

\begin{thebibliography}{}
\makeatletter
\relax
\def\mn@urlcharsother{\let\do\@makeother \do\$\do\&\do\#\do\^\do\_\do\%\do\~}
\def\mn@doi{\begingroup\mn@urlcharsother \@ifnextchar [ {\mn@doi@}
  {\mn@doi@[]}}
\def\mn@doi@[#1]#2{\def\@tempa{#1}\ifx\@tempa\@empty \href
  {http://dx.doi.org/#2} {doi:#2}\else \href {http://dx.doi.org/#2} {#1}\fi
  \endgroup}
\def\mn@eprint#1#2{\mn@eprint@#1:#2::\@nil}
\def\mn@eprint@arXiv#1{\href {http://arxiv.org/abs/#1} {{\tt arXiv:#1}}}
\def\mn@eprint@dblp#1{\href {http://dblp.uni-trier.de/rec/bibtex/#1.xml}
  {dblp:#1}}
\def\mn@eprint@#1:#2:#3:#4\@nil{\def\@tempa {#1}\def\@tempb {#2}\def\@tempc
  {#3}\ifx \@tempc \@empty \let \@tempc \@tempb \let \@tempb \@tempa \fi \ifx
  \@tempb \@empty \def\@tempb {arXiv}\fi \@ifundefined
  {mn@eprint@\@tempb}{\@tempb:\@tempc}{\expandafter \expandafter \csname
  mn@eprint@\@tempb\endcsname \expandafter{\@tempc}}}

\bibitem[\protect\citeauthoryear{{Ahumada} et~al.,}{{Ahumada}
  et~al.}{2019}]{DR16}
{Ahumada} R.,  et~al., 2019, arXiv e-prints, \href
  {https://ui.adsabs.harvard.edu/abs/2019arXiv191202905A} {p. arXiv:1912.02905}

\bibitem[\protect\citeauthoryear{{Andrews}, {Weinberg}, {Sch{\"o}nrich}  \&
  {Johnson}}{{Andrews} et~al.}{2017}]{Andrews2017}
{Andrews} B.~H.,  {Weinberg} D.~H.,  {Sch{\"o}nrich} R.,   {Johnson} J.~A.,
  2017, \mn@doi [\apj] {10.3847/1538-4357/835/2/224}, \href
  {https://ui.adsabs.harvard.edu/abs/2017ApJ...835..224A} {835, 224}

\bibitem[\protect\citeauthoryear{{Arentsen} et~al.,}{{Arentsen}
  et~al.}{2020a}]{Arentsen2020b}
{Arentsen} A.,  et~al., 2020a, arXiv e-prints, \href
  {https://ui.adsabs.harvard.edu/abs/2020arXiv200608641A} {p. arXiv:2006.08641}

\bibitem[\protect\citeauthoryear{{Arentsen} et~al.,}{{Arentsen}
  et~al.}{2020b}]{Arentsen2020a}
{Arentsen} A.,  et~al., 2020b, \mn@doi [\mnras] {10.1093/mnrasl/slz156}, \href
  {https://ui.adsabs.harvard.edu/abs/2020MNRAS.491L..11A} {491, L11}

\bibitem[\protect\citeauthoryear{{Astropy Collaboration} et~al.,}{{Astropy
  Collaboration} et~al.}{2013}]{astropy:2013}
{Astropy Collaboration} et~al., 2013, \mn@doi [aap]
  {10.1051/0004-6361/201322068}, \href
  {http://adsabs.harvard.edu/abs/2013A%26A...558A..33A} {558, A33}

\bibitem[\protect\citeauthoryear{{Barbuy}, {Chiappini}  \& {Gerhard}}{{Barbuy}
  et~al.}{2018}]{Barbuy2018}
{Barbuy} B.,  {Chiappini} C.,   {Gerhard} O.,  2018, \mn@doi [\araa]
  {10.1146/annurev-astro-081817-051826}, \href
  {https://ui.adsabs.harvard.edu/abs/2018ARA&A..56..223B} {56, 223}

\bibitem[\protect\citeauthoryear{{Belokurov} et~al.,}{{Belokurov}
  et~al.}{2006}]{Belokurov2006}
{Belokurov} V.,  et~al., 2006, \mn@doi [\apjl] {10.1086/504797}, \href
  {https://ui.adsabs.harvard.edu/abs/2006ApJ...642L.137B} {642, L137}

\bibitem[\protect\citeauthoryear{{Belokurov}, {Erkal}, {Evans}, {Koposov}  \&
  {Deason}}{{Belokurov} et~al.}{2018}]{Belokurov2018}
{Belokurov} V.,  {Erkal} D.,  {Evans} N.~W.,  {Koposov} S.~E.,   {Deason}
  A.~J.,  2018, \mn@doi [\mnras] {10.1093/mnras/sty982}, \href
  {https://ui.adsabs.harvard.edu/abs/2018MNRAS.478..611B} {478, 611}

\bibitem[\protect\citeauthoryear{{Belokurov}, {Sanders}, {Fattahi}, {Smith},
  {Deason}, {Evans}  \& {Grand }}{{Belokurov} et~al.}{2019}]{Belokurov2019}
{Belokurov} V.,  {Sanders} J.~L.,  {Fattahi} A.,  {Smith} M.~C.,  {Deason}
  A.~J.,  {Evans} N.~W.,   {Grand } R. J.~J.,  2019, arXiv e-prints, \href
  {https://ui.adsabs.harvard.edu/abs/2019arXiv190904679B} {p. arXiv:1909.04679}

\bibitem[\protect\citeauthoryear{{Bensby}, {Feltzing}  \& {Oey}}{{Bensby}
  et~al.}{2014}]{Bensby2014}
{Bensby} T.,  {Feltzing} S.,   {Oey} M.~S.,  2014, \mn@doi [\aap]
  {10.1051/0004-6361/201322631}, \href
  {https://ui.adsabs.harvard.edu/abs/2014A&A...562A..71B} {562, A71}

\bibitem[\protect\citeauthoryear{{Bland-Hawthorn} \&
  {Gerhard}}{{Bland-Hawthorn} \& {Gerhard}}{2016}]{BlandHawthorn2016}
{Bland-Hawthorn} J.,  {Gerhard} O.,  2016, \mn@doi [\araa]
  {10.1146/annurev-astro-081915-023441}, \href
  {https://ui.adsabs.harvard.edu/abs/2016ARA&A..54..529B} {54, 529}

\bibitem[\protect\citeauthoryear{{Blanton} et~al.,}{{Blanton}
  et~al.}{2017}]{Blanton2017}
{Blanton} M.~R.,  et~al., 2017, \mn@doi [\aj] {10.3847/1538-3881/aa7567}, \href
  {https://ui.adsabs.harvard.edu/abs/2017AJ....154...28B} {154, 28}

\bibitem[\protect\citeauthoryear{{Bovy}}{{Bovy}}{2015}]{Bovy2015}
{Bovy} J.,  2015, \mn@doi [\apjs] {10.1088/0067-0049/216/2/29}, \href
  {https://ui.adsabs.harvard.edu/abs/2015ApJS..216...29B} {216, 29}

\bibitem[\protect\citeauthoryear{{Bovy}, {Rix}, {Liu}, {Hogg}, {Beers}  \&
  {Lee}}{{Bovy} et~al.}{2012}]{Bovy2012}
{Bovy} J.,  {Rix} H.-W.,  {Liu} C.,  {Hogg} D.~W.,  {Beers} T.~C.,   {Lee}
  Y.~S.,  2012, \mn@doi [\apj] {10.1088/0004-637X/753/2/148}, \href
  {https://ui.adsabs.harvard.edu/abs/2012ApJ...753..148B} {753, 148}

\bibitem[\protect\citeauthoryear{{Bowen} \& {Vaughan}}{{Bowen} \&
  {Vaughan}}{1973}]{BowenVaughan1973}
{Bowen} I.~S.,  {Vaughan} A.~H. J.,  1973, \mn@doi [\ao]
  {10.1364/AO.12.001430}, \href
  {https://ui.adsabs.harvard.edu/abs/1973ApOpt..12.1430B} {12, 1430}

\bibitem[\protect\citeauthoryear{Campello, Moulavi  \& Sander}{Campello
  et~al.}{2013}]{Campelo2013}
Campello R. J. G.~B.,  Moulavi D.,   Sander J.,  2013, in Pei J.,  Tseng V.~S.,
   Cao L.,  Motoda H.,   Xu G.,  eds, Advances in Knowledge Discovery and Data
  Mining. Springer Berlin Heidelberg, Berlin, Heidelberg, pp 160--172

\bibitem[\protect\citeauthoryear{{Chiappini}, {Romano}  \&
  {Matteucci}}{{Chiappini} et~al.}{2003}]{Chiappini2003}
{Chiappini} C.,  {Romano} D.,   {Matteucci} F.,  2003, \mn@doi [\mnras]
  {10.1046/j.1365-8711.2003.06154.x}, \href
  {https://ui.adsabs.harvard.edu/abs/2003MNRAS.339...63C} {339, 63}

\bibitem[\protect\citeauthoryear{{Cioni}}{{Cioni}}{2009}]{Cioni2009}
{Cioni} M. R.~L.,  2009, \mn@doi [\aap] {10.1051/0004-6361/200912138}, \href
  {https://ui.adsabs.harvard.edu/abs/2009A&A...506.1137C} {506, 1137}

\bibitem[\protect\citeauthoryear{{Cirasuolo} \& {MOONS Consortium}}{{Cirasuolo}
  \& {MOONS Consortium}}{2016}]{Cirasuolo2016}
{Cirasuolo} M.,  {MOONS Consortium} 2016, in {Skillen} I.,  {Balcells} M.,
  {Trager} S.,  eds,  Astronomical Society of the Pacific Conference Series
  Vol. 507, Multi-Object Spectroscopy in the Next Decade: Big Questions, Large
  Surveys, and Wide Fields. p.~109

\bibitem[\protect\citeauthoryear{{Conroy} et~al.,}{{Conroy}
  et~al.}{2019}]{Conroy2019}
{Conroy} C.,  et~al., 2019, \mn@doi [\apj] {10.3847/1538-4357/ab38b8}, \href
  {https://ui.adsabs.harvard.edu/abs/2019ApJ...883..107C} {883, 107}

\bibitem[\protect\citeauthoryear{{Crain} et~al.,}{{Crain}
  et~al.}{2015}]{Crain2015}
{Crain} R.~A.,  et~al., 2015, \mn@doi [\mnras] {10.1093/mnras/stv725}, \href
  {https://ui.adsabs.harvard.edu/abs/2015MNRAS.450.1937C} {450, 1937}

\bibitem[\protect\citeauthoryear{{Das}, {Hawkins}  \& {Jofr{\'e}}}{{Das}
  et~al.}{2020}]{Das2020}
{Das} P.,  {Hawkins} K.,   {Jofr{\'e}} P.,  2020, \mn@doi [\mnras]
  {10.1093/mnras/stz3537}, \href
  {https://ui.adsabs.harvard.edu/abs/2020MNRAS.tmp..325D} {}

\bibitem[\protect\citeauthoryear{{Deason}, {Belokurov}  \& {Sanders}}{{Deason}
  et~al.}{2019}]{Deason2019}
{Deason} A.~J.,  {Belokurov} V.,   {Sanders} J.~L.,  2019, \mn@doi [\mnras]
  {10.1093/mnras/stz2793}, \href
  {https://ui.adsabs.harvard.edu/abs/2019MNRAS.490.3426D} {490, 3426}

\bibitem[\protect\citeauthoryear{{Debattista}, {Ness}, {Gonzalez}, {Freeman},
  {Zoccali}  \& {Minniti}}{{Debattista} et~al.}{2017}]{Debattista2017}
{Debattista} V.~P.,  {Ness} M.,  {Gonzalez} O.~A.,  {Freeman} K.,  {Zoccali}
  M.,   {Minniti} D.,  2017, \mn@doi [\mnras] {10.1093/mnras/stx947}, \href
  {https://ui.adsabs.harvard.edu/abs/2017MNRAS.469.1587D} {469, 1587}

\bibitem[\protect\citeauthoryear{{Di Matteo}}{{Di Matteo}}{2016}]{DiMatteo2016}
{Di Matteo} P.,  2016, \mn@doi [\pasa] {10.1017/pasa.2016.11}, \href
  {https://ui.adsabs.harvard.edu/abs/2016PASA...33...27D} {33, e027}

\bibitem[\protect\citeauthoryear{{Eggen}, {Lynden-Bell}  \& {Sandage}}{{Eggen}
  et~al.}{1962}]{ELS}
{Eggen} O.~J.,  {Lynden-Bell} D.,   {Sandage} A.~R.,  1962, \mn@doi [\apj]
  {10.1086/147433}, \href
  {https://ui.adsabs.harvard.edu/abs/1962ApJ...136..748E} {136, 748}

\bibitem[\protect\citeauthoryear{{El-Badry} et~al.,}{{El-Badry}
  et~al.}{2018}]{ElBadry2018}
{El-Badry} K.,  et~al., 2018, \mn@doi [\mnras] {10.1093/mnras/sty1864}, \href
  {https://ui.adsabs.harvard.edu/abs/2018MNRAS.480..652E} {480, 652}

\bibitem[\protect\citeauthoryear{{Fern{\'a}ndez-Trincado}, {Chaves-Velasquez},
  {P{\'e}rez-Villegas}, {Vieira}, {Moreno}, {Ortigoza-Urdaneta}  \&
  {Vega-Neme}}{{Fern{\'a}ndez-Trincado} et~al.}{2020}]{Trincado2020}
{Fern{\'a}ndez-Trincado} J.~G.,  {Chaves-Velasquez} L.,  {P{\'e}rez-Villegas}
  A.,  {Vieira} K.,  {Moreno} E.,  {Ortigoza-Urdaneta} M.,   {Vega-Neme} L.,
  2020, \mn@doi [\mnras] {10.1093/mnras/staa1386}, \href
  {https://ui.adsabs.harvard.edu/abs/2020MNRAS.495.4113F} {495, 4113}

\bibitem[\protect\citeauthoryear{{Forbes}}{{Forbes}}{2020}]{Forbes2020}
{Forbes} D.~A.,  2020, \mn@doi [\mnras] {10.1093/mnras/staa245}, \href
  {https://ui.adsabs.harvard.edu/abs/2020MNRAS.tmp..233F} {}

\bibitem[\protect\citeauthoryear{{Fragkoudi}, {Di Matteo}, {Haywood},
  {Schultheis}, {Khoperskov}, {G{\'o}mez}  \& {Combes}}{{Fragkoudi}
  et~al.}{2018}]{Fragkoudi2018}
{Fragkoudi} F.,  {Di Matteo} P.,  {Haywood} M.,  {Schultheis} M.,  {Khoperskov}
  S.,  {G{\'o}mez} A.,   {Combes} F.,  2018, \mn@doi [\aap]
  {10.1051/0004-6361/201732509}, \href
  {https://ui.adsabs.harvard.edu/abs/2018A&A...616A.180F} {616, A180}

\bibitem[\protect\citeauthoryear{{Fragkoudi} et~al.,}{{Fragkoudi}
  et~al.}{2020}]{Fragkoudi2020}
{Fragkoudi} F.,  et~al., 2020, \mn@doi [\mnras] {10.1093/mnras/staa1104}, \href
  {https://ui.adsabs.harvard.edu/abs/2020MNRAS.494.5936F} {494, 5936}

\bibitem[\protect\citeauthoryear{{Fulbright}, {McWilliam}  \&
  {Rich}}{{Fulbright} et~al.}{2007}]{Fulbright2007}
{Fulbright} J.~P.,  {McWilliam} A.,   {Rich} R.~M.,  2007, \mn@doi [\apj]
  {10.1086/513710}, \href
  {https://ui.adsabs.harvard.edu/abs/2007ApJ...661.1152F} {661, 1152}

\bibitem[\protect\citeauthoryear{{Gaia Collaboration} et~al.,}{{Gaia
  Collaboration} et~al.}{2018}]{Gaiadr2}
{Gaia Collaboration} et~al., 2018, \mn@doi [\aap]
  {10.1051/0004-6361/201833051}, \href
  {https://ui.adsabs.harvard.edu/abs/2018A&A...616A...1G} {616, A1}

\bibitem[\protect\citeauthoryear{{Garc{\'\i}a P{\'e}rez} et~al.,}{{Garc{\'\i}a
  P{\'e}rez} et~al.}{2013}]{GarciaPerez2013}
{Garc{\'\i}a P{\'e}rez} A.~E.,  et~al., 2013, \mn@doi [\apjl]
  {10.1088/2041-8205/767/1/L9}, \href
  {https://ui.adsabs.harvard.edu/abs/2013ApJ...767L...9G} {767, L9}

\bibitem[\protect\citeauthoryear{{Garc{\'\i}a P{\'e}rez} et~al.,}{{Garc{\'\i}a
  P{\'e}rez} et~al.}{2016}]{GarciaPerez2016}
{Garc{\'\i}a P{\'e}rez} A.~E.,  et~al., 2016, \mn@doi [\aj]
  {10.3847/0004-6256/151/6/144}, \href
  {https://ui.adsabs.harvard.edu/abs/2016AJ....151..144G} {151, 144}

\bibitem[\protect\citeauthoryear{{Garc{\'\i}a P{\'e}rez} et~al.,}{{Garc{\'\i}a
  P{\'e}rez} et~al.}{2018}]{GarciaPerez2018}
{Garc{\'\i}a P{\'e}rez} A.~E.,  et~al., 2018, \mn@doi [\apj]
  {10.3847/1538-4357/aa9d88}, \href
  {https://ui.adsabs.harvard.edu/abs/2018ApJ...852...91G} {852, 91}

\bibitem[\protect\citeauthoryear{{Gilmore} et~al.,}{{Gilmore}
  et~al.}{2012}]{GES}
{Gilmore} G.,  et~al., 2012, The Messenger, \href
  {https://ui.adsabs.harvard.edu/abs/2012Msngr.147...25G} {147, 25}

\bibitem[\protect\citeauthoryear{{Goddard} et~al.,}{{Goddard}
  et~al.}{2017}]{Goddard2017}
{Goddard} D.,  et~al., 2017, \mn@doi [\mnras] {10.1093/mnras/stw2719}, \href
  {https://ui.adsabs.harvard.edu/abs/2017MNRAS.465..688G} {465, 688}

\bibitem[\protect\citeauthoryear{{Gonzalez} et~al.,}{{Gonzalez}
  et~al.}{2011}]{Gonzalez2011}
{Gonzalez} O.~A.,  et~al., 2011, \mn@doi [\aap] {10.1051/0004-6361/201116548},
  \href {https://ui.adsabs.harvard.edu/abs/2011A&A...530A..54G} {530, A54}

\bibitem[\protect\citeauthoryear{{Gonzalez} et~al.,}{{Gonzalez}
  et~al.}{2020}]{Gonzalez2020}
{Gonzalez} O.~A.,  et~al., 2020, arXiv e-prints, \href
  {https://ui.adsabs.harvard.edu/abs/2020arXiv200900635G} {p. arXiv:2009.00635}

\bibitem[\protect\citeauthoryear{{Gunn} et~al.,}{{Gunn}
  et~al.}{2006}]{Gunn2006}
{Gunn} J.~E.,  et~al., 2006, \mn@doi [\aj] {10.1086/500975}, \href
  {https://ui.adsabs.harvard.edu/abs/2006AJ....131.2332G} {131, 2332}

\bibitem[\protect\citeauthoryear{{Hanke}, {Koch}, {Prudil}, {Grebel}  \&
  {Bastian}}{{Hanke} et~al.}{2020}]{Hanke2020}
{Hanke} M.,  {Koch} A.,  {Prudil} Z.,  {Grebel} E.~K.,   {Bastian} U.,  2020,
  \mn@doi [\aap] {10.1051/0004-6361/202037853}, \href
  {https://ui.adsabs.harvard.edu/abs/2020A&A...637A..98H} {637, A98}

\bibitem[\protect\citeauthoryear{{Hasselquist} et~al.,}{{Hasselquist}
  et~al.}{2020}]{Hasselquist2020}
{Hasselquist} S.,  et~al., 2020, arXiv e-prints, \href
  {https://ui.adsabs.harvard.edu/abs/2020arXiv200803603H} {p. arXiv:2008.03603}

\bibitem[\protect\citeauthoryear{{Hawkins}, {Jofr{\'e}}, {Masseron}  \&
  {Gilmore}}{{Hawkins} et~al.}{2015}]{Hawkins2015}
{Hawkins} K.,  {Jofr{\'e}} P.,  {Masseron} T.,   {Gilmore} G.,  2015, \mn@doi
  [\mnras] {10.1093/mnras/stv1586}, \href
  {https://ui.adsabs.harvard.edu/abs/2015MNRAS.453..758H} {453, 758}

\bibitem[\protect\citeauthoryear{{Hayden} et~al.,}{{Hayden}
  et~al.}{2015}]{Hayden2015}
{Hayden} M.~R.,  et~al., 2015, \mn@doi [\apj] {10.1088/0004-637X/808/2/132},
  \href {https://ui.adsabs.harvard.edu/abs/2015ApJ...808..132H} {808, 132}

\bibitem[\protect\citeauthoryear{{Hayes} et~al.,}{{Hayes}
  et~al.}{2018}]{Hayes2018}
{Hayes} C.~R.,  et~al., 2018, \mn@doi [\apj] {10.3847/1538-4357/aa9cec}, \href
  {https://ui.adsabs.harvard.edu/abs/2018ApJ...852...49H} {852, 49}

\bibitem[\protect\citeauthoryear{{Hayes} et~al.,}{{Hayes}
  et~al.}{2020}]{Hayes2020}
{Hayes} C.~R.,  et~al., 2020, \mn@doi [\apj] {10.3847/1538-4357/ab62ad}, \href
  {https://ui.adsabs.harvard.edu/abs/2020ApJ...889...63H} {889, 63}

\bibitem[\protect\citeauthoryear{{Haywood}, {Di Matteo}, {Lehnert}, {Snaith},
  {Khoperskov}  \& {G{\'o}mez}}{{Haywood} et~al.}{2018}]{Haywood2018}
{Haywood} M.,  {Di Matteo} P.,  {Lehnert} M.~D.,  {Snaith} O.,  {Khoperskov}
  S.,   {G{\'o}mez} A.,  2018, \mn@doi [\apj] {10.3847/1538-4357/aad235}, \href
  {https://ui.adsabs.harvard.edu/abs/2018ApJ...863..113H} {863, 113}

\bibitem[\protect\citeauthoryear{{Helmi}}{{Helmi}}{2020}]{Helmi2020}
{Helmi} A.,  2020, arXiv e-prints, \href
  {https://ui.adsabs.harvard.edu/abs/2020arXiv200204340H} {p. arXiv:2002.04340}

\bibitem[\protect\citeauthoryear{{Helmi}, {White}, {de Zeeuw}  \&
  {Zhao}}{{Helmi} et~al.}{1999}]{Helmi1999}
{Helmi} A.,  {White} S. D.~M.,  {de Zeeuw} P.~T.,   {Zhao} H.,  1999, \mn@doi
  [\nat] {10.1038/46980}, \href
  {https://ui.adsabs.harvard.edu/abs/1999Natur.402...53H} {402, 53}

\bibitem[\protect\citeauthoryear{{Helmi}, {Babusiaux}, {Koppelman}, {Massari},
  {Veljanoski}  \& {Brown}}{{Helmi} et~al.}{2018}]{Helmi2018}
{Helmi} A.,  {Babusiaux} C.,  {Koppelman} H.~H.,  {Massari} D.,  {Veljanoski}
  J.,   {Brown} A. G.~A.,  2018, \mn@doi [\nat] {10.1038/s41586-018-0625-x},
  \href {https://ui.adsabs.harvard.edu/abs/2018Natur.563...85H} {563, 85}

\bibitem[\protect\citeauthoryear{{Holtzman} et~al.,}{{Holtzman}
  et~al.}{2015}]{Holtzman2015}
{Holtzman} J.~A.,  et~al., 2015, \mn@doi [\aj] {10.1088/0004-6256/150/5/148},
  \href {https://ui.adsabs.harvard.edu/abs/2015AJ....150..148H} {150, 148}

\bibitem[\protect\citeauthoryear{{Holtzman} et~al.,}{{Holtzman}
  et~al.}{2018}]{Holtzman2018}
{Holtzman} J.~A.,  et~al., 2018, \mn@doi [\aj] {10.3847/1538-3881/aad4f9},
  \href {https://ui.adsabs.harvard.edu/abs/2018AJ....156..125H} {156, 125}

\bibitem[\protect\citeauthoryear{{Horta} et~al.,}{{Horta}
  et~al.}{2020a}]{Horta2020b}
{Horta} D.,  et~al., 2020a, arXiv e-prints, \href
  {https://ui.adsabs.harvard.edu/abs/2020arXiv200801097H} {p. arXiv:2008.01097}

\bibitem[\protect\citeauthoryear{{Horta} et~al.,}{{Horta}
  et~al.}{2020b}]{Horta2020a}
{Horta} D.,  et~al., 2020b, \mn@doi [\mnras] {10.1093/mnras/staa478}, \href
  {https://ui.adsabs.harvard.edu/abs/2020MNRAS.493.3363H} {493, 3363}

\bibitem[\protect\citeauthoryear{{Howard}, {Rich}, {Reitzel}, {Koch}, {De
  Propris}  \& {Zhao}}{{Howard} et~al.}{2008}]{Howard2008}
{Howard} C.~D.,  {Rich} R.~M.,  {Reitzel} D.~B.,  {Koch} A.,  {De Propris} R.,
   {Zhao} H.,  2008, \mn@doi [\apj] {10.1086/592106}, \href
  {https://ui.adsabs.harvard.edu/abs/2008ApJ...688.1060H} {688, 1060}

\bibitem[\protect\citeauthoryear{Hunter}{Hunter}{2007}]{Hunter:2007}
Hunter J.~D.,  2007, \mn@doi [Computing In Science \& Engineering]
  {10.1109/MCSE.2007.55}, 9, 90

\bibitem[\protect\citeauthoryear{{Ibata}, {Gilmore}  \& {Irwin}}{{Ibata}
  et~al.}{1994}]{Ibata1994}
{Ibata} R.~A.,  {Gilmore} G.,   {Irwin} M.~J.,  1994, \mn@doi [\nat]
  {10.1038/370194a0}, \href
  {https://ui.adsabs.harvard.edu/abs/1994Natur.370..194I} {370, 194}

\bibitem[\protect\citeauthoryear{{Iorio}, {Belokurov}, {Erkal}, {Koposov},
  {Nipoti}  \& {Fraternali}}{{Iorio} et~al.}{2018}]{Iorio2018}
{Iorio} G.,  {Belokurov} V.,  {Erkal} D.,  {Koposov} S.~E.,  {Nipoti} C.,
  {Fraternali} F.,  2018, \mn@doi [\mnras] {10.1093/mnras/stx2819}, \href
  {https://ui.adsabs.harvard.edu/abs/2018MNRAS.474.2142I} {474, 2142}

\bibitem[\protect\citeauthoryear{{Jean-Baptiste}, {Di Matteo}, {Haywood},
  {G{\'o}mez}, {Montuori}, {Combes}  \& {Semelin}}{{Jean-Baptiste}
  et~al.}{2017}]{JeanBaptiste2017}
{Jean-Baptiste} I.,  {Di Matteo} P.,  {Haywood} M.,  {G{\'o}mez} A.,
  {Montuori} M.,  {Combes} F.,   {Semelin} B.,  2017, \mn@doi [\aap]
  {10.1051/0004-6361/201629691}, \href
  {https://ui.adsabs.harvard.edu/abs/2017A&A...604A.106J} {604, A106}

\bibitem[\protect\citeauthoryear{{Jones}, {Olyphant}  \& {Peterson}}{{Jones}
  et~al.}{01  }]{SciPy}
{Jones} E.,  {Olyphant} T.,   {Peterson} P.,  2001--, {SciPy}: Open source
  scientific tools for {Python}, \url {http://www.scipy.org/}

\bibitem[\protect\citeauthoryear{{J{\"o}nsson} et~al.,}{{J{\"o}nsson}
  et~al.}{2018}]{Jonsson2018}
{J{\"o}nsson} H.,  et~al., 2018, \mn@doi [\aj] {10.3847/1538-3881/aad4f5},
  \href {https://ui.adsabs.harvard.edu/abs/2018AJ....156..126J} {156, 126}

\bibitem[\protect\citeauthoryear{{J{\"o}nsson} et~al.,}{{J{\"o}nsson}
  et~al.}{2020}]{Jonsson2020}
{J{\"o}nsson} H.,  et~al., 2020, arXiv e-prints, \href
  {https://ui.adsabs.harvard.edu/abs/2020arXiv200705537J} {p. arXiv:2007.05537}

\bibitem[\protect\citeauthoryear{{Kobayashi} \& {Nakasato}}{{Kobayashi} \&
  {Nakasato}}{2011}]{Kobayashi2011}
{Kobayashi} C.,  {Nakasato} N.,  2011, \mn@doi [\apj]
  {10.1088/0004-637X/729/1/16}, \href
  {https://ui.adsabs.harvard.edu/abs/2011ApJ...729...16K} {729, 16}

\bibitem[\protect\citeauthoryear{{Koch}, {Grebel}  \& {Martell}}{{Koch}
  et~al.}{2019}]{Koch2019}
{Koch} A.,  {Grebel} E.~K.,   {Martell} S.~L.,  2019, \mn@doi [\aap]
  {10.1051/0004-6361/201834825}, \href
  {https://ui.adsabs.harvard.edu/abs/2019A&A...625A..75K} {625, A75}

\bibitem[\protect\citeauthoryear{{Koppelman}, {Helmi}, {Massari},
  {Price-Whelan}  \& {Starkenburg}}{{Koppelman} et~al.}{2019}]{Koppelman2019}
{Koppelman} H.~H.,  {Helmi} A.,  {Massari} D.,  {Price-Whelan} A.~M.,
  {Starkenburg} T.~K.,  2019, \mn@doi [\aap] {10.1051/0004-6361/201936738},
  \href {https://ui.adsabs.harvard.edu/abs/2019A&A...631L...9K} {631, L9}

\bibitem[\protect\citeauthoryear{{Koppelman}, {Bos}  \& {Helmi}}{{Koppelman}
  et~al.}{2020}]{Koppelman2020}
{Koppelman} H.~H.,  {Bos} R. O.~Y.,   {Helmi} A.,  2020, arXiv e-prints, \href
  {https://ui.adsabs.harvard.edu/abs/2020arXiv200607620K} {p. arXiv:2006.07620}

\bibitem[\protect\citeauthoryear{{Kroupa}}{{Kroupa}}{2001}]{Kroupa2001}
{Kroupa} P.,  2001, \mn@doi [\mnras] {10.1046/j.1365-8711.2001.04022.x}, \href
  {https://ui.adsabs.harvard.edu/abs/2001MNRAS.322..231K} {322, 231}

\bibitem[\protect\citeauthoryear{{Kruijssen}, {Pfeffer}, {Crain}  \&
  {Bastian}}{{Kruijssen} et~al.}{2019a}]{Kruijssen2019a}
{Kruijssen} J.~M.~D.,  {Pfeffer} J.~L.,  {Crain} R.~A.,   {Bastian} N.,  2019a,
  \mn@doi [\mnras] {10.1093/mnras/stz968}, \href
  {https://ui.adsabs.harvard.edu/abs/2019MNRAS.486.3134K} {486, 3134}

\bibitem[\protect\citeauthoryear{{Kruijssen}, {Pfeffer}, {Reina-Campos},
  {Crain}  \& {Bastian}}{{Kruijssen} et~al.}{2019b}]{Kruijssen2019b}
{Kruijssen} J.~M.~D.,  {Pfeffer} J.~L.,  {Reina-Campos} M.,  {Crain} R.~A.,
  {Bastian} N.,  2019b, \mn@doi [\mnras] {10.1093/mnras/sty1609}, \href
  {https://ui.adsabs.harvard.edu/abs/2019MNRAS.486.3180K} {486, 3180}

\bibitem[\protect\citeauthoryear{{Kruijssen} et~al.,}{{Kruijssen}
  et~al.}{2020}]{Kruijssen2020}
{Kruijssen} J.~M.~D.,  et~al., 2020, arXiv e-prints, \href
  {https://ui.adsabs.harvard.edu/abs/2020arXiv200301119K} {p. arXiv:2003.01119}

\bibitem[\protect\citeauthoryear{{Kunder} et~al.,}{{Kunder}
  et~al.}{2012}]{Kunder2012}
{Kunder} A.,  et~al., 2012, \mn@doi [\aj] {10.1088/0004-6256/143/3/57}, \href
  {https://ui.adsabs.harvard.edu/abs/2012AJ....143...57K} {143, 57}

\bibitem[\protect\citeauthoryear{{Kunder} et~al.,}{{Kunder}
  et~al.}{2020}]{Kunder2020}
{Kunder} A.,  et~al., 2020, \mn@doi [\aj] {10.3847/1538-3881/ab8d35}, \href
  {https://ui.adsabs.harvard.edu/abs/2020AJ....159..270K} {159, 270}

\bibitem[\protect\citeauthoryear{{Leung} \& {Bovy}}{{Leung} \&
  {Bovy}}{2019a}]{Leung2019a}
{Leung} H.~W.,  {Bovy} J.,  2019a, \mn@doi [\mnras] {10.1093/mnras/sty3217},
  \href {https://ui.adsabs.harvard.edu/abs/2019MNRAS.483.3255L} {483, 3255}

\bibitem[\protect\citeauthoryear{{Leung} \& {Bovy}}{{Leung} \&
  {Bovy}}{2019b}]{Leung2019b}
{Leung} H.~W.,  {Bovy} J.,  2019b, \mn@doi [\mnras] {10.1093/mnras/stz2245},
  \href {https://ui.adsabs.harvard.edu/abs/2019MNRAS.489.2079L} {489, 2079}

\bibitem[\protect\citeauthoryear{{Mackereth} \& {Bovy}}{{Mackereth} \&
  {Bovy}}{2018}]{MackerethBovy2018}
{Mackereth} J.~T.,  {Bovy} J.,  2018, \mn@doi [\pasp]
  {10.1088/1538-3873/aadcdd}, \href
  {https://ui.adsabs.harvard.edu/abs/2018PASP..130k4501M} {130, 114501}

\bibitem[\protect\citeauthoryear{{Mackereth} \& {Bovy}}{{Mackereth} \&
  {Bovy}}{2020}]{MackerethBovy2020}
{Mackereth} J.~T.,  {Bovy} J.,  2020, \mn@doi [\mnras] {10.1093/mnras/staa047},
  \href {https://ui.adsabs.harvard.edu/abs/2020MNRAS.492.3631M} {492, 3631}

\bibitem[\protect\citeauthoryear{{Mackereth} et~al.,}{{Mackereth}
  et~al.}{2017}]{Mackereth2017}
{Mackereth} J.~T.,  et~al., 2017, \mn@doi [\mnras] {10.1093/mnras/stx1774},
  \href {https://ui.adsabs.harvard.edu/abs/2017MNRAS.471.3057M} {471, 3057}

\bibitem[\protect\citeauthoryear{{Mackereth}, {Crain}, {Schiavon}, {Schaye},
  {Theuns}  \& {Schaller}}{{Mackereth} et~al.}{2018}]{Mackereth2018}
{Mackereth} J.~T.,  {Crain} R.~A.,  {Schiavon} R.~P.,  {Schaye} J.,  {Theuns}
  T.,   {Schaller} M.,  2018, \mn@doi [\mnras] {10.1093/mnras/sty972}, \href
  {https://ui.adsabs.harvard.edu/abs/2018MNRAS.477.5072M} {477, 5072}

\bibitem[\protect\citeauthoryear{{Mackereth} et~al.,}{{Mackereth}
  et~al.}{2019}]{Mackereth2019}
{Mackereth} J.~T.,  et~al., 2019, \mn@doi [\mnras] {10.1093/mnras/sty2955},
  \href {https://ui.adsabs.harvard.edu/abs/2019MNRAS.482.3426M} {482, 3426}

\bibitem[\protect\citeauthoryear{{Majewski}, {Skrutskie}, {Weinberg}  \&
  {Ostheimer}}{{Majewski} et~al.}{2003}]{Majewski2003}
{Majewski} S.~R.,  {Skrutskie} M.~F.,  {Weinberg} M.~D.,   {Ostheimer} J.~C.,
  2003, \mn@doi [\apj] {10.1086/379504}, \href
  {https://ui.adsabs.harvard.edu/abs/2003ApJ...599.1082M} {599, 1082}

\bibitem[\protect\citeauthoryear{{Majewski} et~al.,}{{Majewski}
  et~al.}{2017}]{Majewski2017}
{Majewski} S.~R.,  et~al., 2017, \mn@doi [\aj] {10.3847/1538-3881/aa784d},
  \href {https://ui.adsabs.harvard.edu/abs/2017AJ....154...94M} {154, 94}

\bibitem[\protect\citeauthoryear{{Maoz}, {Mannucci}  \& {Brandt}}{{Maoz}
  et~al.}{2012}]{Maoz2012}
{Maoz} D.,  {Mannucci} F.,   {Brandt} T.~D.,  2012, \mn@doi [\mnras]
  {10.1111/j.1365-2966.2012.21871.x}, \href
  {https://ui.adsabs.harvard.edu/abs/2012MNRAS.426.3282M} {426, 3282}

\bibitem[\protect\citeauthoryear{{Martell} \& {Grebel}}{{Martell} \&
  {Grebel}}{2010}]{Martell2010}
{Martell} S.~L.,  {Grebel} E.~K.,  2010, \mn@doi [\aap]
  {10.1051/0004-6361/201014135}, \href
  {https://ui.adsabs.harvard.edu/abs/2010A&A...519A..14M} {519, A14}

\bibitem[\protect\citeauthoryear{{Martell} et~al.,}{{Martell}
  et~al.}{2016}]{Martell2016}
{Martell} S.~L.,  et~al., 2016, \mn@doi [\apj] {10.3847/0004-637X/825/2/146},
  \href {https://ui.adsabs.harvard.edu/abs/2016ApJ...825..146M} {825, 146}

\bibitem[\protect\citeauthoryear{{Martell} et~al.,}{{Martell}
  et~al.}{2017}]{GALAH}
{Martell} S.~L.,  et~al., 2017, \mn@doi [\mnras] {10.1093/mnras/stw2835}, \href
  {https://ui.adsabs.harvard.edu/abs/2017MNRAS.465.3203M} {465, 3203}

\bibitem[\protect\citeauthoryear{{Massari}, {Koppelman}  \& {Helmi}}{{Massari}
  et~al.}{2019}]{Massari2019}
{Massari} D.,  {Koppelman} H.~H.,   {Helmi} A.,  2019, \mn@doi [\aap]
  {10.1051/0004-6361/201936135}, \href
  {https://ui.adsabs.harvard.edu/abs/2019A&A...630L...4M} {630, L4}

\bibitem[\protect\citeauthoryear{{Matteucci}, {Grisoni}, {Spitoni},
  {Zulianello}, {Rojas-Arriagada}, {Schultheis}  \& {Ryde}}{{Matteucci}
  et~al.}{2019}]{Matteucci2019}
{Matteucci} F.,  {Grisoni} V.,  {Spitoni} E.,  {Zulianello} A.,
  {Rojas-Arriagada} A.,  {Schultheis} M.,   {Ryde} N.,  2019, \mn@doi [\mnras]
  {10.1093/mnras/stz1647}, \href
  {https://ui.adsabs.harvard.edu/abs/2019MNRAS.487.5363M} {487, 5363}

\bibitem[\protect\citeauthoryear{{Matteucci}, {Vasini}, {Grisoni}  \&
  {Schultheis}}{{Matteucci} et~al.}{2020}]{Matteucci2020}
{Matteucci} F.,  {Vasini} A.,  {Grisoni} V.,   {Schultheis} M.,  2020, arXiv
  e-prints, \href {https://ui.adsabs.harvard.edu/abs/2020arXiv200410133M} {p.
  arXiv:2004.10133}

\bibitem[\protect\citeauthoryear{{McMillan}}{{McMillan}}{2017}]{McMillan2017}
{McMillan} P.~J.,  2017, \mn@doi [\mnras] {10.1093/mnras/stw2759}, \href
  {https://ui.adsabs.harvard.edu/abs/2017MNRAS.465...76M} {465, 76}

\bibitem[\protect\citeauthoryear{{McWilliam} \& {Rich}}{{McWilliam} \&
  {Rich}}{1994}]{McWilliam1994}
{McWilliam} A.,  {Rich} R.~M.,  1994, \mn@doi [\apjs] {10.1086/191954}, \href
  {https://ui.adsabs.harvard.edu/abs/1994ApJS...91..749M} {91, 749}

\bibitem[\protect\citeauthoryear{{Minniti}}{{Minniti}}{1996}]{Minniti1996}
{Minniti} D.,  1996, \mn@doi [\apj] {10.1086/176879}, \href
  {https://ui.adsabs.harvard.edu/abs/1996ApJ...459..175M} {459, 175}

\bibitem[\protect\citeauthoryear{{Molloy}, {Smith}, {Evans}  \&
  {Shen}}{{Molloy} et~al.}{2015}]{Molloy2015}
{Molloy} M.,  {Smith} M.~C.,  {Evans} N.~W.,   {Shen} J.,  2015, \mn@doi [\apj]
  {10.1088/0004-637X/812/2/146}, \href
  {https://ui.adsabs.harvard.edu/abs/2015ApJ...812..146M} {812, 146}

\bibitem[\protect\citeauthoryear{{Myeong}, {Vasiliev}, {Iorio}, {Evans}  \&
  {Belokurov}}{{Myeong} et~al.}{2019}]{Myeong2019}
{Myeong} G.~C.,  {Vasiliev} E.,  {Iorio} G.,  {Evans} N.~W.,   {Belokurov} V.,
  2019, \mn@doi [\mnras] {10.1093/mnras/stz1770}, \href
  {https://ui.adsabs.harvard.edu/abs/2019MNRAS.488.1235M} {488, 1235}

\bibitem[\protect\citeauthoryear{{Naab} \& {Ostriker}}{{Naab} \&
  {Ostriker}}{2017}]{NaabOstriker2017}
{Naab} T.,  {Ostriker} J.~P.,  2017, \mn@doi [\araa]
  {10.1146/annurev-astro-081913-040019}, \href
  {https://ui.adsabs.harvard.edu/abs/2017ARA&A..55...59N} {55, 59}

\bibitem[\protect\citeauthoryear{{Naidu}, {Conroy}, {Bonaca}, {Johnson},
  {Ting}, {Caldwell}, {Zaritsky}  \& {Cargile}}{{Naidu}
  et~al.}{2020}]{Naidu2020}
{Naidu} R.~P.,  {Conroy} C.,  {Bonaca} A.,  {Johnson} B.~D.,  {Ting} Y.-S.,
  {Caldwell} N.,  {Zaritsky} D.,   {Cargile} P.~A.,  2020, arXiv e-prints,
  \href {https://ui.adsabs.harvard.edu/abs/2020arXiv200608625N} {p.
  arXiv:2006.08625}

\bibitem[\protect\citeauthoryear{{Nataf}}{{Nataf}}{2017}]{Nataf2017}
{Nataf} D.~M.,  2017, \mn@doi [\pasa] {10.1017/pasa.2017.32}, \href
  {https://ui.adsabs.harvard.edu/abs/2017PASA...34...41N} {34, e041}

\bibitem[\protect\citeauthoryear{{Ness} et~al.,}{{Ness}
  et~al.}{2013a}]{Ness2013a}
{Ness} M.,  et~al., 2013a, \mn@doi [\mnras] {10.1093/mnras/sts629}, \href
  {https://ui.adsabs.harvard.edu/abs/2013MNRAS.430..836N} {430, 836}

\bibitem[\protect\citeauthoryear{{Ness} et~al.,}{{Ness}
  et~al.}{2013b}]{Ness2013b}
{Ness} M.,  et~al., 2013b, \mn@doi [\mnras] {10.1093/mnras/stt533}, \href
  {https://ui.adsabs.harvard.edu/abs/2013MNRAS.432.2092N} {432, 2092}

\bibitem[\protect\citeauthoryear{{Nidever} et~al.,}{{Nidever}
  et~al.}{2012}]{Nidever2012}
{Nidever} D.~L.,  et~al., 2012, \mn@doi [\apjl] {10.1088/2041-8205/755/2/L25},
  \href {https://ui.adsabs.harvard.edu/abs/2012ApJ...755L..25N} {755, L25}

\bibitem[\protect\citeauthoryear{{Nidever} et~al.,}{{Nidever}
  et~al.}{2014}]{Nidever2014}
{Nidever} D.~L.,  et~al., 2014, \mn@doi [\apj] {10.1088/0004-637X/796/1/38},
  \href {https://ui.adsabs.harvard.edu/abs/2014ApJ...796...38N} {796, 38}

\bibitem[\protect\citeauthoryear{{Nidever} et~al.,}{{Nidever}
  et~al.}{2015}]{Nidever2015}
{Nidever} D.~L.,  et~al., 2015, \mn@doi [\aj] {10.1088/0004-6256/150/6/173},
  \href {https://ui.adsabs.harvard.edu/abs/2015AJ....150..173N} {150, 173}

\bibitem[\protect\citeauthoryear{{Nissen} \& {Schuster}}{{Nissen} \&
  {Schuster}}{2010}]{NissenSchuster2010}
{Nissen} P.~E.,  {Schuster} W.~J.,  2010, \mn@doi [\aap]
  {10.1051/0004-6361/200913877}, \href
  {https://ui.adsabs.harvard.edu/abs/2010A&A...511L..10N} {511, L10}

\bibitem[\protect\citeauthoryear{{Nomoto}, {Kobayashi}  \& {Tominaga}}{{Nomoto}
  et~al.}{2013}]{Nomoto2013}
{Nomoto} K.,  {Kobayashi} C.,   {Tominaga} N.,  2013, \mn@doi [\araa]
  {10.1146/annurev-astro-082812-140956}, \href
  {https://ui.adsabs.harvard.edu/abs/2013ARA&A..51..457N} {51, 457}

\bibitem[\protect\citeauthoryear{Oliphant}{Oliphant}{06  }]{NumPy}
Oliphant T.,  2006--, {NumPy}: A guide to {NumPy}, USA: Trelgol Publishing,
  \url {http://www.numpy.org/}

\bibitem[\protect\citeauthoryear{{Pfeffer}, {Kruijssen}, {Crain}  \&
  {Bastian}}{{Pfeffer} et~al.}{2018}]{Pfeffer2018}
{Pfeffer} J.,  {Kruijssen} J.~M.~D.,  {Crain} R.~A.,   {Bastian} N.,  2018,
  \mn@doi [\mnras] {10.1093/mnras/stx3124}, \href
  {https://ui.adsabs.harvard.edu/abs/2018MNRAS.475.4309P} {475, 4309}

\bibitem[\protect\citeauthoryear{{Pfeffer}, {Trujillo-Gomez}, {Kruijssen},
  {Crain}, {Hughes}, {Reina-Campos}  \& {Bastian}}{{Pfeffer}
  et~al.}{2020}]{Pfeffer2020}
{Pfeffer} J.~L.,  {Trujillo-Gomez} S.,  {Kruijssen} J.~M.~D.,  {Crain} R.~A.,
  {Hughes} M.~E.,  {Reina-Campos} M.,   {Bastian} N.,  2020, arXiv e-prints,
  \href {https://ui.adsabs.harvard.edu/abs/2020arXiv200300076P} {p.
  arXiv:2003.00076}

\bibitem[\protect\citeauthoryear{{Pietrukowicz} et~al.,}{{Pietrukowicz}
  et~al.}{2015}]{Pietrukowicz2015}
{Pietrukowicz} P.,  et~al., 2015, \mn@doi [\apj] {10.1088/0004-637X/811/2/113},
  \href {https://ui.adsabs.harvard.edu/abs/2015ApJ...811..113P} {811, 113}

\bibitem[\protect\citeauthoryear{{Price-Whelan} et~al.,}{{Price-Whelan}
  et~al.}{2018}]{astropy:2018}
{Price-Whelan} A.~M.,  et~al., 2018, \mn@doi [aj] {10.3847/1538-3881/aabc4f},
  \href {https://ui.adsabs.harvard.edu/#abs/2018AJ....156..123T} {156, 123}

\bibitem[\protect\citeauthoryear{{Queiroz} et~al.,}{{Queiroz}
  et~al.}{2020a}]{Queiroz2020b}
{Queiroz} A.~B.~A.,  et~al., 2020a, arXiv e-prints, \href
  {https://ui.adsabs.harvard.edu/abs/2020arXiv200712915Q} {p. arXiv:2007.12915}

\bibitem[\protect\citeauthoryear{{Queiroz} et~al.,}{{Queiroz}
  et~al.}{2020b}]{Queiroz2020a}
{Queiroz} A.~B.~A.,  et~al., 2020b, \mn@doi [\aap]
  {10.1051/0004-6361/201937364}, \href
  {https://ui.adsabs.harvard.edu/abs/2020A&A...638A..76Q} {638, A76}

\bibitem[\protect\citeauthoryear{{Rich}}{{Rich}}{1988}]{Rich1988}
{Rich} R.~M.,  1988, \mn@doi [\aj] {10.1086/114681}, \href
  {https://ui.adsabs.harvard.edu/abs/1988AJ.....95..828R} {95, 828}

\bibitem[\protect\citeauthoryear{{Rich}}{{Rich}}{2013}]{Rich2013}
{Rich} R.~M.,  2013, {The Galactic Bulge}.
p.~271, \mn@doi{10.1007/978-94-007-5612-0_6}

\bibitem[\protect\citeauthoryear{{Rich}, {Origlia}  \& {Valenti}}{{Rich}
  et~al.}{2012}]{RichOriglia2012}
{Rich} R.~M.,  {Origlia} L.,   {Valenti} E.,  2012, \mn@doi [\apj]
  {10.1088/0004-637X/746/1/59}, \href
  {https://ui.adsabs.harvard.edu/abs/2012ApJ...746...59R} {746, 59}

\bibitem[\protect\citeauthoryear{{Rojas-Arriagada} et~al.,}{{Rojas-Arriagada}
  et~al.}{2014}]{Rojas2014}
{Rojas-Arriagada} A.,  et~al., 2014, \mn@doi [\aap]
  {10.1051/0004-6361/201424121}, \href
  {https://ui.adsabs.harvard.edu/abs/2014A&A...569A.103R} {569, A103}

\bibitem[\protect\citeauthoryear{{Rojas-Arriagada} et~al.,}{{Rojas-Arriagada}
  et~al.}{2017}]{Rojas2017}
{Rojas-Arriagada} A.,  et~al., 2017, \mn@doi [\aap]
  {10.1051/0004-6361/201629160}, \href
  {https://ui.adsabs.harvard.edu/abs/2017A&A...601A.140R} {601, A140}

\bibitem[\protect\citeauthoryear{{Rojas-Arriagada}, {Zoccali}, {Schultheis},
  {Recio-Blanco}, {Zasowski}, {Minniti}, {J{\"o}nsson}  \&
  {Cohen}}{{Rojas-Arriagada} et~al.}{2019}]{Rojas2019}
{Rojas-Arriagada} A.,  {Zoccali} M.,  {Schultheis} M.,  {Recio-Blanco} A.,
  {Zasowski} G.,  {Minniti} D.,  {J{\"o}nsson} H.,   {Cohen} R.~E.,  2019,
  \mn@doi [\aap] {10.1051/0004-6361/201834126}, \href
  {https://ui.adsabs.harvard.edu/abs/2019A&A...626A..16R} {626, A16}

\bibitem[\protect\citeauthoryear{{Rojas-Arriagada} et~al.,}{{Rojas-Arriagada}
  et~al.}{2020}]{Rojas2020}
{Rojas-Arriagada} A.,  et~al., 2020, \mn@doi [\mnras] {10.1093/mnras/staa2807},
  \href {https://ui.adsabs.harvard.edu/abs/2020MNRAS.tmp.2647R} {}

\bibitem[\protect\citeauthoryear{{Ryde}, {Schultheis}, {Grieco}, {Matteucci},
  {Rich}  \& {Uttenthaler}}{{Ryde} et~al.}{2016}]{Ryde2016}
{Ryde} N.,  {Schultheis} M.,  {Grieco} V.,  {Matteucci} F.,  {Rich} R.~M.,
  {Uttenthaler} S.,  2016, \mn@doi [\aj] {10.3847/0004-6256/151/1/1}, \href
  {https://ui.adsabs.harvard.edu/abs/2016AJ....151....1R} {151, 1}

\bibitem[\protect\citeauthoryear{{Savino} \& {Posti}}{{Savino} \&
  {Posti}}{2019}]{Savino2019}
{Savino} A.,  {Posti} L.,  2019, \mn@doi [\aap] {10.1051/0004-6361/201935417},
  \href {https://ui.adsabs.harvard.edu/abs/2019A&A...624L...9S} {624, L9}

\bibitem[\protect\citeauthoryear{{Savino}, {Koch}, {Prudil}, {Kunder}  \&
  {Smolec}}{{Savino} et~al.}{2020}]{Savino2020}
{Savino} A.,  {Koch} A.,  {Prudil} Z.,  {Kunder} A.,   {Smolec} R.,  2020,
  arXiv e-prints, \href {https://ui.adsabs.harvard.edu/abs/2020arXiv200612507S}
  {p. arXiv:2006.12507}

\bibitem[\protect\citeauthoryear{{Schaye} et~al.,}{{Schaye}
  et~al.}{2015}]{Schaye2015}
{Schaye} J.,  et~al., 2015, \mn@doi [\mnras] {10.1093/mnras/stu2058}, \href
  {https://ui.adsabs.harvard.edu/abs/2015MNRAS.446..521S} {446, 521}

\bibitem[\protect\citeauthoryear{{Schiavon} et~al.,}{{Schiavon}
  et~al.}{2017}]{Schiavon2017}
{Schiavon} R.~P.,  et~al., 2017, \mn@doi [\mnras] {10.1093/mnras/stw2162},
  \href {https://ui.adsabs.harvard.edu/abs/2017MNRAS.465..501S} {465, 501}

\bibitem[\protect\citeauthoryear{{Schiavon}, {Mackereth}, {Pfeffer}, {Crain}
  \& {Bovy}}{{Schiavon} et~al.}{2020}]{Schiavon2020}
{Schiavon} R.~P.,  {Mackereth} J.~T.,  {Pfeffer} J.,  {Crain} R.~A.,   {Bovy}
  J.,  2020, in {Bragaglia} A.,  {Davies} M.,  {Sills} A.,   {Vesperini} E.,
  eds,  IAU Symposium Vol. 351, IAU Symposium. pp 170--173 (\mn@eprint {arXiv}
  {2002.08380}), \mn@doi{10.1017/S1743921319007889}

\bibitem[\protect\citeauthoryear{{Schultheis} et~al.,}{{Schultheis}
  et~al.}{2017}]{Schultheis2017}
{Schultheis} M.,  et~al., 2017, \mn@doi [\aap] {10.1051/0004-6361/201630154},
  \href {https://ui.adsabs.harvard.edu/abs/2017A&A...600A..14S} {600, A14}

\bibitem[\protect\citeauthoryear{{Searle} \& {Zinn}}{{Searle} \&
  {Zinn}}{1978}]{SearleZinn1978}
{Searle} L.,  {Zinn} R.,  1978, \mn@doi [\apj] {10.1086/156499}, \href
  {https://ui.adsabs.harvard.edu/abs/1978ApJ...225..357S} {225, 357}

\bibitem[\protect\citeauthoryear{{Shetrone}, {Venn}, {Tolstoy}, {Primas},
  {Hill}  \& {Kaufer}}{{Shetrone} et~al.}{2003}]{Shetrone2003}
{Shetrone} M.,  {Venn} K.~A.,  {Tolstoy} E.,  {Primas} F.,  {Hill} V.,
  {Kaufer} A.,  2003, \mn@doi [\aj] {10.1086/345966}, \href
  {https://ui.adsabs.harvard.edu/abs/2003AJ....125..684S} {125, 684}

\bibitem[\protect\citeauthoryear{{Stott} et~al.,}{{Stott}
  et~al.}{2014}]{Stott2014}
{Stott} J.~P.,  et~al., 2014, \mn@doi [\mnras] {10.1093/mnras/stu1343}, \href
  {https://ui.adsabs.harvard.edu/abs/2014MNRAS.443.2695S} {443, 2695}

\bibitem[\protect\citeauthoryear{{Taylor}}{{Taylor}}{2005}]{Taylor2005}
{Taylor} M.~B.,  2005, in {Shopbell} P.,  {Britton} M.,   {Ebert} R.,  eds,
  Astronomical Society of the Pacific Conference Series Vol. 347, Astronomical
  Data Analysis Software and Systems XIV. p.~29

\bibitem[\protect\citeauthoryear{{Tremaine}, {Ostriker}  \&
  {Spitzer}}{{Tremaine} et~al.}{1975}]{Tremaine1975}
{Tremaine} S.~D.,  {Ostriker} J.~P.,   {Spitzer} L. J.,  1975, \mn@doi [\apj]
  {10.1086/153422}, \href
  {https://ui.adsabs.harvard.edu/abs/1975ApJ...196..407T} {196, 407}

\bibitem[\protect\citeauthoryear{{Tumlinson}}{{Tumlinson}}{2010}]{Tumlinson2010}
{Tumlinson} J.,  2010, \mn@doi [\apj] {10.1088/0004-637X/708/2/1398}, \href
  {https://ui.adsabs.harvard.edu/abs/2010ApJ...708.1398T} {708, 1398}

\bibitem[\protect\citeauthoryear{{Wang} et~al.,}{{Wang}
  et~al.}{2019}]{Wang2019}
{Wang} X.,  et~al., 2019, \mn@doi [\apj] {10.3847/1538-4357/ab3861}, \href
  {https://ui.adsabs.harvard.edu/abs/2019ApJ...882...94W} {882, 94}

\bibitem[\protect\citeauthoryear{{Whitten} et~al.,}{{Whitten}
  et~al.}{2019}]{Whitten2019}
{Whitten} D.~D.,  et~al., 2019, \mn@doi [\apj] {10.3847/1538-4357/ab4269},
  \href {https://ui.adsabs.harvard.edu/abs/2019ApJ...884...67W} {884, 67}

\bibitem[\protect\citeauthoryear{{Wilson} et~al.,}{{Wilson}
  et~al.}{2019}]{Wilson2019}
{Wilson} J.~C.,  et~al., 2019, \mn@doi [\pasp] {10.1088/1538-3873/ab0075},
  \href {https://ui.adsabs.harvard.edu/abs/2019PASP..131e5001W} {131, 055001}

\bibitem[\protect\citeauthoryear{{Zasowski}, {Ness}, {Garc{\'\i}a P{\'e}rez},
  {Martinez-Valpuesta}, {Johnson}  \& {Majewski}}{{Zasowski}
  et~al.}{2016}]{Zasowski2016}
{Zasowski} G.,  {Ness} M.~K.,  {Garc{\'\i}a P{\'e}rez} A.~E.,
  {Martinez-Valpuesta} I.,  {Johnson} J.~A.,   {Majewski} S.~R.,  2016, \mn@doi
  [\apj] {10.3847/0004-637X/832/2/132}, \href
  {https://ui.adsabs.harvard.edu/abs/2016ApJ...832..132Z} {832, 132}

\bibitem[\protect\citeauthoryear{{Zasowski} et~al.,}{{Zasowski}
  et~al.}{2017}]{Zasowski2017}
{Zasowski} G.,  et~al., 2017, \mn@doi [\aj] {10.3847/1538-3881/aa8df9}, \href
  {https://ui.adsabs.harvard.edu/abs/2017AJ....154..198Z} {154, 198}

\bibitem[\protect\citeauthoryear{{Zasowski} et~al.,}{{Zasowski}
  et~al.}{2019}]{Zasowski2019}
{Zasowski} G.,  et~al., 2019, \mn@doi [\apj] {10.3847/1538-4357/aaeff4}, \href
  {https://ui.adsabs.harvard.edu/abs/2019ApJ...870..138Z} {870, 138}

\bibitem[\protect\citeauthoryear{{Zoccali} et~al.,}{{Zoccali}
  et~al.}{2006}]{Zoccali2006}
{Zoccali} M.,  et~al., 2006, \mn@doi [\aap] {10.1051/0004-6361:20065659}, \href
  {https://ui.adsabs.harvard.edu/abs/2006A&A...457L...1Z} {457, L1}

\makeatother
\end{thebibliography}


\appendix

\section{High- and Low-$\alpha$ disks in action space}

\begin{figure}
	\includegraphics[width=\columnwidth]{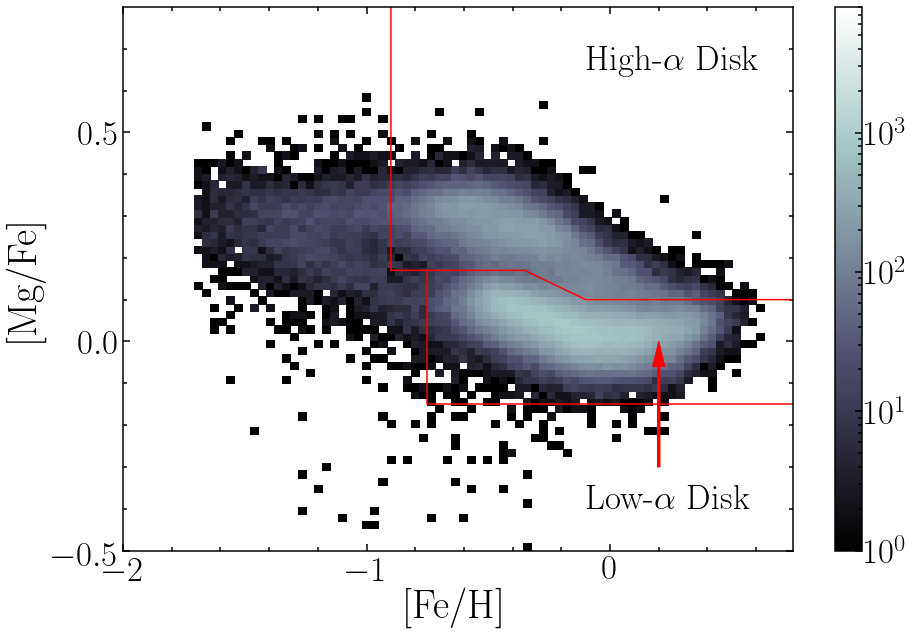}
\caption{Definition of high- and low-$\alpha$ disks on the Mg-Fe plane.  We
added a $\pm$0.04~dex ``cushion'' in [Mg/Fe] around the dividing line in
order to minimize inter-contamination between the two samples.
}
    \label{fig:alpha_disk}
\end{figure}

In this appendix we briefly present the distribution of disk stars
in action space, to inform the discussion of accreted and {\it in
situ} populations in Section~\ref{sec:subiom}.  We start by
distinguishing high- and low-$\alpha$ disk stars.  For that purpose
we display our parent sample on the Mg-Fe plane in
Figure~\ref{fig:alpha_disk}, where the lines defining each sub-sample
are drawn on top of the data.  Metal-poor stars are removed in order
to minimize contamination by halo stars.

\begin{figure}
	\includegraphics[width=\columnwidth]{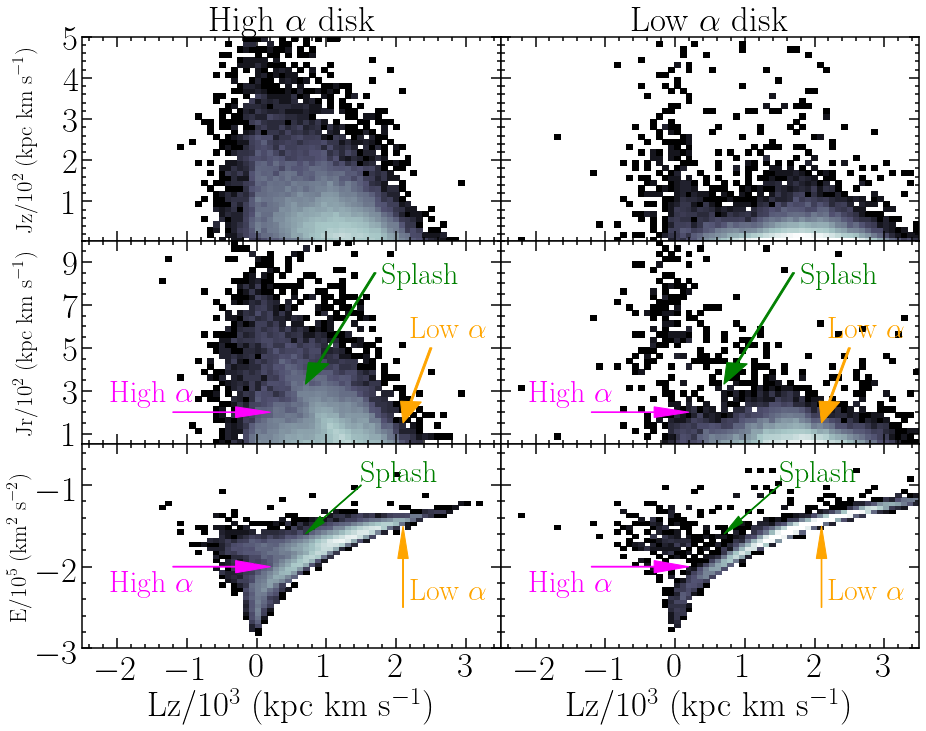}
\caption{Distribution of high- and low-$\alpha$ disk populations
in action space.  The orange arrow indicates the rough position of
the low-$\alpha$/thin disk populations, the magenta arrow that of the
standard high-$\alpha$/thick disk populations, and the green arrow
that of stars associated with the {\it Splash}.  }
    \label{fig:action_alpha}
\end{figure}

High- and low-$\alpha$ disk stars are displayed in action and energy
space in Figure~\ref{fig:action_alpha}.  The arrows indicate the
approximate loci of stars belonging to the each sub-sample.  The
orange arrow points to the location of low-$\alpha$ disk stars
\citep[e.g.,][]{Bovy2012,Bensby2014,Nidever2014,Hayden2015,Mackereth2017,Queiroz2020a},
which are characterised by high angular momentum, small energy
scatter, and low radial ($J_{\rm r}$) and vertical ($J_{\rm z}$)
actions.  There are, however, outliers towards higher $E$, $J_{r}$,
and $J{\rm z}$, which are due to contaminants from the GE/S system
and possibly other accreted systems.

The high-$\alpha$ disk is dominated by two major structures.  For
constant $L_{\rm z}$, these two high-$\alpha$ subgroups occupy
significantly different and well separated values of $J_{\rm r}$
and energy.  The most important for the purposes of this paper is
strongly rotational, overlapping substantially with the low-$\alpha$
disk at high $L_{\rm z}$, but also extending towards $L_{\rm z}\sim
0$ and presenting larger scatter towards higher $E$, $J_{r}$, and
$J_{\rm z}$ (magenta arrow).  The other component presents higher
energy at $L_{\rm z}/10^3\simless1$~kpc~km~s$^{-1}$ and merges with
the other components at higher $L_{\rm z}$, being associated
with the {\it Splash} population \cite[see][]{Belokurov2019} (green
arrow).

\bsp	
\label{lastpage}
\end{document}